\documentclass{aa}
\usepackage[varg]{txfonts}
\usepackage{natbib}
\usepackage{graphicx}
\usepackage{mathabx}
\usepackage{amsmath}

\usepackage{color}
\usepackage{multirow}
\usepackage{longtable}

\bibpunct{(}{)}{;}{a}{}{,}

\begin{document}

\title{CSI~2264: Investigating rotation and its connection with disk accretion in the young open cluster NGC~2264\thanks{Based on observations obtained with the {\it CoRoT} space telescope, and with the wide-field imager MegaCam at the Canada-France-Hawaii Telescope (CFHT).}\fnmsep\thanks{Table 4, reported in its entirety after the References, is also available in electronic form at the CDS via anonymous ftp to cdsarc.u-strasbg.fr (130.79.128.5) or via http://cdsweb.u-strasbg.fr/cgi-bin/qcat?J/A+A/}}

\author{L. Venuti\inst{1,2} \and J. Bouvier\inst{2} \and A.~M. Cody\inst{3} \and J.~R. Stauffer\inst{4} \and G. Micela\inst{1} \and L.~M. Rebull\inst{4} \and S.~H.~P. Alencar\inst{5} \and A.~P. Sousa\inst{5} \and L.~A. Hillenbrand\inst{6} \and E. Flaccomio\inst{1}}

\institute{Istituto Nazionale di Astrofisica, Osservatorio Astronomico di Palermo G.\,S. Vaiana, Piazza del Parlamento 1, 90134 Palermo, Italy\\ e-mail: lvenuti@astropa.unipa.it \and Univ. Grenoble Alpes, CNRS, IPAG, F-38000 Grenoble, France \and NASA Ames Research Center, Kepler Science Office, Mountain View, CA 94035, USA \and Spitzer Science Center, California Institute of Technology, 1200 E California Blvd., Pasadena, CA 91125, USA \and Departamento de F\'isica - ICEx - UFMG, Av. Ant\^onio Carlos, 6627, 30270-901 Belo Horizonte, MG, Brazil \and Astronomy Department, California Institute of Technology, Pasadena, CA 91125, USA}
\date{Received 16 August 2016 / Accepted 25 October 2016}

\abstract {The low spin rates measured for solar-type stars at an age of a few Myr ($\sim$10\% of the break-up velocity) indicate that some mechanism of angular momentum regulation must be at play in the early pre-main sequence. This may be associated with magnetospheric accretion and star-disk interaction, as suggested by observations that disk-bearing objects (CTTS) are slower rotators than diskless sources (WTTS) in young star clusters.}{We characterize the rotation properties for members of the star-forming region NGC~2264 ($\sim$3~Myr) as a function of mass, and investigate the accretion-rotation connection at an age where about 50\% of the stars have already lost their disks.}{We examined a sample of 500 cluster members (40\% with disks, 60\% without disks), distributed in mass between $\sim$0.15 and 2~M$_\odot$, whose photometric variations were monitored in the optical for 38 consecutive days with the {\it CoRoT} space observatory. Light curves were analyzed for periodicity using three different techniques: the Lomb-Scargle periodogram, the autocorrelation function and the string-length method. Periods were searched in the range between 0.17~days (i.e., 4~hours, twice the data sampling adopted) and 19~days (half the total time span). Period detections were confirmed using a variety of statistical tools (false alarm probability, Q-statistics), as well as visual inspection of the direct and phase-folded light curves.}{About 62\% of sources in our sample were found to be periodic; the period detection rate is 70\% among WTTS and 58\% among CTTS. The vast majority of periodic sources exhibit rotational periods shorter than 13~d. The period distribution obtained for the cluster consists of a smooth distribution centered around P=5.2~d with two peaks, located respectively at P=1-2~d and at P=3-4~d. A separate analysis of the rotation properties for CTTS and WTTS indicates that the P=1-2~d peak is associated with the latter, while both groups contribute to the P=3-4~d peak. The comparison between CTTS and WTTS supports the idea of a rotation-accretion connection: their respective rotational properties are statistically different, and CTTS rotate on average more slowly than WTTS. We also observe that CTTS with the strongest signatures of accretion (largest UV flux excesses) tend to exhibit slow rotation rates; a clear dearth of fast rotators with strong accretion signatures emerges from our sample. This connection between rotation properties and accretion traced via UV excess measurements is consistent with earlier findings, revealed by IR excess measurements, that fast rotators in young star clusters are typically devoid of dusty disks. On the other hand, WTTS span the whole range of rotation periods detected across the cluster. We also investigated whether the rotation properties we measure for NGC~2264 members show any dependence on stellar mass or on stellar inner structure (radiative core mass to total mass ratio). No statistically significant correlation emerged from our analysis regarding the second issue; however, we did infer some evidence of a period--mass trend, lower-mass stars spinning on average faster than higher-mass stars, although our data did not allow us to assess the statistical significance of such a trend beyond the 10\% level.}{This study confirms that disks impact the rotational properties of young stars and influence their rotational evolution. The idea of disk-locking, recently tested in numerical models of the rotational evolution of young stars between 1 and 12~Myr, may be consistent with the pictures of rotation and rotation-accretion connection that we observe for the NGC~2264 cluster. However, the origin of the several substructures that we observe in the period distribution, notably the multiple peaks, deserves further investigation.}


\maketitle

\section{Introduction} \label{sec:introduction}

In spite of a substantial effort devoted to the subject over recent decades, the evolution of stellar angular momentum during the pre-main sequence (PMS) remains a somewhat controversial issue. The so-called angular momentum problem \citep[e.g.,][]{bodenheimer1995} is a long-standing conundrum in star formation theories. At an age of a few Myr, low-mass solar type stars (T~Tauri stars, TTS; \citealp{joy1945}) are known from observations to have a spin rate of merely a fraction of their break-up velocity \citep[e.g.,][]{vogel1981, bouvier1986, hartmann1989}. However, if their early rotational evolution was simply governed by conservation of angular momentum as these objects contract toward the zero age main sequence (ZAMS), by an age of $\lesssim$~1~Myr they ought to rotate much faster. This indicates that some mechanism of angular momentum regulation must be at play during the early PMS that effectively counteracts the spin-up effect linked to stellar contraction. In addition, various observational studies of rotation rates in young stellar clusters \citep[e.g.,][]{edwards1993, bouvier1993, herbst2002, lamm2005, littlefair2010, henderson2012, affer2013} have reported evidence of statistically distinct rotational behaviors between classical T~Tauri stars (CTTS; \citealp{herbig1962}) and weak-lined T~Tauri stars (WTTS; \citealp{herbig_bell1988}) within the same region. The former, which are young stellar objects (YSOs) still interacting with an active accretion disk, rotate on average more slowly than WTTS, which are more evolved young stars with no signatures of circumstellar material. These results indicate that angular momentum regulation in YSOs is intimately connected with the star-disk interaction. 

The main idea behind the current model of disk accretion in T~Tauri stars was first examined in \citet{konigl1991}. Based on the formalism developed in \citet{ghosh1978} for neutron stars, the author suggested that disks around TTS do not reach down to the stellar surface, but are truncated at a distance R$_T$ of a few stellar radii from the central source by the strong magnetic field of the star ($\sim$1~kG at the stellar surface). The accretion of matter from the inner disk to the star therefore occurs in a magnetically controlled fashion: material is lifted from the inner disk and channeled along the magnetic field lines, forming accretion columns that impact the stellar surface at near free-fall velocities and thus generate hot shocks close to the magnetic poles. This initial picture, which assumed a stable, funnel-flow accretion process driven by a dipolar magnetosphere aligned with the rotation axis of the star, only provides a basic sketch of the far more complex and dynamic star-disk environment \citep[see, e.g.,][]{romanova2004, kurosawa2013}. Nevertheless, the concept of magnetospheric accretion now defines the widely accepted paradigm for disk accretion in TTS, and finds strong support in its capability to explain many observational features associated with YSOs, such as the emission line profiles, large infrared and UV excesses, spectral veiling, presence of warps in the inner disk, strong photometric variability (see, e.g., \citealp{bouvier2007} for a review).

In the framework of magnetically controlled star-disk interaction, several scenarios have been proposed to solve the angular momentum problem. \citet{konigl1991} suggested that the magnetospheric star-disk coupling may effectively lock the star to the disk. Magnetic field lines threading the disk in the region between R$_T$ and the co-rotation radius $R_C$ (where the orbital velocity in the disk is higher than the angular velocity of the star) transmit a spin-up torque to the star; this is balanced by the spin-down torque ensuing from magnetosphere-disk coupling beyond $R_C$. The result is a braking action on the star. A somewhat different mechanism was proposed in \citet{shu1994}, who identified the main source of angular momentum removal from the system in magnetocentrifugally driven winds launched from the diskplane at distances $r > R_C$. Inside $R_C$, near-corotation of disk material with the star is enforced by the strong magnetic field. Both models assume that the dipolar component of the stellar magnetic field, which dominates the star-disk interaction (it decays more slowly with distance from the star than higher-order components), has a strength of a few kG. However, recent studies suggest that the dipole intensity may be a factor of 10 smaller \citep{gregory2012}. This may imply that the actual stellar dipole is not strong enough to act as an efficient braking source on the star. Alternative models of star-disk interaction have suggested that other mechanisms, such as accretion-powered stellar winds \citep{matt2005} or magnetospheric ejections of material \citep{zanni2013}, may play a more decisive role in extracting angular momentum from the systems.

Although the idea of disk-locking in PMS stars has been standing since the early '90s, evidence for this mechanism is still controversial. One consequence of the disk-locking scenario is that, once the disk accretion phase is over, a young star is relieved of the braking effect and can start to spin up freely as it contracts. Therefore, from an observational perspective, an association between the measured rotation periods for young stars and the presence/absence of disk accretion is to be expected if the model is valid. Systematic surveys of rotation in young clusters are of utmost interest to shed some light on these issues: intra-cluster studies enable investigations of the link between rotational properties and other stellar properties and disk indicators; exploring how the distribution of rotation periods varies between clusters of different age traces the evolution of angular momentum in the PMS. The recent review of \citet{bouvier2014} on the matter well illustrates the current state of the debate. Evidence of statistically distinct rotational behaviors for WTTS and CTTS, reported in several studies, is not supported by others; in some cases, contrasting conclusions are drawn by different authors on the same clusters. External factors such as observational biases or sample completeness, or physical effects such as differing behaviors in different mass ranges, may play a role in this ambiguity.

The typical approach to explore a connection between rotation and disks in young clusters consists of combining optical monitoring surveys, to measure the rotational periods, with near- and mid-IR photometry, to detect a flux excess linked to thermal emission by dust in the stellar surroundings. This approach has been pursued, for instance, in \citet{edwards1993} for a composite sample of T~Tauri stars, \citet{xiao2012} in Taurus, \citet{herbst2002} for the Orion Nebula Cluster (ONC), \citet{rebull2006} in Orion; these studies have shown a statistical association between the amount of IR excess and rotation, in the sense that YSOs with large IR excesses tend to be slow rotators, while young stars with little or no IR excess are spread over a broader range of periods, including both slow and fast rotators. While being indicative of dusty disks, the IR excess diagnostics does not enable direct assessment of whether an accretion process is actually ongoing in the system and at which rate mass is being transferred from the inner disk to the star. This is most directly investigated by detecting and measuring the UV flux excess produced in the accretion shock at the stellar surface. A comparison between UV excess measurements and spin rates derived for large populations of young stars is therefore of great interest to investigate the impact of different accretion regimes on the rotational properties of these sources (see, e.g., \citealp{rebull2001} and \citealp{makidon04}). The possibility of an association between UV excess and rotation in young stars was tested by \citet{fallscheer2006} for a sample of about 100 objects in the 3~Myr-old cluster NGC~2264. Sources with active accretion disks, characterized by a flux excess in the $U$-band and thus smaller (bluer) $U-V$ colors than non-accreting objects, were shown to be slow rotators, whereas fast rotators in the sample did not exhibit significant emission in the $U$-band above the photospheric level. This result supports the view that the angular momentum regulation in TTS is related to the process of mass accretion from the disk. As underlined by the authors, large samples of objects with UV excess measurements and rotational periods are needed to explore in detail this rotation-accretion connection. Unfortunately, the challenging and time-consuming nature of UV observations has often limited the use of this diagnostics in studies of young stars and of their evolution.

In this paper, we present a new study of rotation and of its connection with accretion disks across the young open cluster NGC~2264. This investigation has been conducted in the framework of the ``Coordinated Synoptic Investigation of NGC~2264'' project (CSI~2264; \citealp{cody2014}); this consisted of a multi-site exploration of YSOs variability in the NGC~2264 cluster, from the X-ray domain to UV, optical and mid-IR wavelengths, on timescales from $<$~hours to several weeks. About fifteen observing sites, both space-borne (e.g., the {\it Spitzer} and {\it CoRoT} space observatories) and ground-based (e.g., the Canada-France-Hawaii Telescope, CFHT), were employed in the course of the campaign. The NGC~2264 cluster has long been a benchmark for star formation studies; its young age ($\sim$3~Myr), relative proximity (distance of about 760~pc, in the local spiral arm of the Galaxy), rich population of young stars ($\sim$1\,000 known members), low average foreground extinction (A$_V$ $\sim$ 0.4 mag), are some of the reasons of the long-standing interest toward this star-forming region (see \citealp{dahm08} for a review). The study reported here is centered on the set of optical light curves obtained with the {\it CoRoT} satellite \citep{baglin2006}, which cover a period of 38 consecutive days almost continuously, with a photometric accuracy of $\lesssim$0.01~mag. The effectiveness of {\it CoRoT} light curves for accurate period determinations was well illustrated in \citet{affer2013}, who examined the rotation properties of about 300 NGC~2264 members based on a first, 23 day-long observing run performed with the {\it CoRoT} satellite on the cluster in March 2008. In this study, we use the new, more extensive dataset from the second {\it CoRoT} run on NGC~2264 to derive accurate rotational periods for a larger sample of cluster members, both CTTS and WTTS, and combine these results with other information from the CSI~2264 campaign to investigate how the rotational properties of young stars depend on stellar parameters like mass and on the presence of disks and active accretion. 

The paper is organized as follows. Section~\ref{sec:sample} provides a brief description of the {\it CoRoT} observations and of the selection of the sample of cluster members investigated in this study. Sect.\,\ref{sec:methods_analysis} describes the methods used to derive rotational periods from the light curves and their implementation. Sect.\,\ref{sec:results} illustrates the different variable classes identified across the sample (introduced in more detail in Appendix~\ref{app:light_curve_types}) and presents the period distribution derived for the cluster; the rotational properties of cluster members are then discussed as a function of stellar mass and of their CTTS vs. WTTS classification; some considerations on the similarity in nature of WTTS and CTTS periods are also reported. In Sect.\,\ref{sec:disklocking}, the rotation-disk connection in NGC~2264 is explored: rotational periods derived for CTTS and WTTS in our sample are combined with UV excesses from \citet{venuti2014} to investigate the association between different accretion regimes and the rotational properties of young stars; these results are discussed with reference to the scenarios of magnetospheric accretion and disk-locking, with particular focus on the R$_T$/R$_C$ ratio estimated following theoretical predictions. In Sect.\,\ref{sec:rot_evol}, the case of NGC~2264 is discussed in the context of PMS rotational evolution: its period distribution is compared to those of various clusters between 2 and 13~Myr, and their respective features are discussed as a function of mass with reference to the timescales of evolution and dispersal of protoplanetary disks; these observational results are then compared to recent semi-empirical models of rotational evolution of young stellar clusters in the presence of disk locking. Our results and conclusions are summarized in Sect.\,\ref{sec:conclusions}. Spatial coordinates, classification and rotation parameters for all sources investigated in this study are collected in Table~4, reported after the end of the main paper text. Appendix~\ref{app:hist_bin} illustrates the impact that a specific choice of bin size and/or phase may have on a histogram representation of the period distribution derived for the cluster. Cases of objects with discrepant period measurements between this study and \citet{affer2013}, objects with different periods reported in this study with respect to \citet{cieza2007}, and the cases of objects periodic in the first {\it CoRoT} run \citep{affer2013} but aperiodic here or vice~versa are discussed in Appendices~\ref{app:V16_A13}, \ref{app:V16_C07}, and \ref{app:per_aper}, respectively.

\section{Observations and sample selection} \label{sec:sample}
The {\it CoRoT} monitoring survey of NGC~2264 extended over 38 consecutive days from December 1, 2011 to January 9, 2012. The instrument specifications, as well as a detailed description of the observing run and of the subsequent photometry reduction, were provided in \citet{cody2014}. Observations were carried out using one of the two CCDs originally dedicated to exoplanetary science for the main scientific program of the {\it CoRoT} mission. The instrument has a field of view (FOV) of $1.3^\circ$$\times$$1.3^\circ$, quite adequate to fit NGC~2264 in its entire spatial extent. Time series aperture photometry is downloaded from the satellite only for objects in a pre-determined list of targets in the FOV; the final {\it CoRoT} sample obtained within CSI~2264 contains about 500 sources with robust evidence of membership, 1600 candidate members and 2000 field stars. All of the NGC~2264 light curves we have utilized in producing this
paper are available as part of the CSI~2264 public data release in the IRSA archive\footnote{http://irsa.ipac.caltech.edu/data/SPITZER/CSI2264/}. The website provides users to both view the light curves and to download them, either individually or as the complete set of light curves in a single tar file. In addition, all scientific data issued from the {\it CoRoT} campaign can be found at the IAS CoRoT Public Archive\footnote{http://idoc-corot.ias.u-psud.fr/}.

The magnitudes of monitored objects range from 11 to 17 in the R-band. A time sampling of 512 s was adopted for most targets in the FOV, hence resulting in over 6300 datapoints per light curve. For a subset of objects, a high-cadence observing mode was adopted, with luminosity measurements every 32 s (corresponding to over 100,800 datapoints along the whole observing run). Extracted light curves were preliminarily inspected and corrected for systematic effects such as isolated outliers (flagged by the {\it CoRoT} pipeline and discarded in the analysis) or abrupt flux discontinuities due to detector temperature jumps (which occur in about 10\% of light curves in our sample at the same observing epochs). 

The sample of cluster members investigated in this study was built following primarily the member list provided in \citet{venuti2014}. Membership and classification (WTTS vs. CTTS) criteria are listed in that paper and comprise photometric or spectroscopic H$\alpha$, X-ray emission, radial velocity and UV/IR excess diagnostics. The {\it CoRoT} counterparts of these sources were identified as their closest match within a radius of 1~arcsec around their spatial coordinates. The sample of members studied in \citet[][Table~2]{venuti2014} includes 757 objects; among these, only 433 have a {\it CoRoT} counterpart from the CSI~2264 observing run. A good fraction of the sources not matched in the {\it CoRoT} sample are fainter than R=17. Additional members, not included in \citeauthor{venuti2014}'s (\citeyear{venuti2014}) sample (mainly objects that fell outside the FOV probed in that paper, or brighter than the magnitude range explored), were selected from the CSI~2264 master catalog\footnote{Available at http://irsa.ipac.caltech.edu/data/SPITZER/CSI2264/} of the region, upon the condition of being flagged as ``very likely NGC~2264 member'' \citep[i.e., MEM=1; see][]{cody2014}. The final sample of members thus selected for the present study of rotation comprises 500\footnote{In the process of members selection, we rejected the object CSIMon-000661, that had been classified as member in \citet{cody2014} and \citet{venuti2014}. The reasons for its exclusion are the following: this star is located beyond the periphery of the molecular cloud in an RA-Dec diagram of the region; the {\it CoRoT} light curve amplitude for the source is of $\sim$0.02~mag, significantly lower than the typical amount of variability measured for these young stars; VLT/FLAMES data acquired for this object show no signs of Lithium absorption, which suggests that this star is more evolved; the previous classification as cluster member was based on \citeauthor{furesz06}'s (\citeyear{furesz06}) radial velocity survey of the cluster, but the specific measurement for this object stood about 3-4~$\sigma$ away from the typical cluster locus.} objects, distributed in mass between $\sim$0.1 and 2 M$_\odot$; the ratio of disk-bearing (CTTS) to disk-free (WTTS) members is of about 40\% to 60\%. 

\begin{figure}
\resizebox{\hsize}{!}{\includegraphics{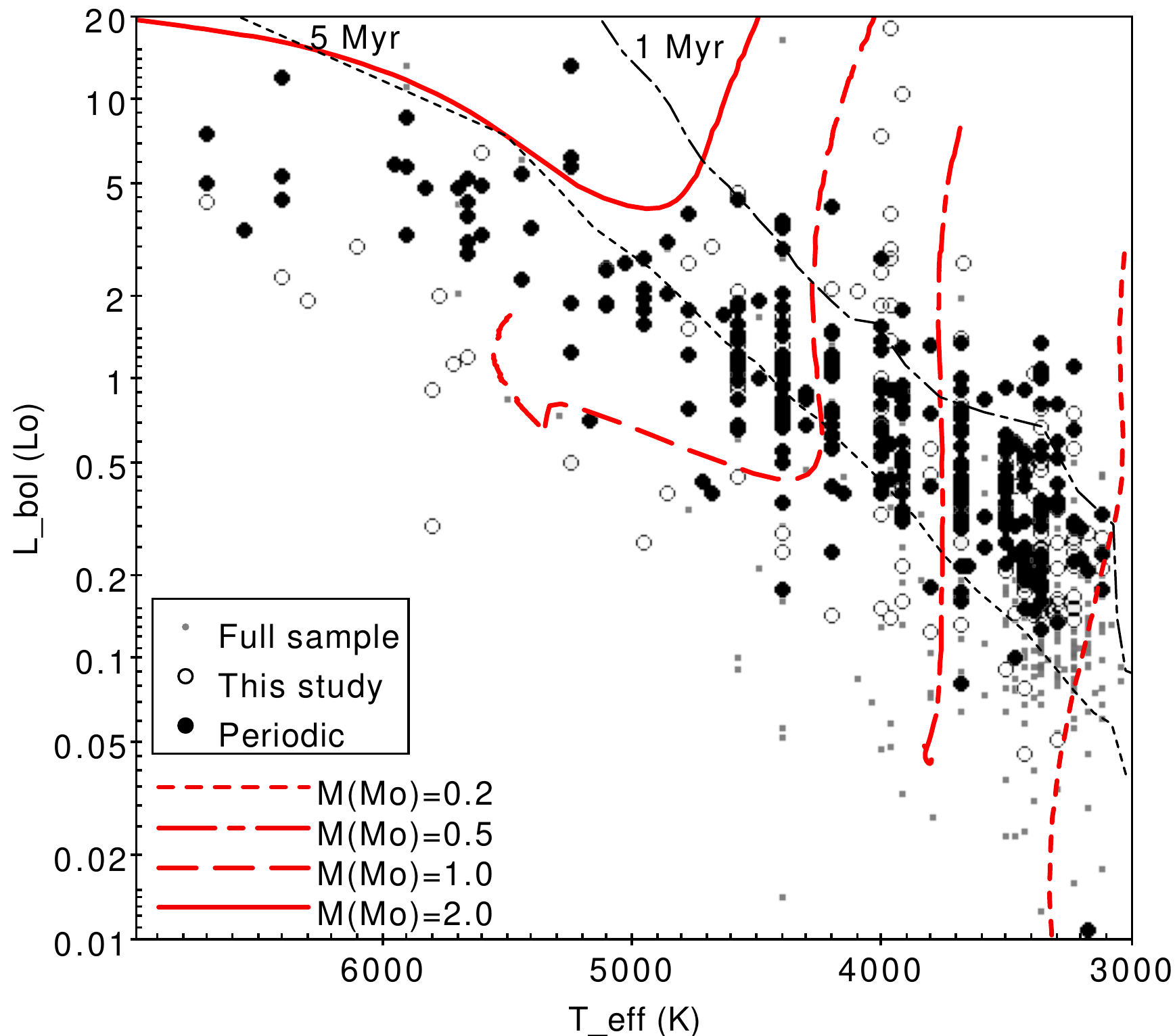}}
\caption{The properties of NGC~2264 members considered in this study (open and filled circles) are compared to those of the full sample of NGC~2264 members available from CSI~2264 (grey dots). Filled circles identify the periodic sources emerging from the analysis in Section~\ref{sec:methods_analysis}. Mass tracks shown on the diagram and isochrones at 1~Myr (dash-dotted line) and 5~Myr (dotted line) are from \citeauthor{siess2000}'s (\citeyear{siess2000}) models.}
\label{fig:HRD_full_sample_per}
\end{figure}

In Figure~\ref{fig:HRD_full_sample_per}, the properties of the subsample of objects examined in this study (periodic or aperiodic as resulting from the analysis in Sect.\,\ref{sec:methods_analysis}) are compared to those of the full NGC~2264 population on a H-R diagram.  
Effective temperatures T$_{eff}$ were assigned to each object based on their spectral type SpT and the SpT--T$_{eff}$ conversion scale provided in \citet[][see also \citealp{luhman2003}]{cohen1979}; bolometric luminosities L$_{bol}$ were derived from the dereddened 2MASS J-band magnitudes, using T$_{eff}$--dependent bolometric corrections from the scales of \citet{pecaut2013} and \citet{bessell98}. The reader is referred to \citet{venuti2014} for further details.

It can be seen that objects appear to be spread over a broad range of L$_{bol}$ at any given T$_{eff}$. At first sight, this would indicate a significant age/evolutionary spread among cluster members, although a number of recent studies have stressed how separate effects such as individual accretion histories \citep{baraffe2009} or observational uncertainties \citep[e.g.,][]{pecaut2016} may result in an artificial apparent age spread on a H-R diagram. Notably, Fig.\,\ref{fig:HRD_full_sample_per} shows a population of about 25 objects, classified as cluster members and with no detectable periodic behavior, that define a lower boundary to the distribution of the NGC~2264 population on the diagram, well below the 5~Myr isochrone. About 55\% of them are non-variable according to the \citeauthor{stetson96}'s (\citeyear{stetson96}) $J$-index \citep[see][]{venuti2015}, while the remaining 45\% are variable but exhibit an irregular light curve pattern. For some objects in this group, the classification as cluster member derives from literature studies based on a limited number of parameters (e.g., radial velocity), and the small additional information available for these cases from our campaign did not enable a reassessment of the membership issue. Therefore, some of them, amounting to a small percent of the total sample, may actually be field contaminants. However, sound evidence of being a young star (presence of disk, Lithium absorption) is available for other sources in this group; their position on the H-R diagram may then be affected by uncertainties on the derived stellar parameters, or by strong light attenuation by the circumstellar environment.

\section{Period search} \label{sec:methods_analysis}

Due to their intense chromospheric activity, as well as magnetospheric accretion on disk-bearing objects, the surface of young stars looks far from homogeneous. In fact, a significant fraction (up to a few tenths) of it appears covered by uneven spots of different temperature relative to the photosphere. As the star rotates, different portions of the stellar surface appear on the line of sight to the observer, hence resulting in a modulation effect of the photospheric flux by surface spot seen at different phases. When objects are monitored during several rotational cycles, provided that the lifetimes of surface spots are longer than the timescales of interest, a periodicity is therefore expected to appear in the light curves, corresponding to the rotation period of the stars. 

Several methods have been proposed to extract a periodic component from a time-ordered series of signal measurements. Some, more analytic, consist in decomposing the signal in waves at a given frequency; others, more empirical, are based on a comparative examination of the morphology of different segments of light curve, in units of trial period. We adopt here 3 different methods to search for periodic signals in our sample of light curves: the Lomb-Scargle periodogram (LSP), the auto-correlation function (ACF) and the string-length method (SL). A brief description and comparison of advantages and disadvantages of these methods is provided in Sect.\,\ref{sec:methods}, while details on their application to our analysis are provided in Sect.\,\ref{sec:implementation}.

\subsection{Methods} \label{sec:methods}

\subsubsection{Lomb-Scargle periodogram}
This approach \citep{scargle1982, horne1986} is a revised version of the discrete Fourier transform method, applicable to datasets with uneven temporal sampling and invariant to a shift of time origin. The method is equivalent to least-squares fitting of sinusoidal waves at a test frequency $\omega$ to the observed light curve: the power spectrum of the light curve is reconstructed by varying the test frequency $\omega$ in the range of investigation. 

\begin{figure}
\resizebox{\hsize}{!}{\includegraphics{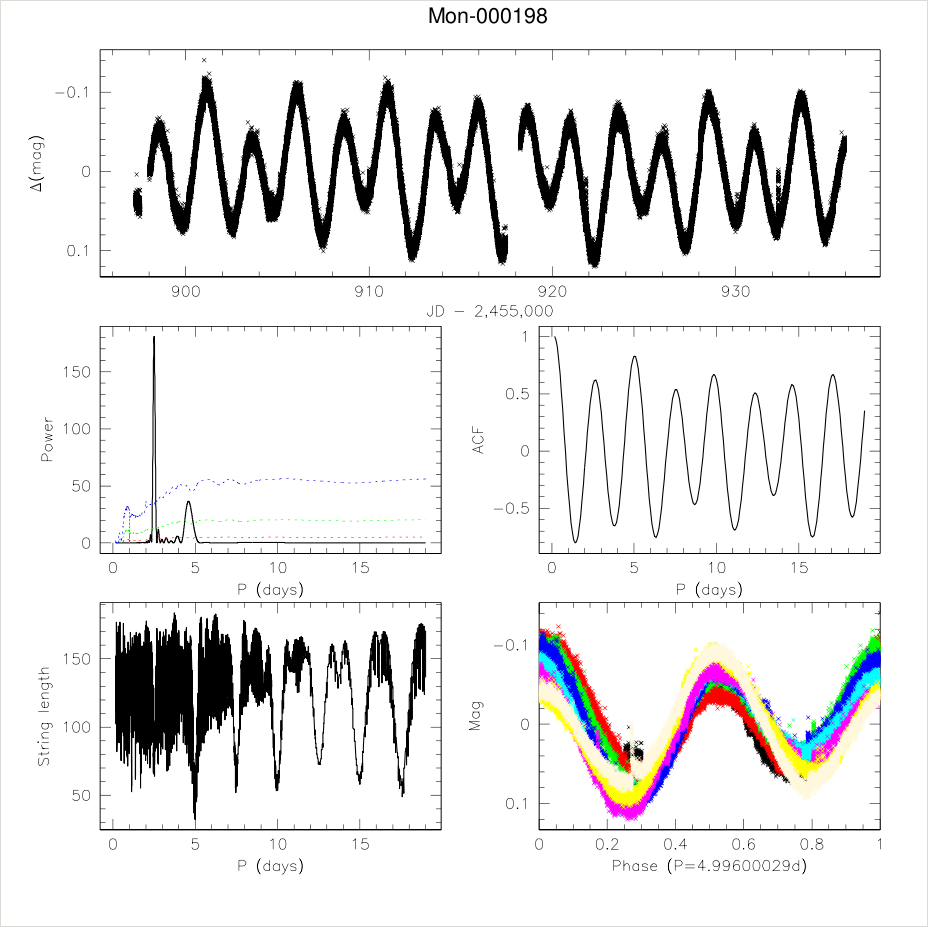}}
\caption{Period analysis for the object with multiple spots CSIMon-000198: original light curve after removal of spurious points in the upper panel, periodogram in the middle left panel, auto-correlation function in the middle right panel, string-length method in the lower left panel, phased light curve with the final period in the lower right panel. Red, green and blue dotted curves superimposed over the periodogram in the middle left panel correspond respectively to the mean, mean+3\,$\sigma$ and mean+10\,$\sigma$ periodogram expected in case the mag fluctuations about the mean were due to random noise. Different colors in the lower right panel correspond to different rotational cycles. Due to the nearly symmetric shape of the main periodic feature around a less deep minimum at about the half-period, this last is erroneously detected as the highest power (best rotational period) when using the periodogram analysis; on the other hand, the correct periodicity is indicated by both the other methods.}
\label{fig:half_per}
\end{figure}

The LSP method ensures more accurate period measurement than ACF and SL, as it is less sensitive to spurious points and long-term trends; in addition, it provides a straightforward estimate of the uncertainty on the derived periodicity, by measuring the Gaussian width of the highest peak in the periodogram. On the other hand, the explicit assumption that the periodic luminosity component can be described in terms of sine curves may contrast with the actual light curve shape observed for these young stellar objects. Moreover, this analytic approximation may lead to incorrect period identifications when the folded light curve is nearly symmetrical with respect to phase $\phi = 0.5$, as illustrated by the case in Fig.\,\ref{fig:half_per}.

\subsubsection{Auto-correlation function}
Contrary to the LSP method, the ACF method \citep{box1976, mcquillan2013} does not introduce any assumptions on the shape of the light curve. The method consists in exploring a range of trial periods and computing, for each of them, the autocorrelation coefficient $r_k$ of the light curve at lag $k$, where $k$ is the tested period $P$ in units of time cadence $\Delta t$ ($P = k \Delta t$):
\begin{equation}
r_k = \frac{\sum_{i=1}^{N-k} (m_i - \overline{m})(m_{i+k}-\overline{m})}{\sum_{i=1}^N (m_i - \overline{m})^2}\,,
\end{equation}
where $m_i$ is the magnitude at time $t_i$, $m_{i+k}$ is the magnitude at time $t_i + P$, $\overline{m}$ is the average light curve magnitude and $N$ is the total number of light curve points (equally spaced in time with step $\Delta t$). The auto-correlation function is reconstructed by plotting $r_k$ as a function of $P$. When the $P$ value tested matches the actual periodicity of the light curve, similar behaviors relative to the typical luminosity state $\overline{m}$ are expected at epochs $i$ and $i+k$, hence resulting in large values of $r_k$ and a maximum in the ACF; conversely, small values of $r_k$ are found if matched epochs $i$ and $i+k$ are out-of-phase.

By virtue of its direct reference to the actual shape of the light curve, the ACF technique allows more reliable identification of the peak corresponding to the true periodicity (as shown in Fig.\,\ref{fig:half_per}); however, the need for an even time binning and for adopting a $\Delta P$ step for period investigation that is the same as (or a multiple of) the light curve $\Delta t$ reduces the accuracy of the extracted period value compared with the value of the corresponding peak in the periodogram.

\subsubsection{String-length method} \label{sec:SL}
As with ACF, the SL method \citep{SL1983} does not introduce any assumptions on the shape of the light curve. The method explores a range of trial periods; for each $P$, light curve points are ordered in phase and the string-length $L$ is defined as the sum of the lengths of line segments that connect successive points of the phased light curve on the ($m, \phi$) diagram:
\begin{multline}
L = \sum_{i=1}^{N-1} \sqrt{(m_i - m_{i-1})^2 + (\phi_i - \phi_{i-1})^2}\, + \\ + \sqrt{(m_1 - m_N)^2 + (\phi_1 - \phi_N +1)^2}
\end{multline}
A preliminary rescaling of the observed magnitude range to the phase range is needed in order to assign equal weight to variations in the two variables for the computation of $L$. The best period $P$ is the one that minimizes the value of $L$, while large $L$ values will be found when the light curve is phased with an arbitrary period, hence resulting in a scattered cloud of points with no specific pattern on the phase diagram.

This technique combines the advantages of being conceptually straightforward, directly related to the actual variability pattern of the light curve, and of being applicable to any temporal sampling; due to the greater sensitivity to spurious points and long-term trends, the overall SL curve as a function of $P$ tends to be more noisy, notably at lower period values.

\subsection{Implementation} \label{sec:implementation}

A preliminary light curve ``cleaning'' routine was performed to reject all points with flag $\neq$ 0 from the {\it CoRoT} reduction pipeline\footnote{The meaning of different flag values is detailed in the manual on {\it CoRoT} N2 data stored at http://idoc-corot.ias.u-psud.fr}. A 10\,$\sigma$-clipping selection was subsequently performed to discard isolated discrepant points. 

Resulting light curves were then rebinned to 2\,h. This time step choice was on one side motivated by the 1.7\,h-long orbital period of the {\it CoRoT} satellite, smoothed out in this new data binning (see also \citealp{affer2012, affer2013}); this also represents a suitable choice with respect to computation efficiency. We computed the expected break-up velocity $\mathcal{V}$\footnote{The break-up velocity is defined as the tangential velocity at which the centrifugal force perceived by an element of mass at the surface of the star equals the gravitational force that keeps that mass element bound to the star: $\mathcal{V} = (2/3)^{1/2}\,\sqrt{G M_\star / R_\star}$, where G is the gravitational constant and the factor (2/3)$^{1/2}$ accounts for the polar-to-equatorial radius ratio when the surface rotates with the critical velocity \citep[cf.][]{gallet2013, ekstrom2008}.} for all sources in our sample with estimated mass $M_\star$ and radius $R_\star$ from \citet{venuti2014} (86\%), and ascertained that this would result in minimum rotational periods longer than 2\,h for the vast majority of objects. The typical break-up period we estimated across our sample is 0.5~days; less than 1\% of the considered objects have $\mathcal{V} > 2\pi R_\star/2\,h$, while about 6\% have $\mathcal{V} > 2\pi R_\star/4\,h$ (where 4\,h is the smallest period that can be investigated following the time step definition). Therefore, we assume that no significant bias on the detection of high-frequency periodicities is potentially introduced when performing the data rebinning to 2\,h. Nevertheless, we did adopt a smaller temporal bin and inspected the light curves for periods shorter than 4\,h in cases where no period was found from the procedure described in the following, and when hints of a possible short periodicity were conveyed from the light curve and/or the analysis. Only for one object was a significant period shorter than 4\,h actually detected.

Rebinned light curves are processed afterwards for period investigation. Each light curve is examined three times, using a different technique at each step, as enumerated in Sect.\,\ref{sec:methods}. Periods explored range from 4\,h (i.e., twice the data sampling) to 19~days (i.e., half the total duration of the time series). The upper limit of the period range selection ensues from the assumption that a periodic behavior can be reliably identified if this is repeated at least twice during the monitored time; however, we did extend the range explored to longer periods in cases where the light curve and/or period diagrams obtained provided some hints of a periodicity above 19~d. A step $dP$ of 0.1\,h is adopted to explore the period range with the LSP and the SL methods, while $dP$ is set to the data cadence for the ACF method.

Period diagrams from LSP, ACF and SL methods are visually inspected and compared and the best period is selected as the one toward which the 3 diagnostic tools converge. The ACF and SL diagrams are used as the primary reference to locate the correct periodicity, for the reasons explained in Sect.\,\ref{sec:methods}; the period value is then extracted from the peak displayed in the periodogram at the position predicted from the other two methods. In case no significant features are present in the LSP diagram at this location, the period value is extracted from the SL diagram; an extensive comparison of values measured from different methods for non-ambiguous periodic variables allowed us to conclude that SL estimates are typically more accurate than ACF estimates and in very good agreement with those extracted from the LSP diagram. As illustrated in Fig.\,\ref{fig:half_per} and discussed in Sect.\,\ref{sec:SL}, peaks in the SL curve tend to be less sharp than in the ACF or the LSP. For this reason, to derive a more precise SL period, instead of taking the position of the first minimum, we extracted the position of each minimum in the diagram, then measured the distance between every pair of consecutive minima, and defined the best period as the mean of these distances.

To obtain a first indication of the significance of periodogram peaks, we adopted the following procedure (see also \citealp{affer2013, flaccomio2005, eaton1995}). We segmented the original light curve in blocks of 12\,h, shuffled them and reassembled them in a random order. Every potential periodicity longer than the time length of the segments is destroyed in the process, hence resulting in a test light curve where the main contribution to variability arises from photometric noise or short-lived events like flares and episodic accretion. The periodogram analysis is then performed on this ``stochastic'' light curve and the whole routine is iterated 1\,000 times. The noise periodogram is thus defined point by point as the mean power measured across the 1\,000 simulations at the given $\omega$ value, while the variance is used to define confidence levels above the mean. The true periodogram of the source is therefore compared to these curves in order to get an indication of the confidence associated with the period detection/non-detection. This is illustrated in the middle panel of Fig.\,\ref{fig:half_per}. It is important to note that the confidence levels estimated with this procedure might be somewhat optimistic: irrespective of its periodic or non-periodic nature, the variability of T~Tauri stars exhibits a characteristic coherence timescale of a few to several days (see, for instance, the analysis of variability self-correlation in non-periodic TTS presented in \citealp{percy2006}). When 12\,h-long light curve segments are reshuffled, this coherence in the light curve is destroyed together with the rotational modulation; hence, the procedure may result in an underestimation of the ``noise'' component that affects the periodogram results. 

Another statistical indicator of the degree of periodicity in the light curve is the parameter $Q$ introduced in \citet{cody2014}. This measures how well a periodic trend at the period extracted can describe the original light curve. $Q$ is defined as
\begin{equation} \label{eqn:Q_cody}
Q = \frac{\left(rms_{resid}^2-\sigma^2\right)}{\left(rms_{raw}^2-\sigma^2\right)}\,,
\end{equation}
where $\sigma$ is the photometric measurement uncertainty, $rms_{raw}$ is the level of rms scatter in the original light curve, and $rms_{resid}$ is the level of rms scatter in the light curve after subtraction of the periodic trend (which is reconstructed by generating a smoothed phase-folded curve and overlaying it to the original light curve, repeating it once per period). $Q$ therefore measures how close the light curve points are to the systematic noise floor before and after subtraction of the phased trend from the light curve. Following the scheme of \citet{cody2014}, $Q<0.11$ indicates strictly periodic\footnote{Light curves that exhibit stable, repeating patterns, with shapes that evolve minimally over the monitored time span.} light curves, $0.11<Q<0.61$ corresponds to quasi-periodic\footnote{Light curves that exhibit a stable period, but with varying shape and/or amplitude from one cycle to the next, or light curves where lower-amplitude, irregular flux variations are superimposed over a modulated pattern.} light curves, and $Q>0.61$ indicates likely aperiodic light curves, with spurious period determination.

A careful inspection of phase-folded light curves was performed to reinforce the statistical period validation. This visual inspection is especially useful to decide on borderline cases or to select the correct periodicity in case of multiple peaks, as well as to check the accuracy of periodogram estimates (imprecise values will translate to offsets between different cycles in the phase-folded light curve). On the sole basis of individual diagrams such as Fig.\,\ref{fig:half_per}, we sorted all objects in the sample into periodic or aperiodic; we then combined this visual classification with the results of the $Q$ statistics. In nearly 80\% of the total, the results of the two selections were consistent (i.e., objects visually classified as aperiodic have $Q>0.61$ and vice versa); in the remaining cases, a final decision was taken upon further visual inspection. A few of us (LV, JB, AMC, JRS) examined the period diagrams for these objects independently, and listed what sources they thought were periodic or aperiodic. These results were then put together to assign or discard a period estimate to a given source with a certain degree of confidence.  

\section{Results} \label{sec:results}

At the end of the analysis detailed in Sect.\,\ref{sec:methods_analysis}, a definite periodicity is detected for 309 objects, i.e., for 62\% of the objects in our sample. Three of these periodic sources (CSIMon-000256, 6079, 6465) are eclipsing binaries; another 34 exhibit more than one significant and distinct periodicity (not harmonics). 

\begin{figure*}
\centering
\includegraphics[scale=0.6]{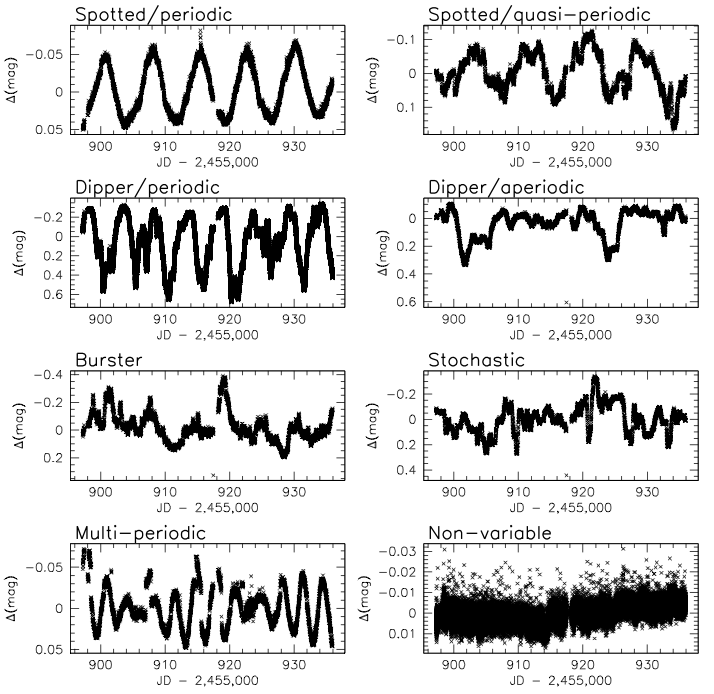}
\caption{Examples of the different morphological classes of light curves identified among the population of NGC~2264, in the optical dataset acquired with {\it CoRoT}. The objects illustrated are, from top to bottom, from left to right: CSIMon-000954 (strictly periodic), CSIMon-001167 (quasi-periodic), CSIMon-000660 (periodic dipper), CSIMon-000126 (aperiodic dipper), CSIMon-000567 (burster), CSIMon-000346 (stochastic), CSIMon-000967 (multi-periodic), CSIMon-000069 (non-variable).}
\label{fig:lcmorph1}
\end{figure*}

Among the 272 objects for which a single periodicity is detected in our analysis, several light curve types can be distinguished. As extensively discussed in \citet{cody2014}, the diversity in light curve morphology, very nicely highlighted in the NGC~2264 sample thanks to the accuracy and time coverage of the {\it CoRoT} dataset, likely reflects a variety of dominant physical mechanisms. Following the scheme of \citet{cody2014}, in Appendix~\ref{app:light_curve_types} we provide a brief list of the main variable classes identified among cluster members analyzed in this study, and of the relevant physical processes which may dominate the detected light variations. An illustration of the various classes is shown in Fig.\,\ref{fig:lcmorph1} in the main text; Table~\ref{tab:morph_class} provides a synthetic view of the statistical occurrence of each variable class across the NGC~2264 population, of their distribution between CTTS and WTTS, and of what fraction of objects belonging to each class is found to be periodic.

\begin{table}
\caption{Morphology types in the light curve sample of NGC~2264 members.}
\label{tab:morph_class}
\centering
\begin{tabular}{l r r r r}
\hline\hline
\textit{Morphology class} & \textit{Count} & c/w & \textit{\% tot.} & \% Per \\
\hline
Burster & 21 & 21/0 & 4.2 & 38.1 \\
Dipper & 37 & 35/0 & 7.4 & 68.6 \\
Spotted & 187 & 44/153 & 37.3 & 100.0 \\
Multi-periodic & 34 & 6/24 & 7.2 & 100.0\\
Eclipsing binary & 3 & 2/1 & 0.6 & 100.0\\
Stochastic & 22 & 19/0 & 4.4 & 31.6\\
Long-timescale variable & 3 & 2/1 & 0.6 & 33.3\\
Non-variable & 82 & 20/40 & 16.3 & 0.0\\
Others\tablefootmark{*} & 114 & 27/58 & 22.7 & 28.7\\
\hline
\end{tabular}
\tablefoot{{\it Morphology class} = classification of the light curve variability type; each class is briefly defined in Appendix~\ref{app:light_curve_types} of this work. {\it Count} = number of objects in the sample with light curves falling in the corresponding category (a given object is here attributed to a single category, namely the one that best describes the dominant variability features of its light curve, although some cases may also exhibit traits of a different morphology class). {\it c/w} = number of CTTS / number of WTTS among objects in the corresponding morphology class. {\it \% tot.} = percentage of objects in the sample that fall in the corresponding morphology class. {\it \% Per} = percentage of objects in the corresponding morphology class which are found to be periodic.
\tablefoottext{*}{Non-sorted light curves (unclassifiable variable type or affected by, e.g., temperature jumps).}
}
\end{table}

Among WTTS, a periodicity frequency of about $70\% \pm 9\%$ is detected; among CTTS, the frequency is of about $58\% \pm 6\%$. These numbers comprise both light curves that are ``strictly periodic'', i.e., that exhibit a stable pattern with shape that evolves minimally over the monitored span, and those that are ``quasi-periodic'', i.e., light curves with stable period but with changes in shape and/or amplitude from cycle to cycle (see the upper panels of Fig.\,\ref{fig:lcmorph1} for a comparative illustration of the two). The fraction of NGC~2264 members, and, among these, of CTTS found to be periodic from the {\it CoRoT} sample are consistent with the fraction of periodic-to-all sources recovered in the study of \citet{affer2013}. Conversely, the fraction of WTTS with detected periodicity among the sample investigated in \citet{affer2013} is larger ($\sim$88\%) than that recovered here. This may be due to the fact that the sample of cluster members investigated here includes more stars in the low mass range than the sample investigated in \citet{affer2013}; at fainter magnitudes, the impact of photometric noise is more considerable and may blur the intrinsic modulated pattern of the light curve. Indeed, the fraction of sources that would be classified as non-variable above the noise level, based on, e.g., \citeauthor{stetson96}'s (\citeyear{stetson96}) $J$-index (see \citealp{venuti2015}), is higher at fainter magnitudes. Of the 81 WTTS in our sample for which no periodicity is detected, 41 had no data in the previous {\it CoRoT} campaign, 23 had an aperiodic light curve in 2008 as well \citep{affer2013}, 5 (not included in the study of \citealp{affer2013}) exhibit a flat-line light curve in 2008, 7 were classified as periodic in \citet{affer2013}, and 3 (not included in \citealp{affer2013}) show some indications of periodicity in their {\it CoRoT} light curves from 2008. The last 10 cases are discussed in detail in Appendix~\ref{app:per_aper}. 

A direct estimate to the uncertainty on period measurements from the LSP can be derived as
\begin{equation} \label{eqn:P_err}
\delta P = \delta \nu \times P^2 \simeq 0.0096\times P^2\,,
\end{equation}
where $\delta \nu$ = 0.0096 is the average sigma width of a Gaussian fit to the periodogram peak across our sample and P is expressed in days. The resulting typical uncertainty amounts to $\delta P \sim 0.15$~days for periods on the order of 4~days and $\delta P \sim 1$~day for periods on the order of 10~days. This, however, appears to be very conservative, especially for longer period sources, compared to the actual accuracy we can reach thanks to the time coverage and cadence of the {\it CoRoT} light curves. 

A ``theoretical'' estimate of the uncertainty associated with the measured periods, based on the time sampling and number of cycles occurring in the light curve, can be derived following \citet{mighell2013}. For a light curve of length $L$ and period $P$, we can compute the number $M$\,=\,int($L$/$P$) of complete cycles contained in the light curve. The light curve will then contain $M+1$ maxima/minima, at positions $t_1 \dots t_{M+1}$. The light curve period can be defined as the average distance between two consecutive maxima/minima in the light curve:
\begin{equation} \label{eqn:period_definition}
P = \frac{(t_{M+1} - t_M) + (t_M - t_{M-1}) + \dots + (t_2 - t_1)}{M}.
\end{equation}
If the light curve points are evenly spaced with a time step $dt$, the position of each maximum/minimum has an associated uncertainty $\sigma = dt/2$. Therefore, by applying standard error propagation techniques, the uncertainty on P is given by
\begin{multline} \label{eqn:Perr_theory}
\delta P = \sqrt{\left(\frac{\delta t_{M+1}}{M}\right)^2 + \left(\frac{\delta t_{M}}{M}\right)^2 + \dots + \left(\frac{\delta t_{1}}{M}\right)^2} = \\ = \sqrt{2 M \left(\frac{\sigma}{M}\right)^2} =\, \sigma \sqrt{\frac{2}{M}}\,,
\end{multline}
where $\sigma$\,=\,0.0417~d in our case. This corresponds to an uncertainty of 0.02~d for periods of 4~d and of 0.03~d for periods of 10~d. Uncertainties estimated from Eq.\,\ref{eqn:period_definition} are slightly larger than those derived from Eq.\,\ref{eqn:P_err} for periods shorter than 1~d, and significantly smaller for periods of several days. The reasoning adopted to derive Eq.\,\ref{eqn:Perr_theory} assumes that the light curve is strictly periodic, so that the only uncertainty on the position of maxima/minima derives from the non-sampled time interval between two consecutive light curve points. However, this is not the case for the majority of light curves in our sample. Therefore, for illustration purposes we will use here error bars derived from Eq.\,\ref{eqn:Perr_theory}, but we note that realistic uncertainties on the measured periods probably lie between the estimates from Eq.\,\ref{eqn:Perr_theory} and those from Eq.\,\ref{eqn:P_err}.

\begin{figure*}
\centering
\includegraphics[width=\textwidth]{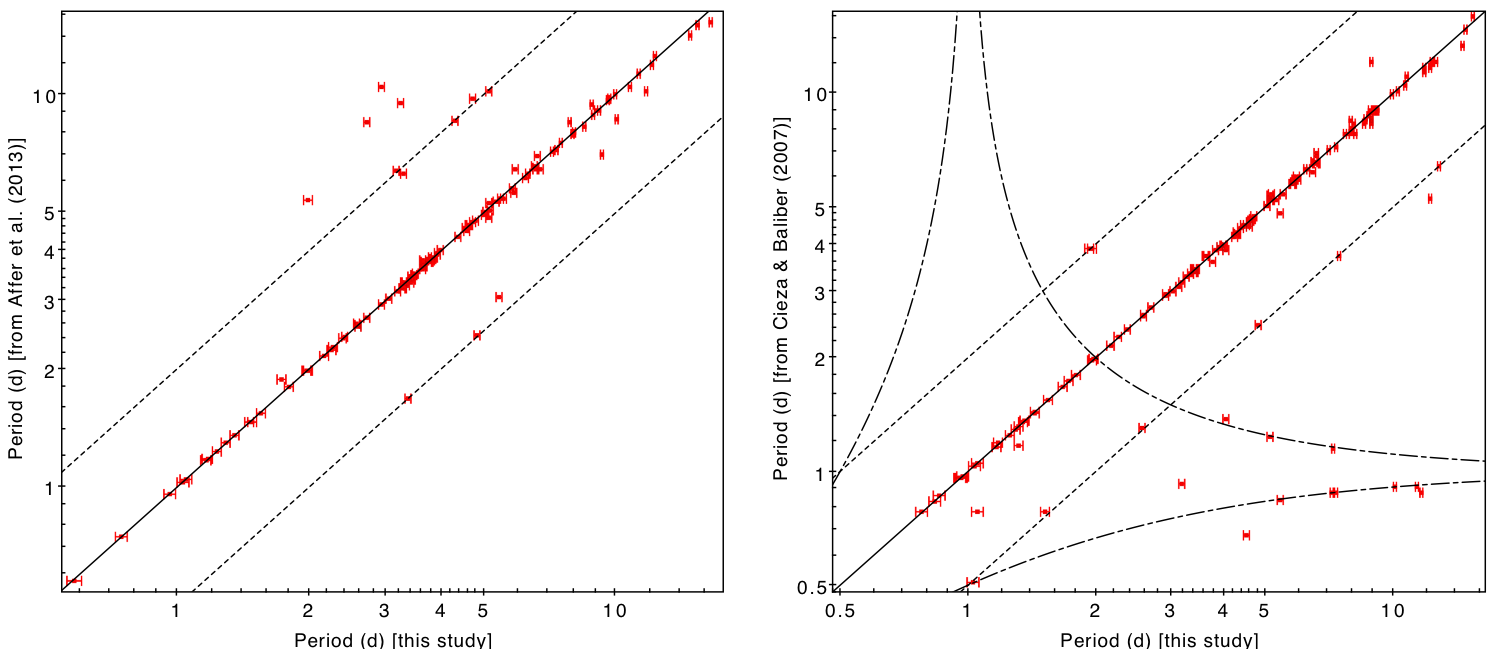}
\caption{Comparison of the period values derived in this study with those obtained in the study of \citet{affer2013} (left panel; 117 sources depicted) and those reported in \citet{cieza2007} (right panel; 145 sources depicted). 91\% of the periods reported in \citeauthor{cieza2007}'s (\citeyear{cieza2007}) sample are from \citet{lamm2005} (after \citealp{lamm04}); the remaining are from \citet{makidon04}. The intersection between the sample shown on the left panel and that shown on the right panel of this Figure amounts to 61 objects. The equality line is traced as a solid line; the half and double values lines are traced as dotted lines. The dash-dot lines trace the beat period with a 1~d sampling interval \citep[e.g.,][]{cieza2006, davies2014}. Error bars along the x-axis are computed following Eq.\,\ref{eqn:Perr_theory}. Multi-periodic objects are not shown on these diagrams.}
\label{fig:P_comp}
\end{figure*}

The period measurements derived in this study are reported in Table \ref{tab:periods}. In Fig.\,\ref{fig:P_comp}, these values are compared to the results of the similar investigation performed in \citet{affer2013} from the first {\it CoRoT} run on NGC~2264, and those reported in \citet{cieza2007}, which in turn refer to the studies performed in the optical by \citet{lamm04} and \citet{makidon04}.

The intersection between the sample of periodic sources with single periodicity found in this study and that of periodic variables listed in \citet{affer2013} consists of 117 sources. Among these, 16 (14\%) have different period estimates between the two studies (i.e., the estimate of period derived here for these objects is not consistent with \citeauthor{affer2013}'s estimate within the error bar associated with our measurements following the prescription of Eq.\,\ref{eqn:P_err}). These cases are individually illustrated and discussed in Appendix \ref{app:V16_A13}; several of them are harmonics at half or double period.

Similar considerations are reached when comparing our period estimates to those reported in \citet{cieza2007}. The common sample in this case comprises 145 objects, out of which 24 (17\%) have different period estimates in the two studies. A significant fraction of these outliers with respect to the equality line on Fig.\,\ref{fig:P_comp} (right panel) lies along a horizontal line at P(\citealp{cieza2007})$\sim$1~d; the corresponding periods derived in this study range from about 1~d to 12~d. As \citeauthor{cieza2007}'s (\citeyear{cieza2007}) period distribution is based on photometric measurements performed from the ground, these 1-day periods may actually be spurious detections (aliases) linked to the limited time sampling and day-night alternance which affect observations from the ground. A similar effect was identified in \citet{affer2013}, who compared their own results to those published by \citet{lamm04}.

In the following, we will assume that, when we detect a single photometric period in the light curve, this corresponds to the rotation rate of the star\footnote{\citet{artemenko2012} argue that, for a fraction of CTTS, the periodicity that we may detect in the brightness variations is not driven by temperature inhomogeneities at the surface of the star, but rather by dust structures in the disk rotating at the Keplerian velocity.}; we will then focus on the subsample of 272 periodic sources mentioned earlier to investigate the rotation properties of cluster members. A detailed analysis of EBs (see, e.g., \citealp{gillen2014}) and multi-periodic members of NGC~2264 is deferred to subsequent papers.

\subsection{Period distribution} \label{sec:per_dist}

\begin{figure*}
\centering
\includegraphics[scale=0.8]{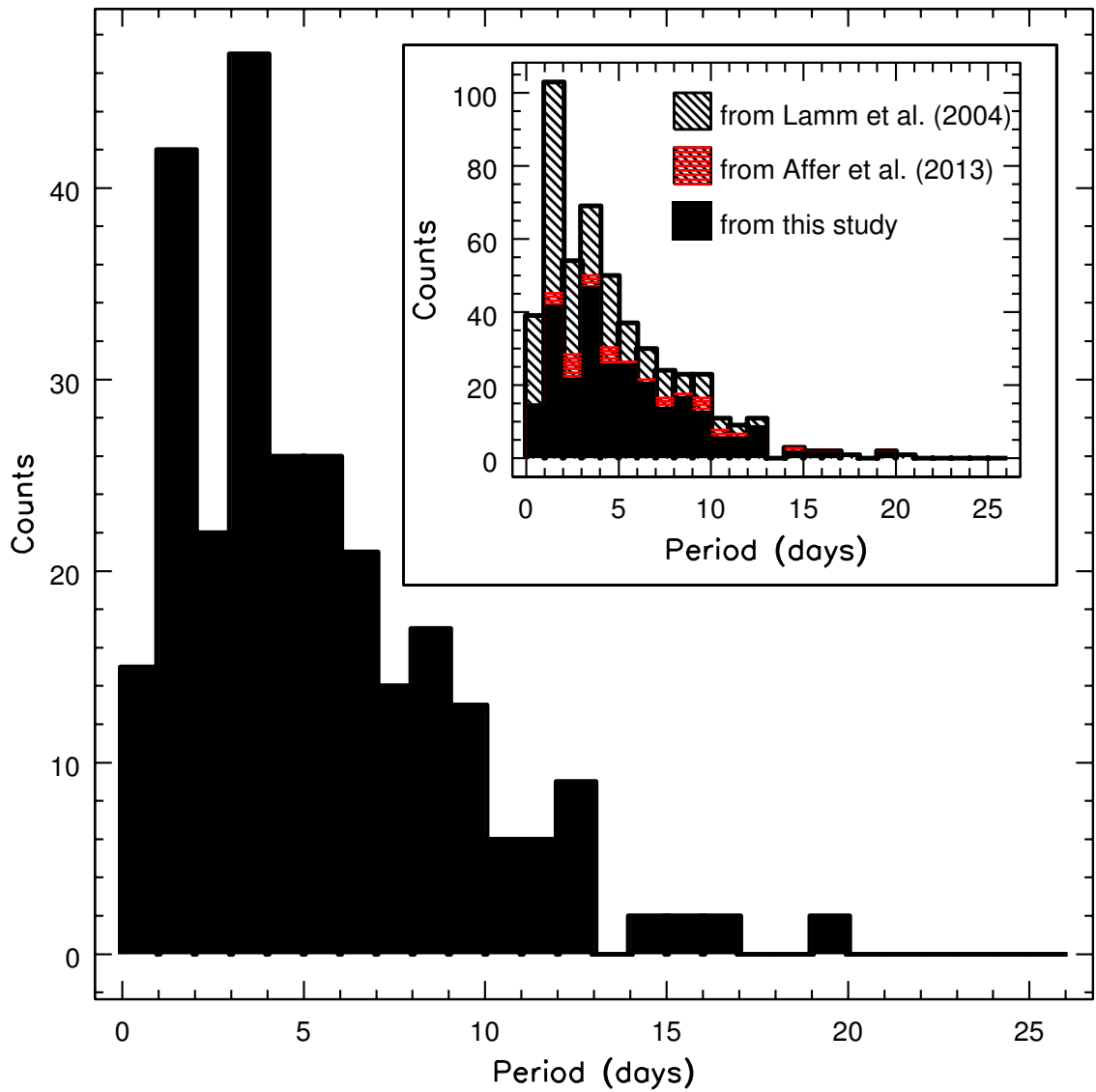}
\caption{Period distribution for the NGC~2264 cluster, as inferred from the population investigated in this study (black filled bars). Only single-period objects are considered here to build the histogram. The inset box to the right of the main period histogram shows the period distribution that would be derived, for the cluster, when completing our sample with the period detections obtained, in order of preference, by \citet[][shaded in red]{affer2013} or \citet[][cross-hatched in black]{lamm04} for additional cluster members with no light curves in the {\it CoRoT} 2011 dataset.}
\label{fig:hist}
\end{figure*}

Figure~\ref{fig:hist} illustrates the period distribution inferred here for the NGC~2264 population. A bin size of 1~d was adopted for the histogram, slightly smaller than the 1.4~d value that would be derived when applying \citeauthor{freedman1981}'s (\citeyear{freedman1981}) rule. This appears to be a reasonable choice in terms of resolution and statistics, and it is also the bin size commonly used in the literature, which enables direct comparison of the resulting period distribution with those inferred from analogous studies. Although the baseline of 38 days covered by the {\it CoRoT} light curves would enable robust detection of periodicities of up to 19~days, the vast majority of periodic cluster members have periods shorter than 13~days. This is similar to the period distribution inferred in the study of \citet{affer2013} from the previous, 23-day long {\it CoRoT} observing run on NGC~2264. 

A prominent feature of the period distribution found here is the presence of two peaks on top of a smooth distribution. The overall shape of the distribution can be described as a gamma or a Weibull distribution\footnote{The curve fitting was performed using the {\it fitdistrplus} package built for the R environment, and employing the maximum likelihood estimation technique to assess which probability distribution provides the best fit to the observed distribution of values.} \citep{weibull}, with a mean of 5.2$\pm$0.6~d and a variance of 13$\pm$3~d. These density distributions have an asymmetric shape, with a concave, rapidly rising profile between zero and the peak of the distribution, and a convex, gradually decreasing profile afterwards. Regarding the two peaks observed in Fig.\,\ref{fig:hist}, the first occurs around P$\sim$1-2~days, the second at P$\sim$3-4~days. This is followed by a gradually decreasing tail of longer periods. When reducing the bin size from 1~d to 0.5~d to better resolve the structure of the peaks, these appear to be centered around P$\sim$1.3~d (1-1.5~d) and P$\sim$3.3~d (3-3.5~d) respectively; the computed $\pm \sigma$ width for each of the peaks is 0.8~d. These numbers imply that the two peaks would no longer be resolved, but start to merge, if we adopted a bigger bin size (e.g., 1.5~d). Similarly, if we retained a bin size of 1~d but shifted the bin centers along the x-axis on Fig.\,\ref{fig:hist}, the global shape and properties of the histogram would be maintained for shifts of 0.1 or 0.2~d, but shifts of 0.3 or 0.4~d would determine the two peaks in Fig.\,\ref{fig:hist} to ``disappear'' as they redistribute into two neighboring channels. This is illustrated in Fig.\,\ref{fig:hist_bin_size_phase} of Appendix~\ref{app:hist_bin}.

\citet{attridge1992} were the first to report evidence of a similar feature (two peaks) in the period distribution of a very young stellar cluster (the 2~Myr-old Orion Nebula Cluster, ONC). To explain this observational feature, the authors suggested that, contrary to the rapid rotators, the slow rotators may be experiencing magnetic braking through interaction with their circumstellar disk. This suggestion was later confirmed by \citet{edwards1993}. The bimodal period distribution for the ONC, as defined by \citet{attridge1992}, was strengthened in the follow-up study of \citet{choi1996}, and its nature was later explored by \citet{herbst2001, herbst2002} with special reference to stellar mass. To assess a possible bimodal nature of the period distribution we obtain here for NGC~2264, we computed several statistical parameters, such as the kurtosis of the distribution \citep[e.g.,][]{decarlo97} and the bimodality coefficient BC, which is in turn based on the kurtosis and skewness of the distribution \citep{sas2012}. However, the high degree of asymmetry of the distribution and the presence of a heavy tail of slow rotators strongly affect the numerical values of these parameters, thus rendering the test inconclusive. Similarly, we attempted to apply the dip test statistics formulated in \citet{2hartigan1985} and \citet{hartigan1985}\footnote{We used here the version of the test implemented in the {\it diptest} package for the R environment.}, which consists in computing the unimodal distribution that best approximates the empirical cumulative distribution function observed in our sample and measuring the maximum difference between this fitting unimodal distribution and the empirical distribution. Again, the results of the test do not allow us to formally reject the null hypothesis that the observed distribution is unimodal, with a $p$-value of 0.26. Nevertheless, the presence of two peaks appears to be sound in the empirical period distribution for the cluster in Fig.\,\ref{fig:hist}; we will therefore abstain from calling this distribution bimodal, but retain the observational evidence of two peaks, and examine its possible implications in the context of young stars and rotational evolution in the following.

The inset panel in Fig.\,\ref{fig:hist} shows a more populated period distribution for the cluster that would be obtained if we integrate our sample (reported in Table \ref{tab:periods}) with the periods derived for additional cluster members by, in order of preference, \citet{affer2013} or \citet{lamm04}\footnote{No additional measurements from \citet{makidon04} are reported here because they amount to only a few sources, hence with no significant impact on the period distribution shown in the inset panel of Fig.\,\ref{fig:hist}.}. To complete the sample, we cross-correlated the full list of confirmed members from the CSI~2264 campaign \citep{cody2014} with the lists of periodic sources provided by \citet{affer2013} and \citet{lamm04}, and selected common objects with no {\it CoRoT} light curve from the 2011 dataset, hence not included in the period analysis performed in this study. As can be observed, the addition of a few tens of objects (35) from the \citeauthor{affer2013}'s (\citeyear{affer2013}) sample does not significantly modify the overall shape of the period distribution obtained from the present study. A far larger sample of additional periodic sources (190) can be retrieved from \citet{lamm04}; about 75\% of them have R-band magnitude fainter than 17 (limiting magnitude for the {\it CoRoT} sample, as mentioned in Sect.\,\ref{sec:sample}). When adding measurements from \citet{lamm04} to the period distribution obtained here, the first peak (P=1--2~d) is amplified relative to the second (P=3--4~d). However, this may be affected by a non-negligible fraction of aliases among the additional $\sim$1~d rotators detected by \citet{lamm04} in their ground-based campaign; indeed, about 10\% of the objects common to this study and to \citeauthor{lamm04}'s study, with assigned period between 1 and 3 days in the latter, are found to be slower rotators from the analysis of the {\it CoRoT} light curves (as illustrated in the right panel of Fig.\,\ref{fig:P_comp}).

\subsection{Period distribution: CTTS vs. WTTS} \label{sec:tts_per}

\begin{figure*}
\centering
\includegraphics[scale=0.9]{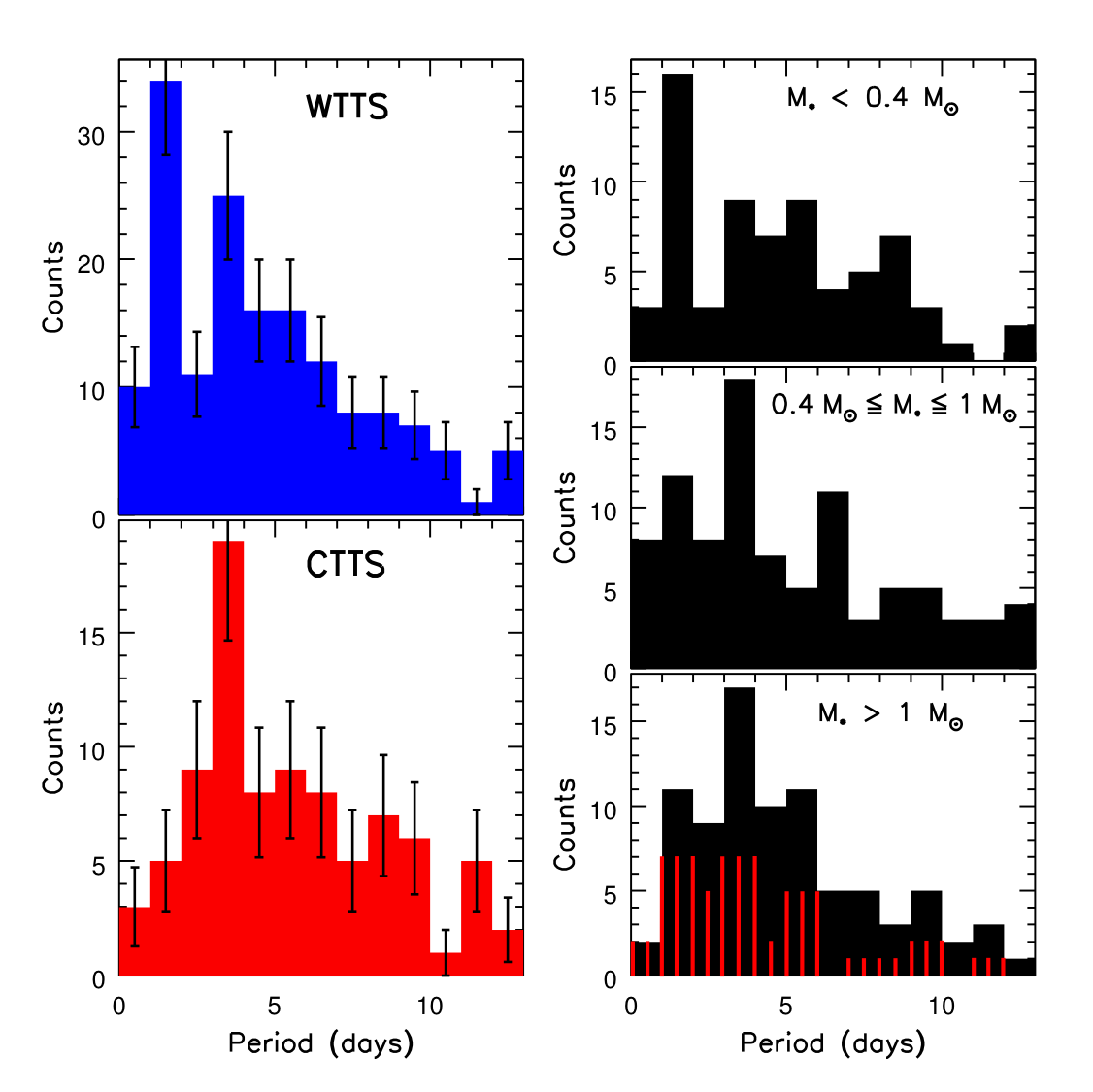}
\caption{{\it Left panel}: Period distributions inferred separately for the populations of disk-free young stars (WTTS; blue) and disk-bearing objects (CTTS; red) in NGC~2264, probed in this study. Only periods comprised between 0~d and 13~d are shown here; in addition, only objects with single periodicity and with an associated WTTS/CTTS flag in \citet{venuti2014} are considered to build the histogram. The error bars trace the Poissonian uncertainty computed on each histogram bin. {\it Right panel}: Period distribution of NGC~2264 members as a function of stellar mass. The sample investigated for periodicity in this study is separated into three different mass bins: M$_\star$$<$0.4\,M$_\odot$ (upper panel), M$_\star$ comprised between 0.4\,M$_\odot$ and 1\,M$_\odot$ (middle panel), M$_\star$$>$1\,M$_\odot$ (lower panel). In the latter, the period distribution of objects with M$_\star$$>$1.4\,M$_\odot$ is highlighted further as a superimposed histogram hatched in red.}
\label{fig:hist_tts_mass}
\end{figure*}

We will now focus on the periods derived in this study for the sample of NGC~2264 members included in the {\it CoRoT} 2011 campaign. To investigate further the nature of the period distribution shown in Fig.\,\ref{fig:hist} (main panel), we follow the member classification proposed in \citet{venuti2014} and explore the rotation properties of cluster members with active accretion disks (CTTS) and of those without evidence of ongoing accretion and circumstellar material (WTTS) separately. The comparison between the period histograms obtained for the two classes is illustrated in Fig.\,\ref{fig:hist_tts_mass} (left panel). 

A visual inspection of the two distributions suggests that CTTS exhibit distinct rotation properties from WTTS. WTTS' distribution peaks at 1-2~d, and steadily decreases toward longer periods, with hints of a second, less prominent peak between 3 and 4~d. Conversely, very few CTTS are found to exhibit rotation periods shorter than 2.5~d; their distribution in periods exhibits a single peak around P$\sim$3-4~d, and then decreases toward longer periods. A Kolmogorov-Smirnov (K-S) test \citep{numerical_recipes} applied to the two populations supports this idea of statistically distinct rotation properties for the two groups. Indeed, the test yields a probability of only $4\times 10^{-3}$ that the two distributions in period corresponding to CTTS and WTTS are extracted from the same parent distribution. This is also consistent with the conclusions presented in previous studies of rotation in NGC~2264 \citep[e.g.,][]{cieza2007, affer2013}.

\begin{figure*}
\centering
\includegraphics[scale=0.72]{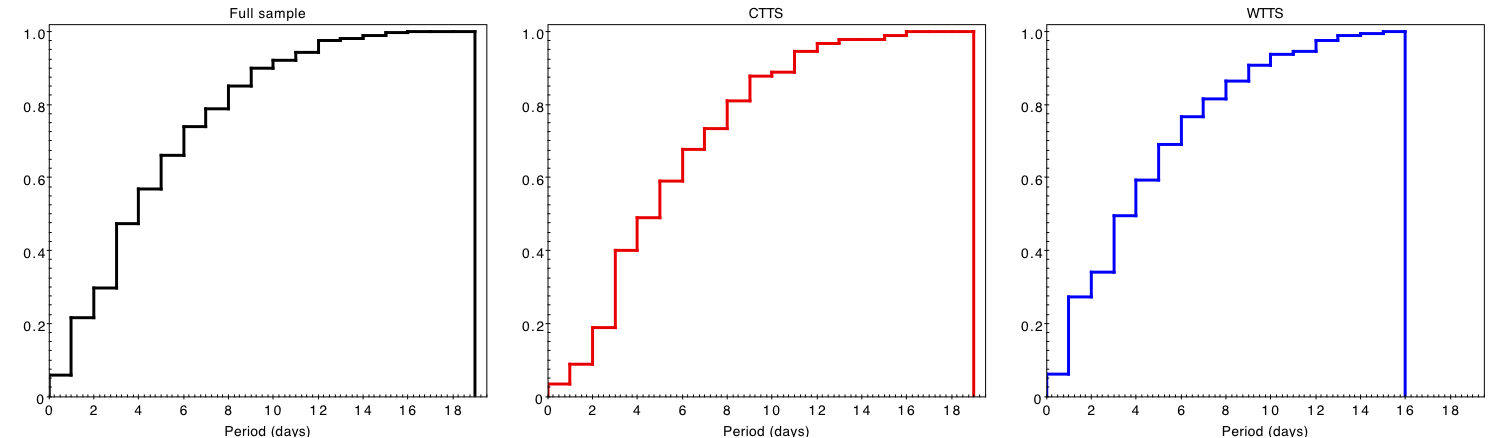}
\caption{Normalized cumulative distributions in period associated with the full sample of NGC~2264 members (left panel), the population of CTTS (middle panel), and the population of WTTS (right panel) analyzed in this study. Two prominent jumps of similar extent (0.16 and 0.17) appear in the cumulative distribution for the full sample at P=1~d and P=3~d, respectively. In the case of the CTTS, only one prominent jump of 0.21 at P=3~d is seen. In the case of the WTTS, two prominent discontinuities can again be observed, one steeper at P=1~d (0.21) and one smaller at P=3~d (0.15).}
\label{fig:cumul_distr_per}
\end{figure*}

If we compare the left panel of Fig.\,\ref{fig:hist_tts_mass} with the overall period distribution of NGC~2264 shown in Fig.\,\ref{fig:hist}, the following inferences can be drawn: i) the peak at short periods (1--2~d) observed in the global distribution is clearly associated with WTTS; ii) the second peak observed in Fig.\,\ref{fig:hist} (P=3--4~d) takes contribution from both CTTS and WTTS. These results provide a clear indication of a statistical connection between disk and rotation properties across the cluster. The vast majority of fast rotators found among cluster members are objects which lack disk signatures. Fig.\,\ref{fig:hist} and Fig.\,\ref{fig:hist_tts_mass} clearly illustrate that the first peak in the overall period distribution is associated with disk-free objects, and contains about 21\% of WTTS with detected rotation period. Conversely, the second peak in Fig.\,\ref{fig:hist} contains about 15\% of the periodic WTTS and 21\% of the periodic CTTS in our sample (6\% and 11\%, respectively, if we count objects in the peak only above the underlying ``continuum''). This is also illustrated in the cumulative distributions in period relevant to the total, CTTS, and WTTS samples, shown in Fig.\,\ref{fig:cumul_distr_per}.

\subsection{Period distribution as a function of stellar mass} \label{sec:mass_dependence}

In their review on the rotation properties and evolution of low-mass stars, \citet{herbst2007} discuss a possible mass-dependence of the measured period distributions for young stars. Observational results in favor of this point were reported by \citet{herbst2001} for the ONC ($\sim$2~Myr) and by \citet{lamm04, lamm2005} in NGC~2264. In both cases, although quantitative differences exist between the rotation properties derived for the two clusters, a bimodal period distribution was derived by the authors for cluster members more massive than $\sim$0.4~M$_\odot$, whereas lower-mass objects appear to spin faster on the average and define a possibly unimodal period distribution (see Fig.\,2 of \citealp{herbst2007}).

To test a possible mass-dependence in our data, we divided our sample into three similarly populated mass\footnote{Mass estimates are from \citet{venuti2014}.} subgroups: i)~M$_\star$$<$0.4\,M$_\odot$ (72~objects); ii)~0.4\,M$_\odot$$\leq$M$_\star$$\leq$1\,M$_\odot$ (95~objects); iii)~M$_\star$$>$1\,M$_\odot$ (86~objects). The separate period distributions for these three mass groups are shown in the right panel of Fig.\,\ref{fig:hist_tts_mass}. A simple visual inspection of these histograms would suggest that the measured period distribution does evolve, to a certain extent, as a function of mass: a peak of fast rotators (P=1-2~d) dominates the distribution in the lowest-mass group, albeit with a significant dispersion of objects at slower rotation rates; as stellar mass increases, this feature becomes less important compared to an emerging peak at slower rotation rates (P=3-4~d). However, no statistical support to this inference arises from the application of a two-sample Kolmogorov-Smirnov (K-S) test, applied to the lowest-mass and highest-mass groups (a $p$-value of 0.3 is obtained from the test, which does not allow us to discard the null hypothesis that the two populations are extracted from the same parent distribution)\footnote{We adopted here the version of the test implemented in the {\it stats} package for the R environment, and assuming as alternative hypothesis that the cumulative distribution function of the first population (the lower-mass group) lies above that of the second population (the higher-mass group).}. This result does not change when considering CTTS and WTTS separately: both groups of objects exhibit similar mass properties, as ascertained by comparing their respective cumulative distributions in mass; no statistically supported evidence of a mass dependence in the rotation properties is inferred in either case\footnote{If we repeated the analysis on a possible mass dependence after integrating our sample with additional periodic sources from \citet{lamm04}, which would mostly fall in the M$_\star$<0.4~M$_\odot$ bin, the null hypothesis that objects below 0.4~M$_\odot$ and objects above 1~M$_\odot$ share the same rotation properties would be rejected to the 5\% level. However, this result may be affected by aliases at P$\sim$1~d detected from the ground, which would artificially increase the strength of the peak of fast rotators.}. 

\begin{figure}
\resizebox{\hsize}{!}{\includegraphics{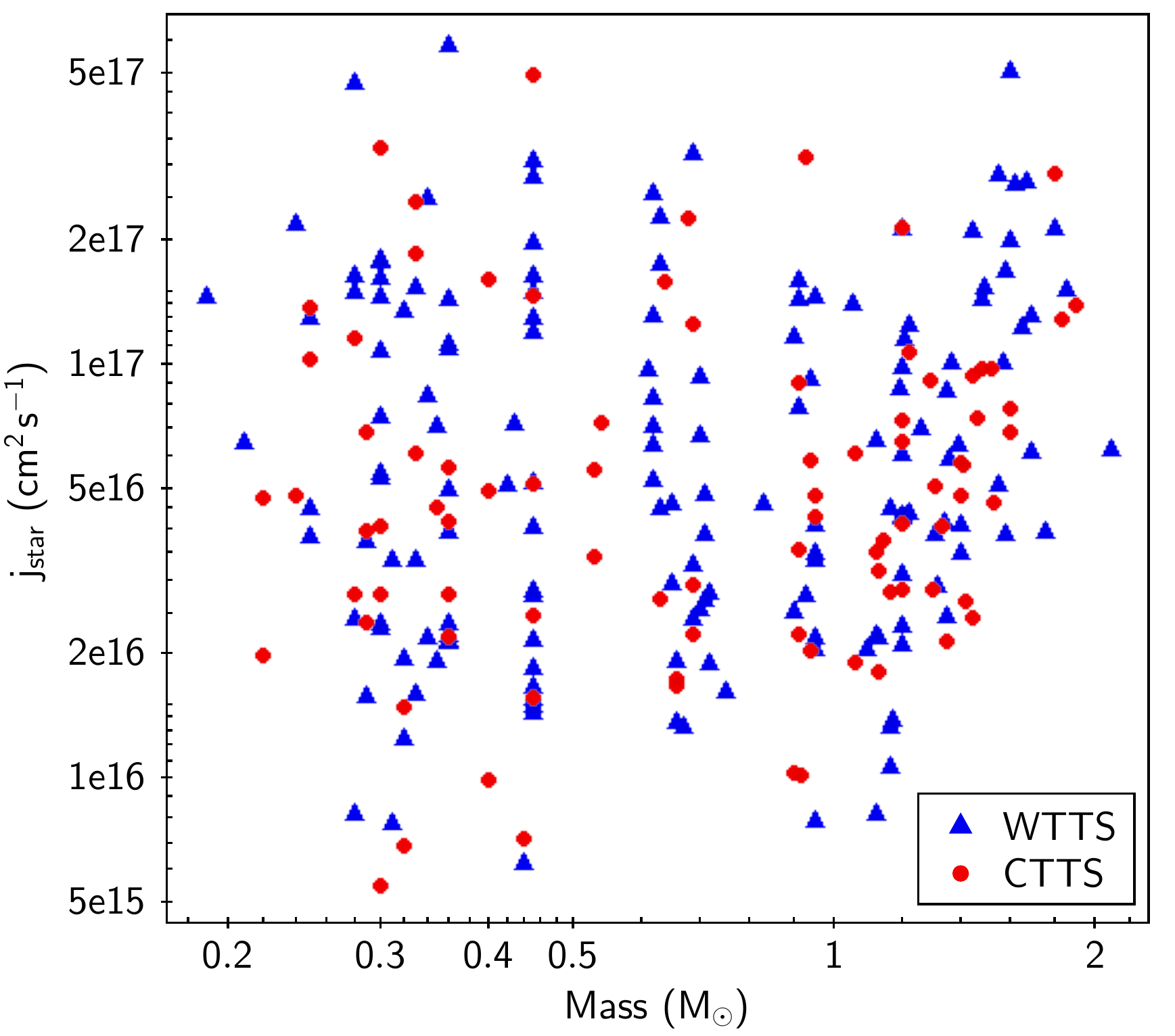}}
\caption{Specific angular momentum, as a function of stellar mass, measured for CTTS (red circles) and WTTS (blue triangles) in our sample. Hints of a break in rotation properties can be observed around M$_\star$$\sim$1.4~M$_\odot$, as discussed in the text.}
\label{fig:jstar}
\end{figure}

A key physical quantity to investigate the rotational evolution of stars is the specific angular momentum {\it j$_{star}$} (see, e.g., \citealp{herbst2005}). This is linked to the period P and radius R$_{\star}$ of the star according to the relation

\begin{equation}\label{eqn:jstar}
j_{star} = k^2 R_\star^2 \omega = \frac{2 \pi k^2 R_\star^2}{P}\,,
\end{equation}

\noindent where $k^2$ is the radius of gyration in units of stellar radius. In Figure~\ref{fig:jstar}, we show the values of $j_{star}$ computed for objects in our sample as a function of mass. In the computation we assumed a constant value of 0.203 for $k^2$, following \citet{vasconcelos2015} and their Figure~12 for a cluster aged of $\sim$3~Myr.
\citet{herbst2001} investigated the $j_{star}$ distribution of objects in the ONC, and found that $j_{star}$ is roughly independent of stellar mass over the mass range 0.1-1~M$_\odot$. Since lower-mass stars are smaller (i.e., have smaller radii) than higher-mass stars, this result suggests that they tend to rotate faster. The $j_{star}$ vs. M$_\star$ distribution we derive here for NGC~2264 is consistent with this picture: no correlation is observed between the two quantities, but the same range of $j_{star}$ values is spanned at any given mass. This would support the visual inference from Fig.\,\ref{fig:hist_tts_mass} of some mass dependence in the rotation properties of NGC~2264 members, with lower-mass objects exhibiting on average shorter rotation periods than higher-mass objects.

The third mass group (M$_\star$$>$1\,M$_\odot$) includes the critical mass (M$_\star$$\sim$1.3~M$_\odot$) at which a break in rotation properties is observed among solar-type stars \citep{kraft1967}. Objects more massive than this threshold have largely radiative interiors, and spend little time along the convective track during their PMS evolution. Convection plays an important role in braking the stars, by powering stellar winds that carry away angular momentum. Massive stars, which are deprived of this mechanism, experience a different rotational evolution from less massive stars with deep convective envelopes, and reach the ZAMS with rotational velocities nearly an order of magnitude higher than the latter. To check whether this effect is already seen at the age of NGC~2264, we selected objects below and above M$_\star$=1.4~M$_\odot$ in this mass group, and compared their respective frequencies in the P<6~d region of the histogram shown in the bottom right panel of Fig.\,\ref{fig:hist_tts_mass}. This period range contains 60\% of objects with 1~M$_\odot$$<$M$_\star$$\leq$1.4~M$_\odot$ and 85\% of objects with M$_\star$$>$1.4~M$_\odot$ (highlighted in red on the period distribution shown in the bottom right panel of Fig.\,\ref{fig:hist_tts_mass}). Moreover, about 70\% of objects in this mass group with P<2~d have M$_\star$$>$1.4~M$_\odot$. The median period measured across objects with mass between 1 and 1.4~M$_\odot$ is 5~d, while the median period measured for objects with mass above 1.4~M$_\odot$ is 3~d. This comparison suggests that some separation in rotation properties below and above M$_\star$$\sim$1.4~M$_\odot$ might already be present at an age of a few Myr, although with no clear break, but a substantial overlap in period distributions. To test whether objects more massive than 1.4~M$_\odot$ may blur the mass trend in rotation properties, we repeated the K-S test between the lowest-mass and highest-mass groups in Fig.\,\ref{fig:hist_tts_mass} after removing objects with M$_\star$\,$\geq$\,1.4~M$_\odot$. This returned a $p$-value of 0.09, lower than that obtained when applying the test to the complete M$_\star$$<$0.4~M$_\odot$ and M$_\star$$>$1.0~M$_\odot$ groups. Thus, the null hypothesis that the period distributions associated with the two mass groups are extracted from the same parent distribution is rejected at the 10\% significance level, but not at the 5\% significance level.

\subsection{Period distribution as a function of variability class}

\begin{figure*}
\centering
\includegraphics[scale=0.65]{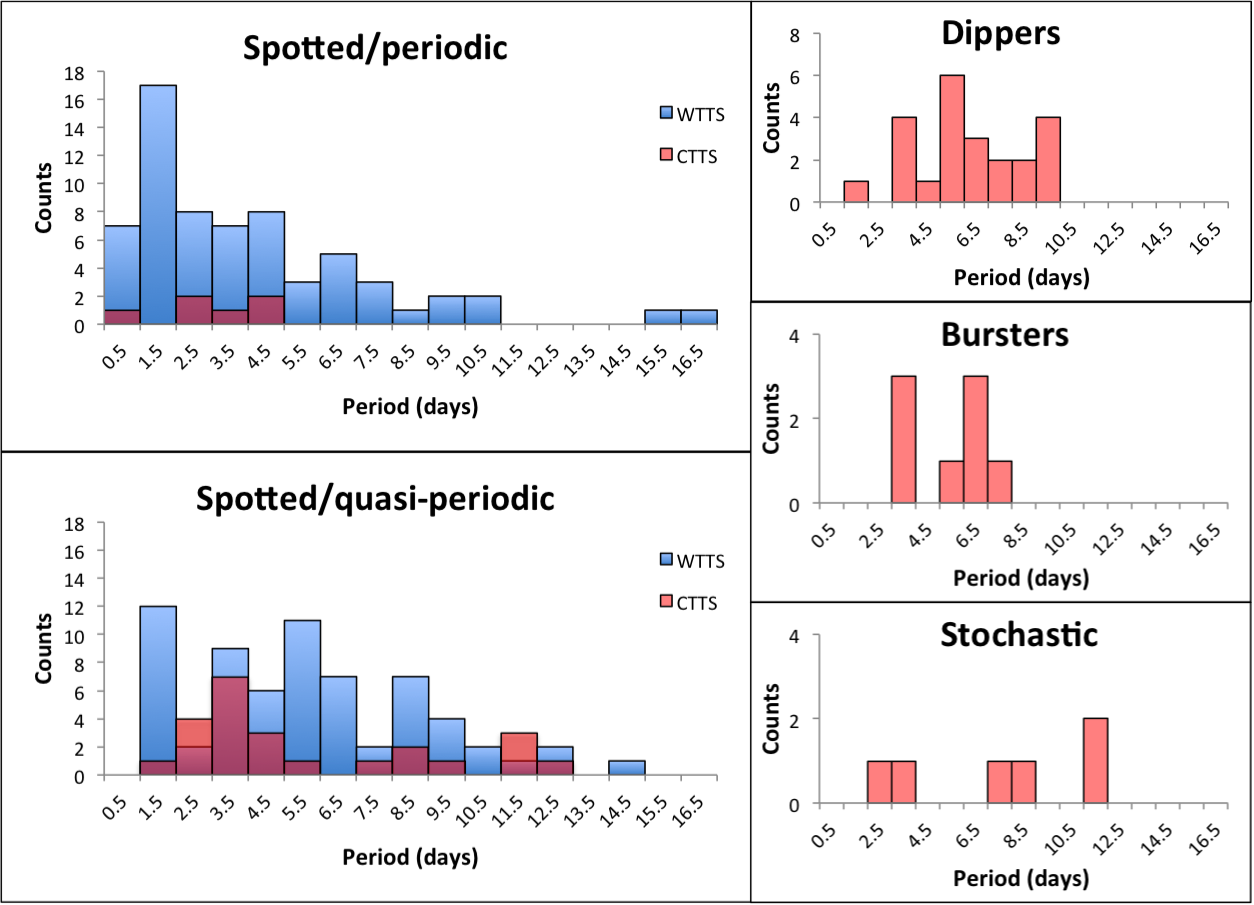}
\caption{Period distributions of different groups of NGC~2264 members, divided according to the morphology of their {\it CoRoT} light curve. WTTS are illustrated in blue, CTTS are overlapped in red.}
\label{fig:per_lctype}
\end{figure*}

Figure \ref{fig:per_lctype} illustrates the period distribution for NGC~2264 members which fall respectively into the strictly periodic (spotted), quasi-periodic (spotted), dipper, burster and stochastic light curve classes, defined based on the morphology of the {\it CoRoT} light curves (Appendix~\ref{app:light_curve_types}). The first two classes include both WTTS and CTTS, whereas the classes of dippers, bursters and stochastic light curves are specific to disk-bearing objects.

Among sources with spot-dominated light curves, a larger fraction is recovered in the quasi-periodic class than in the strictly periodic class (104 vs. 75). In addition, the CTTS/WTTS ratio differs significantly between the two classes: less than 10\% of strictly periodic sources are CTTS, while nearly 30\% of spotted sources with quasi-periodic light curves are CTTS. This can be understood if we consider that the light curves of CTTS typically result from a variety of co-existing mechanisms (spot modulation, spot evolution, accretion); therefore, their morphology may exhibit rapid changes, even when the periodicity does not evolve during the monitored time. Conversely, the light variations of WTTS are driven by cold spot modulation; these spots are often stable and long-lived, thus leading to stable light curve morphology on tens or hundreds of rotational cycles. Nevertheless, a significant fraction of WTTS falls into the definition of quasi-periodic light curves; this may reflect spot evolution or migration on timescales of a few stellar periods.

The different ratio of CTTS to WTTS in the two classes is also reflected in the properties of the relevant period histograms. The period distribution of purely periodic light-curve objects include more fast rotators than the group of quasi-periodic light curves, and conversely, more slow rotators are included in the quasi-periodic sample than in the strictly periodic sample. No fast rotators have light curves in the dipper, burster or stochastic categories; the periods measured among these latter samples span a broad range of values, from a few to several days. Only a few sources with periodic signatures are detected among the burster and the stochastic light curve types; this is due to the episodic nature of the physical processes which dominate their light variations \citep{stauffer2014, stauffer2016}. 

\subsection{Are CTTS periods similar in nature to WTTS periods?}

The derivation of stars' rotation periods from monitoring their photometric variability relies on the assumption that these light variations are dominated by localized temperature inhomogeneities at the stellar surface. These inhomogeneities would then modulate the apparent luminosity of the stars as these spin on their axes, with a characteristic timescale of variability equal to the rotation period of the star. Recently, \citet{artemenko2012} questioned the validity of this assumption as a general rule for CTTS. The authors examined the light curve of about 50 CTTS in the photometric catalog of \citet{rotor_ctts} to derive their rotational periods via power spectrum analysis. The rotational period of a star can be expressed as
\begin{equation} \label{eqn:period}
P=\frac{2\pi R_\star}{v} \equiv \frac{2 \pi R_\star \sin{i}}{v\,\sin{i}}\,
\end{equation}
where $v$ is the equatorial velocity, $i$ is the inclination\footnote{Angle between the rotation axis and the line of sight to the observer.} of the system, and $v\,\sin{i}$ is the projected rotational velocity, measured from spectroscopic observations. An estimate of $\sin{i}$ can then be derived from Eq.\,\ref{eqn:period}, if P, $v\,\sin{i}$ and R$_\star$ are known. If the value of period P we measure is photospheric, then estimates of $\sin{i}\lesssim1$ ought to be derived from Eq.\,\ref{eqn:period}. Instead, $\sin{i}$ estimates larger than 1 were obtained by \citet{artemenko2012} across their sample of CTTS. Objects were found to trace a unique sequence on the $\sin{i}$ vs. P diagram, with $\sin{i}$ tending to increase with P (see their Fig.\,2). The authors suggested that in some cases, the measured photometric periods for CTTS do not arise from surface spot modulation, but from clumps of dust in the disk which periodically occult the stellar photosphere, at a rate corresponding to the Keplerian orbit where they are located in the disk. In this case, the formal application of Eq.\,\ref{eqn:period} to derive $\sin{i}$ will yield values larger than 1, since the measured period originates at distances larger than the stellar radius (unless they arise from the co-rotation radius, or close to it).

Conversely, there should be no ambiguity on the nature of the photometric periods measured for disk-free young stars; therefore, the same test, applied to a sample of WTTS, is expected to produce estimates of $\sin{i}$ systematically lower than 1, conditional upon the accuracy of the stellar parameters determined for those objects.

\begin{figure*}
\centering
\includegraphics[width=\textwidth]{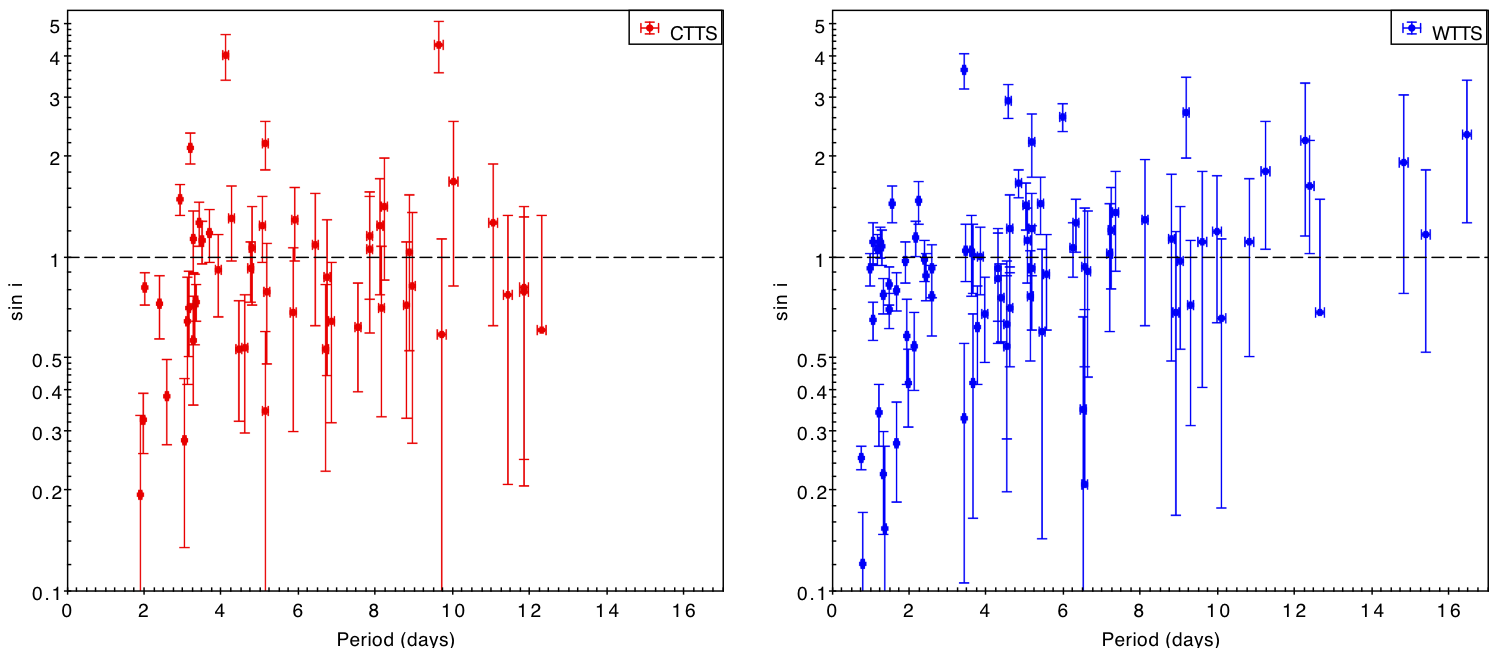}
\caption{Estimates of sin\,i obtained, as a function of the rotation period, for CTTS (red; left panel) and WTTS (blue; right panel) in our sample with v~sin\,i measurements from \citet{baxter2009}, periods from this study and R$_\star$ estimates from \citet{venuti2014}. The sin\,i = 1 level is marked as a black dashed line. Horizontal error bars are derived following Eq.\,\ref{eqn:Perr_theory}. Vertical error bars are derived by standard error propagation on Eq.\,\ref{eqn:sini}. A typical uncertainty of 5~km/s is assumed on $v\,\sin{i}$; this corresponds to the median difference between $v\,\sin{i}$ estimates by \citet{baxter2009} and those derived by S.~Alencar from CSI~2264 VLT/FLAMES observations, for objects common to the two samples. Similarly, a typical uncertainty of 0.2~R$_\odot$ is adopted for R$_\star$, that is the median difference between R$_\star$ estimates obtained by \citet{venuti2014} and those of \citet{rebull2002}, for objects in common.}
\label{fig:per_sini}
\end{figure*}

Here we follow \citeauthor{artemenko2012}'s (\citeyear{artemenko2012}) approach to test the nature of the photometric periods we measured for NGC~2264 members, and invert Eq.\,\ref{eqn:period} to derive $\sin{i}$ estimates for CTTS and WTTS in our sample. Values of $\sin{i}$ are calculated as
\begin{equation} \label{eqn:sini}
\sin{i} = 0.0195\, P\, (v\,\sin{i})/R_\star\,,
\end{equation}
where P is expressed in days, $v\,\sin{i}$ is in km\,s$^{-1}$, and R$_\star$ in solar radii R$_\odot$. We use R$_\star$ estimates from \citet{venuti2014}, while $v\,\sin{i}$ values are retrieved from the study of \citet{baxter2009}. For this test, we selected 51 CTTS and 81 WTTS, common to the samples of \citet{venuti2014} and \citet{baxter2009}, with single periodicity detected in the present study. The results of this computation are shown, as a function of period, in Fig.\,\ref{fig:per_sini}.

When comparing the $v\,\sin{i}$ distributions for CTTS and WTTS in our sample, no statistically significant difference between the two is found, although the average $v\,\sin{i}$ is larger in the WTTS group than in the CTTS group. A similar analysis was presented by \citet{rhode2001} for the PMS population of the ONC. In the majority of cases (84\%) across our sample, values of $\sin{i}\leq 1$ are obtained, within the associated uncertainties, from Eq.\,\ref{eqn:sini}. This result is similar to that obtained by \citet{artemenko2012}. As noted by the latter, only a few points appear at values of $\sin{i}<0.2$; this is a selection effect due to the fact that for low inclinations (nearly pole-on objects), the photometric modulation is difficult to detect. A number of objects fall above the $\sin{i}=1$ line on Fig.\,\ref{fig:per_sini}; interestingly, a fraction of both the CTTS (6/51, or 11.8\%) and the WTTS (15/81, or 18.5\%) groups are found in this region of the diagram. For both CTTS and WTTS, the average $\sin{i}$ computed neglecting sources with $\sin{i}$ estimates larger than 1 is $<\sin{i}> = 0.6 \pm 0.2$; this value is consistent with that found by \citet{rhode2001} for the ONC. In at least a few of the 6 CTTS with $\sin{i}$ estimate larger than 1, this result may be severely affected by an erroneous v\,$\sin{i}$ measurement: largely discrepant (lower) v\,$\sin{i}$ values from \citeauthor{baxter2009}'s (\citeyear{baxter2009}) estimates are derived, for the same sources, from VLT/FLAMES spectra obtained within the CSI~2264 campaign. Among the 15 WTTS with $\sin{i}>1$ on Fig.\,\ref{fig:per_sini}, an assessment of the impact of uncertainties on v\,$\sin{i}$ measurements is more complicated, as not many of them have additional v\,$\sin{i}$ derivations for comparison purposes. In the few cases where such a comparison is possible, the v\,$\sin{i}$ measurements from different sources are not too dissimilar from each other, and the estimate of $\sin{i}$ obtained is only marginally larger than 1; the discrepancy here can then likely be explained in terms of uncertainties on the parameters adopted for the sources. While in a few cases it may be possible that these objects with no sign of accretion might still possess some material in the circumstellar environment, most of these objects exhibit strong evidence of being disk-free young cluster members. Therefore, uncertainties on the nominal parameters used for the $\sin{i}$ computation are likely to affect significantly the results shown in Fig.\,\ref{fig:per_sini}. At any rate, the conclusion we may derive from this Figure is that, for the majority of the objects investigated here, the period measured from the {\it CoRoT} light curves is likely photospheric, and hence corresponds to the spin rate of the star. Thus, the distribution in $\sin{i}$ appears to be independent of the accretion status of the objects.

\section{The rotation -- accretion connection in NGC~2264} \label{sec:disklocking}

\begin{figure*}
\centering
\includegraphics[scale=0.65]{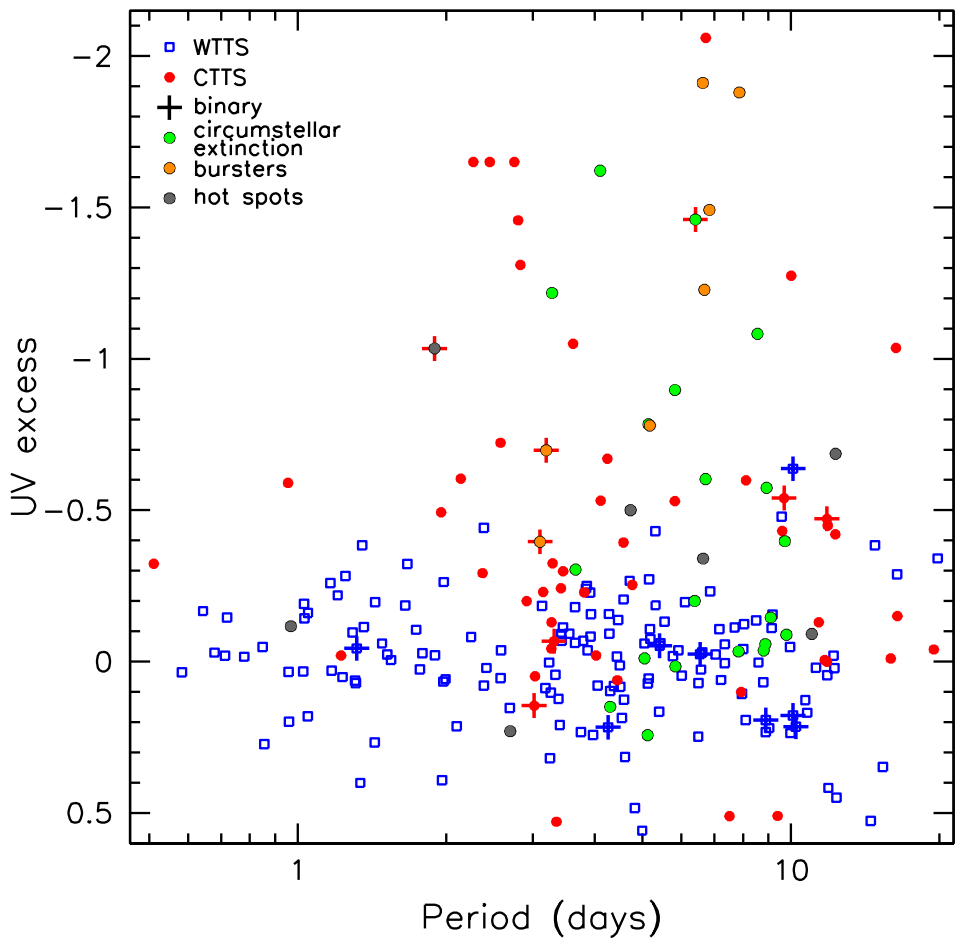}
\caption{UV excesses measured for NGC~2264 cluster members as described in \citet{venuti2014} are compared to their measured rotation periods, reported in Table \ref{tab:periods} of the present study. Filled dots correspond to disk-bearing objects (CTTS), while disk-free WTTS are indicated as empty blue squares. The different color groups among CTTS correspond to different variable classes, as detailed in the legend (green indicates light curves dominated by flux dips; orange corresponds to stochastic bursters; grey is used for CTTS with light curves dominated by spot modulation; red corresponds to other CTTS members whose light curve is not ascribed to any specific class among the ones listed before). A cross superimposed on a dot or square signifies that the corresponding object is a known spectroscopic binary.}
\label{fig:Corot_UV_P_AATau}
\end{figure*}

Fig.\,\ref{fig:hist_tts_mass} illustrates that statistically distinct, though overlapping, behaviors in rotation properties characterize cluster members with accretion disk signatures with respect to disk-free sources. Namely, a pronounced peak of fast rotators (P$\sim$1-2~d) appears in the period distribution pertaining to WTTS, whereas very few fast rotators are found among objects still surrounded by disks. 

As discussed in Sect.\,\ref{sec:introduction}, the impact of accretion disks on the early rotational evolution of young stars is an issue that has long been debated over the past decades. Observational evidence has been gathered in support of a positive connection between the presence of disks and the measured rotation rates within coeval populations of young stars; objects still surrounded by disks, and hence presumably in magnetospheric interaction with them, appear to rotate more slowly, on the average, than young stars whose disks have already disappeared. A number of studies \citep[e.g.,][]{herbst2002, rebull2006, cieza2007} have shown that the frequency of objects with near-infrared excess (indicative of dusty inner disks) increases with the rotation period, hence lending credit to this ``disk-locking'' scenario. Moreover, recent studies using Monte-Carlo simulations to investigate the early rotational evolution of low-mass stars \citep[e.g.,][]{vasconcelos2015} have shown that, starting from the disk-locking assumption, the period distributions observed for young clusters of different ages can be reproduced reasonably well. However, as illustrated in Sect.\,\ref{sec:introduction} (see also \citealp{bouvier2014}), no definite consensus has yet been reached on this issue, and which mechanisms provide an effective source of braking in young stars is still a matter of controversy.

The magnetospheric accretion process plays an important role in regulating the star-disk interaction during the first few Myr of a stellar lifetime. Measurements of the infrared excess of YSOs, although providing diagnostics of the presence of dusty disks around these young stars, are not able to probe the rate of mass accretion onto the central source. UV excess measurements are, instead, a direct indicator of accretion, as they probe the hot excess emission which arises from the accretion shock at the stellar surface. First studies to investigate a possible correlation between UV excesses and rotation periods include \citet{rebull2001} in Orion and \citet{makidon04} in NGC~2264, although no conclusive evidence could be drawn from those analyses. Later, \citet{fallscheer2006} determined UV excess estimates for a sample of 95 NGC~2264 members with known rotation periods from the studies of \citet{makidon04} or \citet{lamm04}, and compared these two quantities to show the presence of an overall association between the two: slowly rotating stars appeared to be more likely to have large UV excesses (and hence, strong ongoing accretion activity) than faster rotators. 

In Fig.\,\ref{fig:Corot_UV_P_AATau}, we report the same comparison for the broader sample available for NGC~2264 from the CSI~2264 campaign. The sample of members and their classification as CTTS or WTTS, as well as their measured UV excesses, follow from the CFHT-based study of \citet{venuti2014}\footnote{See in particular Eq.\,9 of that paper for details on how the UV excess measurements were obtained; see also discussion in Sect.\,3.2 of \citet{venuti2015} regarding the significance of such UV excess measurements for WTTS.}; the rotation periods are those derived in the present study and reported in Table~4. WTTS distribute horizontally across the whole period range, and exhibit no UV excess within an uncertainty of $\pm$0.2 mag. Conversely, very few CTTS are found at periods shorter than 1.5--2~days; furthermore, a dearth of strong accretors (UV excess larger, i.e., more negative, than -0.75 mag) with short rotation periods is clear in the diagram. The features of Fig.\,\ref{fig:Corot_UV_P_AATau} do not exhibit any mass dependence, as the same qualitative diagram is recovered when splitting the sample into the three mass bins indicated on the right panel of Fig.\,\ref{fig:hist_tts_mass}. These features recall the analogous diagram presented in \citet[][Fig.\,3]{rebull2006}, which juxtaposes rotation periods and mid-IR excesses measured from {\it Spitzer/IRAC} data, in the case of Orion. There, the authors found that objects with clear disk signatures stand out with respect to those with no IR excess, and are clustered at periods longer than $\sim$1.8~days; conversely, objects with no evidence of disk were found to span the whole range in periods (from $\sim$0.3 to 10~days), as is the case here. 

When exploring the distribution in $\dot{M}_{acc}$\footnote{Accretion rates are derived from UV excesses shown in Fig.\,\ref{fig:Corot_UV_P_AATau} as described in \citet{venuti2014}.} of CTTS as a function of period, no 1-to-1 correspondence between the two quantities is found; rather, at a given period, $\dot{M}_{acc}$ values can span over an order of magnitude. This variety of $\dot{M}_{acc}$ regimes at a given P may reflect distinct accretion mechanisms \citep{romanova2004, kulkarni2008}, or correspond to different accretion histories \citep{matt2012}, perhaps linked to varying properties of the circumstellar environment for individual objects (e.g., disk masses; see \citealp{manara2016}). The fact that a given rotation period includes objects with diverse accretion rates may indicate that distinct components of the star-disk environment are predominant in determining the two sets of properties: the accretion rate is ultimately regulated by the small-scale magnetic field structure in proximity of the stellar surface; conversely, the star-disk coupling and angular momentum transfer is dominated by the large-scale, ordered dipole component of the stellar magnetosphere \citep[see discussion in][]{gregory2012}. The fact that the magnetic field strength may vary from object to object was also suggested by \citet{muzerolle2001} to explain the lack of an observed correlation between P and $\dot{M}_{acc}$, which would be expected from a theoretical standpoint if higher accretion rates tend to push the disk truncation radius R$_T$ closer to the star (cf. Sect.\,\ref{sec:Rt} and Eq.\,\ref{eqn:RtoverRc}).

\subsection{Disk-locking?}

As discussed in Sect.\,\ref{sec:introduction}, one of the most debated issues regarding angular momentum regulation in young stars concerns the role played by star-disk interaction, and the possible magnetic star-disk locking that would prevent the objects from spinning up during the disk accretion phase. In this Section, we explore some main concepts related to the magnetospheric accretion picture and test their agreement with the observational parameters measured for NGC~2264 disk-bearing objects.

\subsubsection{Truncation radius} \label{sec:Rt}

A critical parameter in the star-disk magnetospheric interaction is the location of the truncation radius R$_T$, i.e., the radial distance from the star at which the inner disk is disrupted by the stellar magnetosphere. Another important distance in the picture of magnetospheric accretion is the corotation radius R$_C$, that is, the radius of the disk annulus where the Keplerian velocity of the disk equals the angular velocity of the star. Disk material orbiting the star at distances R$<$R$_C$ rotates faster than the star, while at radial distances R$>$R$_C$ the magnetosphere threading the disk rotates faster than the corresponding Keplerian orbit in the disk. Thus, the outcome of the magnetosphere-disk interaction will depend on the mutual position of R$_T$ and R$_C$. When R$_T$<R$_C$, at the interaction interface a negative magnetic torque will be exerted on the disk material; this favors its channeling along the magnetic field lines and subsequent accretion onto the star. Conversely, when R$_T$>R$_C$, at the interaction interface the magnetosphere rotates faster than the disk material, which is then accelerated along the azimuthal direction and may eventually be expelled radially from the system (the so-called ``propeller'' regime; \citealp{ustyugova2006}). In the latter scenario, no stable funnel-flow accretion can occur.

\begin{figure*}
\centering
\includegraphics[width=\textwidth]{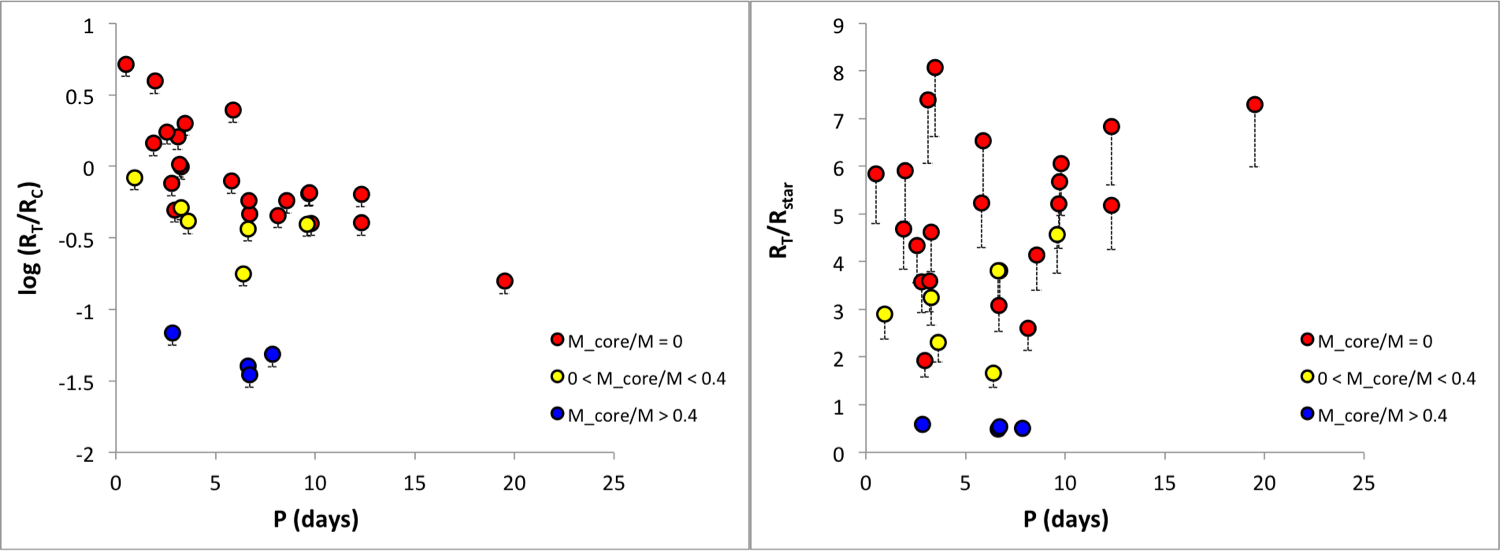}
\caption{Estimated ratios of truncation-to-corotation radius (R$_T$/R$_C$, left) and truncation-to-stellar radius (R$_T$/R$_\star$, right), as a function of rotation period for a subsample of CTTS in NGC~2264 with period measurements from this study and stellar and accretion parameters measured from \citet{venuti2014}. The estimates of R$_T$ and R$_C$ are obtained following \citet{bessolaz2008}, in the assumption that steady accretion funnels are formed. Circles correspond to truncation radius estimates obtained assuming that the sonic Mach number at the disk midplane ($m_S$) is equal to 1; the lower bar associated with each point marks by how much the values of R$_T$ or R$_T$/R$_C$ would vary if we adopted $m_S$ = 0.5 \citep[see][]{bessolaz2008}. Red is used for objects with a mass ratio M$_{core}$/M$_\star$ between the radiative core and the total mass of the star equal to 0 (i.e., fully convective objects); yellow identifies objects in the 0$<$M$_{core}$/M$_\star$$<$0.4 group; blue is for objects with M$_{core}$/M$_\star$$>$0.4. Different dipolar field strengths are used in the computation of R$_T$ for each of these groups (see text).}
\label{fig:Rt_RtRc_P}
\end{figure*}

If the physical conditions for the creation of stable, magnetically driven accretion funnels are met, the R$_T$-to-R$_C$ ratio in a given system is expected, from a theoretical standpoint, to be related to the stellar mass M$_\star$, radius R$_\star$, accretion rate $\dot{M}_{acc}$, rotation period P and magnetic field B$_\star$\footnote{Strength of the dipolar component of the stellar magnetic field at the equator. As discussed in \citet{gregory2012}, the magnetic field hosted by T Tauri stars can strongly depart from a pure dipole; the degree of complexity of the magnetic field at the stellar surface, and the importance of higher-order components relative to the dipolar component, tend to increase, along the HR diagram, from fully convective stars to objects with a radiative core. However, the dipolar component of the magnetic field dominates on the large scale, as it decays more slowly with distance from the central source; hence, it is the dipolar component that regulates the star-disk interaction at the truncation radius.} according to the following equation \citep{bessolaz2008}:
\begin{equation} \label{eqn:RtoverRc}
\frac{R_T}{R_C} \simeq 0.25\,m_S^{2/7} \,B_\star^{4/7} \, \dot{M}_{acc}^{-2/7} \, M_\star^{-10/21} \, R_\star^{12/7} \, P^{-2/3}\, ,
\end{equation} 
where $m_S$ is the Mach number at the disk midplane, B$_\star$ is normalized to 140~G, M$_\star$ is normalized to 0.8~M$_\odot$, R$_\star$ is normalized to 2~R$_\odot$, $\dot{M}_{acc}$ is normalized to 10$^{-8}$~M$_\odot$/yr and P is normalized to 8~d. To test whether this prescription yields R$_T$/R$_C$ estimates consistent with the picture of stable funnel-flow accretion for NGC~2264 members, we selected a subsample of accreting objects for which all parameters listed in Eq.\,\ref{eqn:RtoverRc} are known from this study (rotation period) and from \citet[][stellar parameters and mass accretion rate]{venuti2014}. To each of these objects, a value of B$_\star$ was assigned, following \citet{gregory2012}, based on their estimated radiative core mass to stellar mass ratio (M$_{core}$/M$_\star$), deduced from \citeauthor{siess2000}'s (\citeyear{siess2000}) evolutionary tracks. Specifically, a value of B$_\star$=1.5~kG\footnote{This value quoted here, as the following ones, are average estimates of the polar dipole strength for objects in different M$_{core}$/M$_\star$ regimes, as discussed in \citet{gregory2012}. However, in Eq.\,\ref{eqn:RtoverRc}, the magnetic field is that measured at the equator. For a dipole, this corresponds to half the strength measured at the magnetic pole.} was adopted for objects with M$_{core}$/M$_\star$=0 (i.e., fully convective stars); B$_\star$=0.6~kG was used for objects with 0$<$M$_{core}$/M$_\star$$\lesssim$0.4; B$_\star$=0.1~kG was adopted for objects with a developed radiative core (M$_{core}$/M$_\star$$\gtrsim$0.4). Objects considered for this test are listed in Table~\ref{tab:obj_rtrunc_rcor}, where the relevant R$_\star$, $\dot{M}_{acc}$ and M$_{core}$/M$_\star$ parameters are reported; their masses and rotation periods are instead reported in Table~4. Following \citet{bessolaz2008}, two values of $m_S$, 1 and 0.5, were considered for the computation of R$_T$/R$_C$; both sets of results are illustrated in Fig.\,\ref{fig:Rt_RtRc_P}.

\begin{table}
\caption{Subset of CTTS and relevant parameters used to test the disk locking assumptions.}
\label{tab:obj_rtrunc_rcor}
\centering
\begin{tabular}{c c c r c c}
\hline\hline
{\small CSIMon-\#} & M$_\star$\tablefootmark{1} & R$_\star$\tablefootmark{1} & {\small $\log(\dot{M}_{acc})$}\tablefootmark{1} & M$_{core}$/M$_\star$\tablefootmark{2} & P\,(d)\\
\hline
000007 & 0.69 & 2.4 & -7.22 & 0.00 & 3.192\\
000153 & 0.29 & 1.4 & -8.04 & 0.00 & 1.896\\
000290 & 0.25 & 3.4 & -7.53 & 0.00 & 5.900\\
000314 & 0.29 & 1.4 & -8.02 & 0.00 & 3.279\\
000326 & 0.66 & 1.3 & -8.77 & 0.03 & 6.642\\
000358 & 0.29 & 1.4 & -8.18 & 0.00 & 5.821\\
000406 & 1.13 & 1.3 & -7.32 & 0.81 & 6.631\\
000412 & 0.45 & 2.2 & -6.98 & 0.00 & 6.679\\
000433 & 0.44 & 1.0 & -8.86 & 0.00 & 9.798\\
000484 & 0.13 & 0.3 & -10.04 & 0.00 & 19.50\\
000619 & 0.69 & 1.4 & -7.40 & 0.01 & 6.404\\
000637 & 0.45 & 1.6 & -8.50 & 0.00 & 12.31\\
000717 & 0.53 & 2.0 & -7.55 & 0.00 & 8.558\\
000766 & 0.53 & 1.5 & -7.67 & 0.00 & 2.798\\
000926 & 0.40 & 1.3 & -8.30 & 0.00 & 12.32\\
000964 & 0.95 & 1.5 & -8.43 & 0.23 & 3.289\\
000965 & 0.36 & 1.7 & -7.99 & 0.00 & 9.688\\
001003 & 0.24 & 1.5 & -8.70 & 0.00 & 3.454\\
001054 & 0.36 & 2.2 & -6.68 & 0.00 & 8.142\\
001114 & 0.40 & 2.4 & -7.37 & 0.00 & 2.579\\
001132 & 0.33 & 1.6 & -6.55 & 0.00 & 2.958\\
001187 & 0.40 & 1.5 & -8.72 & 0.00 & 3.102\\
001199 & 1.20 & 1.9 & -7.69 & 0.01 & 3.617\\
001217 & 1.30 & 1.8 & -7.05 & 0.52 & 7.865\\
001234 & 0.94 & 1.7 & -8.84 & 0.10 & 9.606\\
001249 & 0.30 & 3.0 & -7.53 & 0.00 & 1.954\\
001294 & 0.92 & 1.0 & -7.68 & 0.87 & 6.723\\
001296 & 0.69 & 2.0 & -8.10 & 0.00 & 9.725\\
001308 & 0.63 & 1.6 & -7.72 & 0.00 & 6.717\\
006079 & 0.45 & 1.9 & -8.12 & 0.00 & 0.511\\
006325 & 0.54 & 1.0 & -8.59 & 0.06 & 0.956\\
006465 & 0.88 & 1.0 & -7.88 & 0.88 & 2.829\\
\hline
\end{tabular}
\tablefoot{
\tablefoottext{1}{Values from \citet{venuti2014}. Masses and radii are reported in units of M$_\odot$ and R$_\odot$, respectively.}
\tablefoottext{2}{Estimate derived using the temperature and luminosity parameters derived for the object in \citet{venuti2014} and \citeauthor{siess2000}'s (\citeyear{siess2000}) model tracks.}
}
\end{table} 

The results of the R$_T$ and R$_T$/R$_C$ computation for this subset of objects are shown, as a function of P, in Fig.\,\ref{fig:Rt_RtRc_P}.
As ensuing from Eq.\,\ref{eqn:RtoverRc}, stronger dipolar fields tend to disrupt the disks at larger radii; this is reflected in the vertical separation of the three different color groups (corresponding to different B$_\star$ assumptions) on the diagrams. The right panel of Fig.\,\ref{fig:Rt_RtRc_P} shows that typical R$_T$ estimates range from a few to several stellar radii \citep[see also][]{johnstone2014}. Values of R$_T$/R$_\star$$<$1 are obtained from Eq.\,\ref{eqn:RtoverRc} for the group of objects with the largest radiative cores; this may indicate that the average value of B$_\star$ adopted for these sources is an underestimate to the actual magnetic field strength. In the majority of cases, the ratio of the truncation-to-corotation radius, illustrated in the left panel of Fig.\,\ref{fig:Rt_RtRc_P}, is smaller than 1 or close to 1; this is consistent with the expected behavior in the magnetospheric accretion picture, as discussed earlier. At the same time, a R$_T$/R$_C$ ratio larger than 1 is obtained for a small group of objects in our sample, among the stars with M$_{core}$/M$_\star$=0. Although the derived values of R$_T$ may be affected by uncertainties on the stellar parameters and especially on the magnetic field intensity, which is not constrained here on an individual basis, it is interesting that objects with estimated R$_T$/R$_C$$>$1 are clustered at short rotation periods on Fig.\,\ref{fig:Rt_RtRc_P} (left panel). High stellar spin rates push the corotation radius closer to the star, hence favoring a scenario where the disk is truncated beyond the corotation radius. As mentioned earlier, in this scenario the inner disk material may be ejected from the system along open field lines, in the propeller regime \citep{ustyugova2006}, carrying away angular momentum. This mechanism can efficiently spin down the star on timescales of $\lesssim$10$^6$~yr \citep{ustyugova2006}, shorter than the typical CTTS lifetimes of several Myrs. Therefore, this would suggest that objects accreting in the propeller regime at the age of NGC~2264 might be expected to lie on the longer-period side of Fig.\,\ref{fig:Rt_RtRc_P}, rather than on the short-period side. Objects with the largest R$_T$/R$_C$ exhibit small UV excesses (and, therefore, weak $\dot{M}_{acc}$) with respect to the bulk of NGC~2264 disk-bearing members, as inferred when comparing the R$_T$/R$_C$ estimates derived here to the accretion parameters derived in \citet{venuti2014} for the whole NGC~2264 sample. A few of them would be classified as significantly younger ($<$10$^6$~yr) than the other cluster members following \citeauthor{siess2000}'s (\citeyear{siess2000}) model tracks on the H-R diagram of the cluster. This might suggest that they are young objects, accreting in a propeller regime, which have not yet been significantly spun down by the star-disk interaction mechanism. Nevertheless, the global inference we may derive from Fig.\,\ref{fig:Rt_RtRc_P} is that there is no relationship between the truncation-to-corotation radius ratio and the rotation period of disk-bearing sources in NGC~2264, in the sense that values of R$_T$/R$_C$ consistent with the magnetospheric accretion picture are found across the whole period range.

\citet{gregory2012} suggested that magnetic topology and its evolution as the star ages may have a direct impact on the rotational evolution of young stars. Namely, as the dipole component becomes weaker and the field complexity increases when the stars start to develop a radiative core, the magnetic ram pressure close to the truncation radius will decrease and hence the disk may push closer to the star. At this stage, the star would start to spin up due to the combined effects of the magnetosphere-inner disk angular momentum exchange, of the accretion process onto the star, and of the stellar contraction. Therefore, a connection would be expected between the measured rotation period of the star and the strength/complexity of its magnetic field. \citet{johnstone2014} investigated this connection across a sample of 10 CTTS with reconstructed magnetic maps from the MaPP project \citep[e.g.,][]{MaPP}, and found that sources which host a stronger dipolar field tend to be associated with longer rotation periods. Even though no detailed knowledge of the magnetic topology is available for NGC~2264 members investigated here, some indications on a possible connection between stellar rotation and inner structure may be inferred by comparing how objects belonging to different M$_{core}$/M$_\star$ groups distribute in P on Fig.\,\ref{fig:Rt_RtRc_P}. However, no definite evidence of such a relationship appears on this diagram: although the few points at P$>$10-12~d all fall into the M$_{core}$/M$_\star$=0 group (and are therefore associated with larger dipole strengths, following \citealp{gregory2012}), objects belonging to any M$_{core}$/M$_\star$ group are similarly mixed at shorter periods.
\begin{figure*}
\centering
\includegraphics[width=\textwidth]{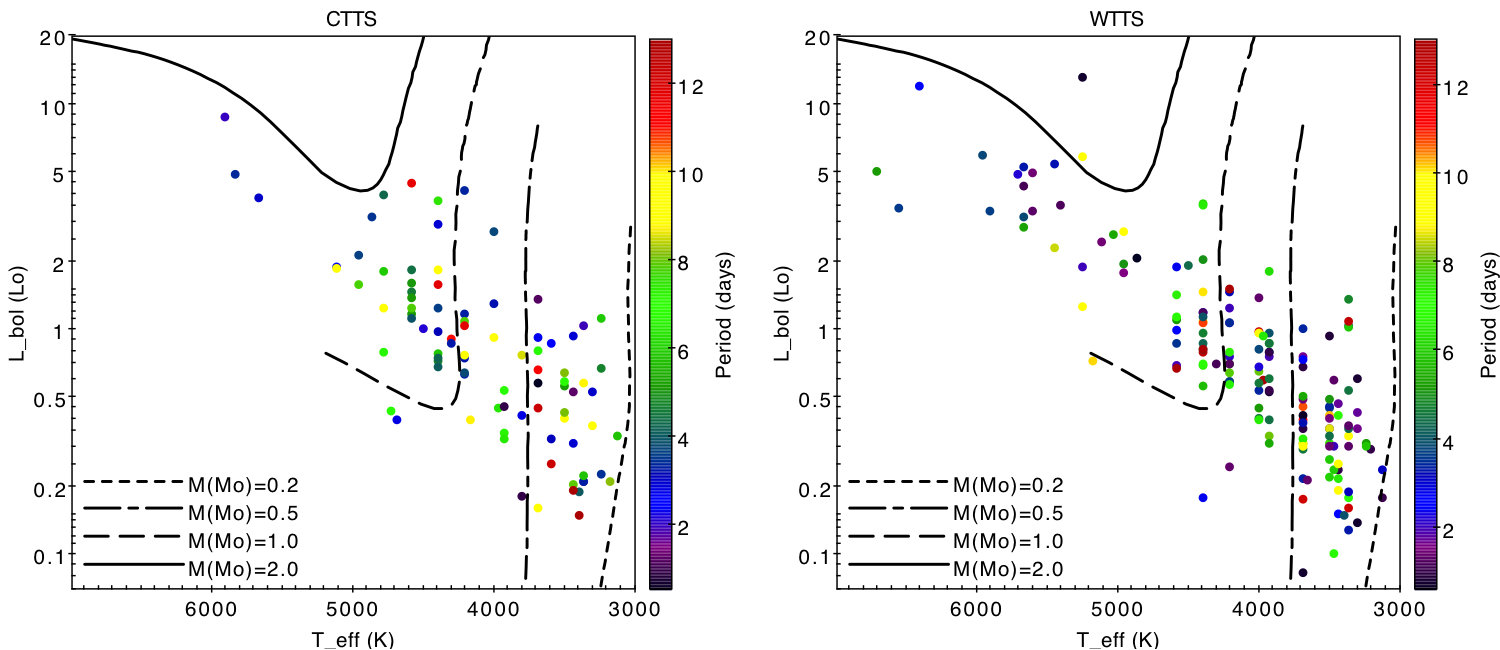}
\caption{Distribution, on the H-R diagram, of CTTS (left panel) and WTTS (right panel) in NGC~2264 with period measurements from the present study. A color scale, proportional to the period as indicated on the side axis to the right of the diagrams, is used to fill the symbols of each source. Mass tracks shown on the plots, truncated at an age of 30~Myr, are from \citeauthor{siess2000}'s (\citeyear{siess2000}) evolutionary models.}
\label{fig:tts_HR_per}
\end{figure*}
To test this scenario further for the NGC~2264 sample, we examined the position of CTTS and WTTS on a H-R diagram with reference to their rotation properties; this is illustrated in Fig.\,\ref{fig:tts_HR_per}. Again, no evolution in rotation properties is observed along the mass tracks, as objects evolve from being fully convective to developing a radiative core; shorter rotation periods (corresponding to violet/blue on the diagrams) and longer rotation periods (yellow/red) are well mixed across the depicted sample. This suggests that the internal structure has no obvious impact on the rotation properties of young stars belonging to the NGC~2264 cluster. This conclusion, in contrast with the picture discussed in \citet{gregory2012}, may be a consequence of the youth of NGC~2264: at an age of $\sim$3~Myr, most of its members are still fully convective; therefore, the sample we are investigating here may be unsuitable for testing the connection between the evolution of magnetic topology and the rotational evolution of young stars.

\subsubsection{Testing magnetospheric accretion models}

Theories of magnetospheric accretion predict definite relationships between the stellar parameters (M$_\star$, R$_\star$), the rotation period P, the magnetic field B$_\star$ and the accretion rate $\dot{M}_{acc}$. These predictions therefore provide an indirect way to test the validity of the magnetospheric accretion picture, when those same correlations are looked for among measured parameters for a given set of accreting young stars. \citet{johns2002} examined different theories of magnetospheric accretion, which in particular cover different assumptions on the geometry of the magnetic field. They collected a sample of a few tens of CTTS with known stellar, rotation, and accretion parameters from the literature, and showed that the trend observed in the data best agrees with the predicted correlation of a modified version of the magnetospheric accretion theory presented in \citet{ostriker1995}, where the magnetic field topology is allowed to divert from a dipole. This predicted correlation is the following:
\begin{equation} \label{eqn:os95}
R_\star^2\, f_{acc}\, \propto\, M_\star^{1/2}\, \dot{M}_{acc}^{1/2}\, P^{1/2}\,,
\end{equation}
where $f_{acc}$ is the filling factor of the accretion spots at the stellar surface, and it is assumed that the magnetic field strength B$_\star$ which participates in the accretion flow at the stellar surface does not vary significantly from object to object. \citet{cauley2012} applied the same analysis to a sample of 36 CTTS in NGC~2264, and reached similar conclusions.

\begin{figure}
\resizebox{\hsize}{!}{\includegraphics{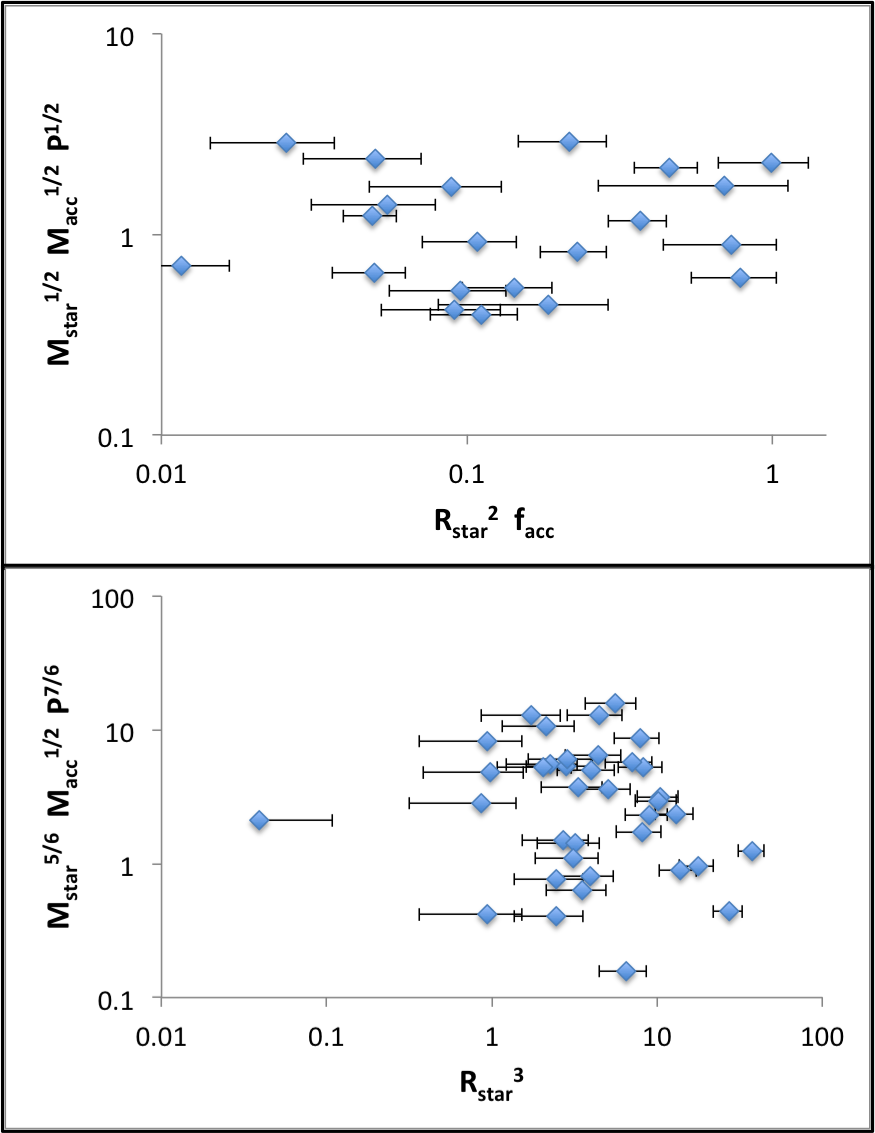}}
\caption{Plot of left and right sides of Eq.\,\ref{eqn:os95} (upper panel) and Eq.\,\ref{eqn:shu94} (lower panel), for a subset of NGC~2264 CTTS with known stellar and accretion parameters from \citet{venuti2014} and rotation periods from this study.}
\label{fig:test_magn_acc}
\end{figure}

Here we perform the same test as in \citet{cauley2012}, for the subsample of NGC~2264 CTTS examined for Sect.\,\ref{sec:Rt}. We follow \citet{johns2002} in assuming that B$_\star$ is the same for all stars in the sample, and adopt M$_\star$, R$_\star$ and $\dot{M}_{acc}$ values derived in \citet{venuti2014} and individual estimates of $f_{acc}$ derived from spot modeling of simultaneous, multi-wavelength light curves as described in \citet{venuti2015}. The values obtained, across our sample, for left and right side of Eq.\,\ref{eqn:os95}, are compared in the upper panel of Fig.\,\ref{fig:test_magn_acc}. 
Clearly, no significant correlation trend between the x-axis and the y-axis is present on the diagram. Most of the objects in the sample shown on Fig.\,\ref{fig:test_magn_acc} are fully convective (M$_{core}$/M$_\star$=0, as deduced from the model tracks of \citealp{siess2000}); therefore, the assumption that the B$_\star$ strength is uniform across the sample likely does not impact significantly this apparent lack of a correlation. A bigger impact on the position of individual points on Fig.\,\ref{fig:test_magn_acc} may arise from the values of $f_{acc}$, which, as discussed in \citet{venuti2015}, may be subject to somewhat large uncertainties when considered on an individual basis.

For completeness, we also tested whether similar prescriptions, extracted from magnetospheric accretion theories where the field is assumed to be purely dipolar \citep{konigl1991, shu1994}, would adapt better to the inferred parameters for NGC~2264 sources. In this case, the form of the expected correlation between R$_\star$ and M$_\star$, $\dot{M}_{acc}$, P is the following:
\begin{equation} \label{eqn:shu94}
R_\star^3\, \propto\, M_\star^{5/6}\, \dot{M}_{acc}^{1/2}\, P^{7/6}\,,
\end{equation}
where, again, we are assuming that the strength of the (dipolar) magnetic field can be considered to be the same for all sources. Left side and right side of Eq.\,\ref{eqn:shu94} are plotted one against the other for NGC~2264 members in the lower panel of Fig.\,\ref{fig:test_magn_acc}. Again, no correlation between the x-axis and the y-axis variables is found; however, the global behavior of objects on this diagram appears to be different from that in the upper panel of Fig.\,\ref{fig:test_magn_acc}. In the first case (upper panel), points distribute along a fairly narrow, horizontal belt: they span a wide range of values along the x-axis, but show no apparent trend relative to the y-axis. Conversely, points form a diffuse cloud on the lower diagram in Fig.\,\ref{fig:test_magn_acc}. This qualitative behavior is also observed in the analogous test diagrams shown in \citet{johns2002}: data points tend to show a scatter plot on a $M_\star^{5/6}\, \dot{M}_{acc}^{1/2}\, P^{7/6}$ vs. $R_\star^3$ plane (whereas a correlation is predicted by \citealp{konigl1991} and \citealp{shu1994}). On the other hand, a correlation trend between $R_\star^2\, f_{acc}$ and $M_\star^{1/2}\, \dot{M}_{acc}^{1/2}\, P^{1/2}$ is generally found by \citet{johns2002} when testing the predictions of \citeauthor{ostriker1995}'s (\citeyear{ostriker1995}) theory. 

Error bars shown on Fig.\,\ref{fig:test_magn_acc} were obtained via error propagation on the relevant quantities, using a typical uncertainty of 0.2~R$_\odot$ for R$_\star$, 0.05~M$_\odot$ for M$_\star$, 0.5~dex for $\dot{M}_{acc}$, and uncertainties derived in \citet{venuti2015} for $f_{acc}$ and from Eq.\,\ref{eqn:Perr_theory} of this paper for the rotation period. Stellar radii appear to be the dominant source of uncertainty on these diagrams. The values of R$_\star$ adopted here were derived in \citet{venuti2014} based on the effective temperatures T$_{eff}$ and bolometric luminosities L$_{bol}$ of the sources. T$_{eff}$ were assigned based on the spectral type of the objects and \citeauthor{cohen1979}'s (\citeyear{cohen1979}) SpT - T$_{eff}$ conversion scale; L$_{bol}$ were computed from the dereddened J-band photometry of the sources. An order-of-magnitude estimate of the uncertainty on the derived R$_\star$, taking into account typical uncertainties on the parameters involved in the computation of R$_\star$, is $\sim$0.1~R$_\odot$. However, a comparison between the R$_\star$ estimates derived in \citet{venuti2014} and those published in \citet{rebull2002} results in a somewhat higher average discrepancy for objects in common. Therefore, we chose to adopt a more conservative value of 0.2~R$_\odot$ for the error on R$_\star$ here. At any rate, this does not seem to impact significantly the trends observed on the diagrams in Fig.\,\ref{fig:test_magn_acc} and discussed in this Section. Notably, uncertainties on R$_\star$ would affect the exact position of each point along the x-axis, while the overall range of values spanned by the point distribution along the two axes would remain mostly unaffected.

\section{The rotational evolution of young stars} \label{sec:rot_evol}

\subsection{The period distribution of the NGC~2264 cluster: multiple populations?} \label{sec:gaussian_peaks}

As noted in Sect.\,\ref{sec:results}, the period distribution derived for the NGC~2264 cluster consists of a smooth distribution with two peaks (P$\sim$1-2~d and P$\sim$3-4~d). Similar features are observed when analyzing the rotation periods separately for WTTS and CTTS, with the exception that, while WTTS do exhibit signatures of two peaks corresponding to those observed for the total period distribution, only the longer period peak (P$\sim$3-4~d) is found in the case of CTTS. A fit to the different groups with a gamma distribution (see Sect.\,\ref{sec:per_dist}) provides the following parameters:

\begin{itemize}
\item cluster $\rightarrow$ mean P = 5.2$\pm$0.6~d; variance = 13$\pm$3~d;
\item CTTS \mbox{ }$\rightarrow$ mean P = 6.1$\pm$1.3~d; variance = 13$\pm$4~d;
\item WTTS $\rightarrow$ mean P = 4.9$\pm$0.8~d; variance = 12$\pm$3~d.
\end{itemize}

The mean value measured for the whole sample falls in between the CTTS' mean period and the WTTS' mean period, and is closer to the latter than to the former, reflecting their relative contributions to the entire period distribution of the cluster. This shift between CTTS' and WTTS' period distributions can be understood if we consider the impact of star-disk interaction on the rotational properties and evolution of young stars. Given an ensemble of stars, it is reasonable to assume that their initial periods may be normally distributed \citep[e.g.,][]{tinker2002}: the center of the distribution will reflect the average evolutionary status of the population; the dispersion of values around the center of the distribution will reflect varying initial conditions from object to object (e.g., varying star/disk mass, or different accretion history), which translate to varying rotational properties. As time progresses, if all objects follow similar evolutionary paths, we would then expect the period distribution of the ensemble of stars to evolve accordingly: the position of the center will change following the laws that govern angular momentum evolution in young stars; conversely, the overall shape of the distribution will remain unchanged \citep[e.g.,][]{vasconcelos2015}. As WTTS represent a later evolutionary stage than CTTS, where the disk has disappeared and the systems are no longer braked by the star-disk interaction, the center of their period distribution is shifted toward shorter P.

The origin of the two peaks in Fig.\,\ref{fig:hist} is more difficult to comprehend in this picture. As illustrated in the recent study of \citet{vasconcelos2015}, who simulated the rotational evolution of young low-mass stars from an age of 1~Myr to 12~Myr, multiple peaks in the period distribution of a cluster at a given age can only be obtained if statistically distinct rotational behaviors are present {\it ab~initio} in the investigated population. If a homogeneous, Gaussian-shaped distribution of periods is assumed for the whole sample at time t$_0$ and this is let to evolve, a continuous distribution with a single peak will be obtained at time t. Therefore, the fact that we do observe two separate peaks may suggest that the sample of objects we are investigating comprises several subpopulations with different histories and rotational properties. This distinction goes beyond the separation between disk-bearing and disk-free sources, as one or two peaks on a smooth distribution of periods characterize both CTTS and WTTS individually.

Several studies have now assessed that star formation in NGC~2264 did not occur in a single event, but rather in a sequential fashion. Indeed, the presence of Herbig-Haro objects \citep[e.g.,][]{reipurth2004}, molecular outflows \citep[e.g.,][]{margulis1988} and embedded sources \citep[e.g.,][]{wolfchase2003, teixeira2006} attests that active star formation is still ongoing within the region. \citet{sung09} combined multiple investigations of the NGC~2264 cluster in the optical \citep{sung08} and mid-IR (their study; \citealp{teixeira2006}) to derive a map of the different subclusterings within the region, based on the spatial density of protostars and YSOs in various evolutionary stages (see Figure~13 of their paper). In particular, they identified two embedded regions (the Spokes cluster and the core of the Cone nebula region, Cone~(C)) in the southern part of the cloud, surrounded by a less embedded halo (Cone~(H)) dominated by more evolved YSOs. In the northern part of the cloud, another subclustering of disk-bearing YSOs is identified around the massive star S~Mon. The S~Mon and Cone~(H) subclusterings are in turn surrounded by the Halo region, which encompasses the periphery of the cloud and is prevalently populated by disk-free cluster members. As discussed in \citet{sung2010}, objects located in the Halo were formed earlier, followed by objects in the S~Mon region and in Cone~(H); objects in the Cone~(C) and in the Spokes subclustering are the most recently formed. 

\begin{figure*}
\centering
\includegraphics[width=\textwidth]{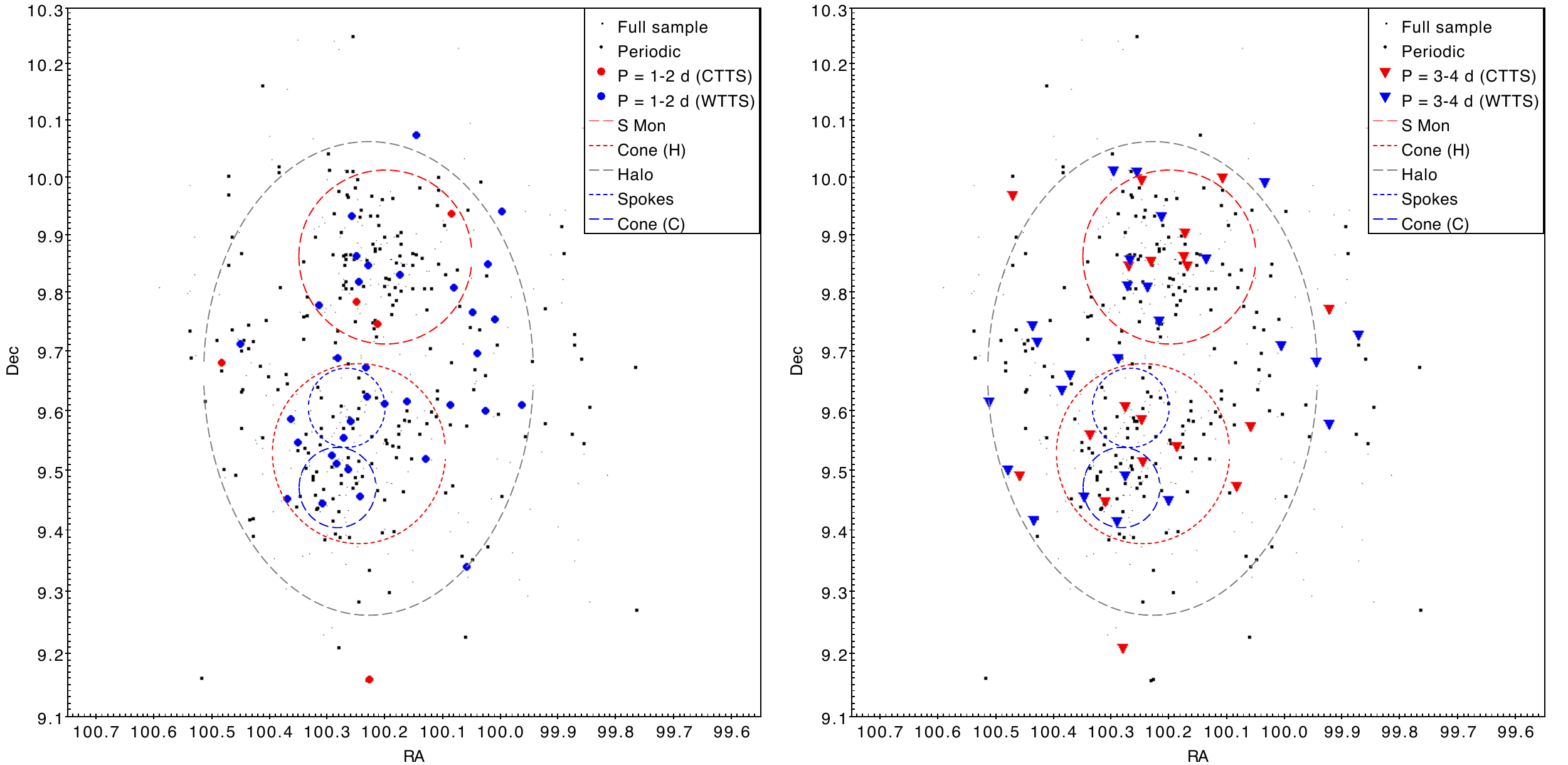}
\caption{Spatial distribution of NGC~2264 members investigated in this study. Small grey dots represent all objects in the sample (Table~4); periodic sources with single periodicity detected are highlighted as black dots; blue/red circles (on the left-side diagram) and triangles (right-side diagram) identify periodic sources with periods in the 1-2~d range and in the 3-4~d range, respectively; blue corresponds to WTTS, red to CTTS. The dashed lines delimit the various regions/subclusters identified, within the cloud, by \citet{sung09} after \citet{sung08} and \citet{teixeira2006}. North is up and east is left on the spatial diagram.}
\label{fig:radec_per_subgroups}
\end{figure*}

To investigate the nature of objects that populate the P=1-2~d and the P=3-4~d peaks, in Figure~\ref{fig:radec_per_subgroups} we compare the spatial distribution of objects belonging to these two period groups to that of the full sample of periodic members, and to the location of the various subclusters identified across NGC~2264 by \citet{sung09} and discussed above.
The two period groups (P=1-2~d and P=3-4~d, hereafter $P^{[1,2]}$ and $P^{[3,4]}$) contain about the same number of objects (39 and 44, respectively). About 48\% of objects in $P^{[1,2]}$ and 55\% of objects in $P^{[3,4]}$ are in excess of the underlying continuum in the corresponding period bins (Fig.\,\ref{fig:hist}). To derive some quantitative indications of the spatial properties of objects in $P^{[1,2]}$ and of those in $P^{[3,4]}$ relative to the full sample of members investigated in this study and to the full sample of periodic sources identified in this study, for each of these groups we measured the fraction of objects projected onto the various subregions identified by \citet{sung09} within NGC~2264. These frequencies are compared in Table~\ref{tab:per_subclusters}.     
\begin{table}
\caption{Distribution of periodic sources among the various NGC~2264 subregions.}
\label{tab:per_subclusters}
\centering
\begin{tabular}{l c c c c}
\hline\hline
 & Cone (H) & S~Mon & Halo & external \\
\hline
Full sample & 29.8\% & 28.0\% & 27.8\% & 14.4\% \\
Periodic & 30.9\% & 29.8\% & 29.1\% & 10.2\% \\
$P^{[1,2]}$ & 38.5\% & 25.6\% & 28.2\% & 7.7\% \\
$P^{[3,4]}$ & 22.7\% & 27.3\% & 34.1\% & 15.9\%\\
\hline
\end{tabular}
\tablefoot{
Each column corresponds to a different subclustering identified in the NGC~2264 region by \citet{sung09} and illustrated here in Fig.\,\ref{fig:radec_per_subgroups}. Rows indicate, in the order, the percentage of objects located on top of each of the subregions among i) the full sample investigated in this study, ii) objects found to be periodic (single period) in this study, iii) periodic sources with period between 1 and 2 days, and iv) periodic sources with period between 3 and 4 days. The percentage indicated for Cone~(H) also accounts for objects projected onto Cone~(C) and onto the Spokes region.  
}
\end{table} 
About the same fraction of cluster members (30\%) are found projected on the S~Mon, Cone (core + halo + Spokes cluster) and Halo regions. This percentage does not vary significantly if we restrict our sample of members to periodic sources only. Conversely, some quantitative difference in the spatial distribution of objects across the cloud can be observed when comparing the full sample of sources with objects in $P^{[1,2]}$ and $P^{[3,4]}$. Objects in the $P^{[1,2]}$ group appear to be more numerous at the RA-Dec location corresponding to the Cone region, and populate preferentially the sourthern part of the S~Mon region and the western part of the Halo region. On the other hand, a smaller-than-average fraction of objects in the $P^{[3,4]}$ group is projected on the Cone region, whereas they populate the periphery of the cloud and the Halo region more densely, and distribute predominantly along the eastern side of the latter. Interestingly, these are also the regions of the cluster where disks might be expected to last longer compared to more embedded regions, where the impact of the ionizing radiation from the OB stars contained in the cluster is stronger (see, e.g., the study of \citealp{mann2010} on the ONC).

These features may support the view that several populations of stars are mingled with one another in the sample of cluster members that we are investigating here. However, no conclusive evidence in this respect can be drawn from this analysis. No obvious difference emerges between $P^{[1,2]}$, $P^{[3,4]}$ and the full sample of objects when comparing their respective distributions on the H-R diagram of the cluster, nor on their isochronal age distributions.

\subsection{NGC~2264 as a benchmark cluster in the scenario of PMS rotational evolution}

\begin{figure*}
\centering
\includegraphics[scale=0.7]{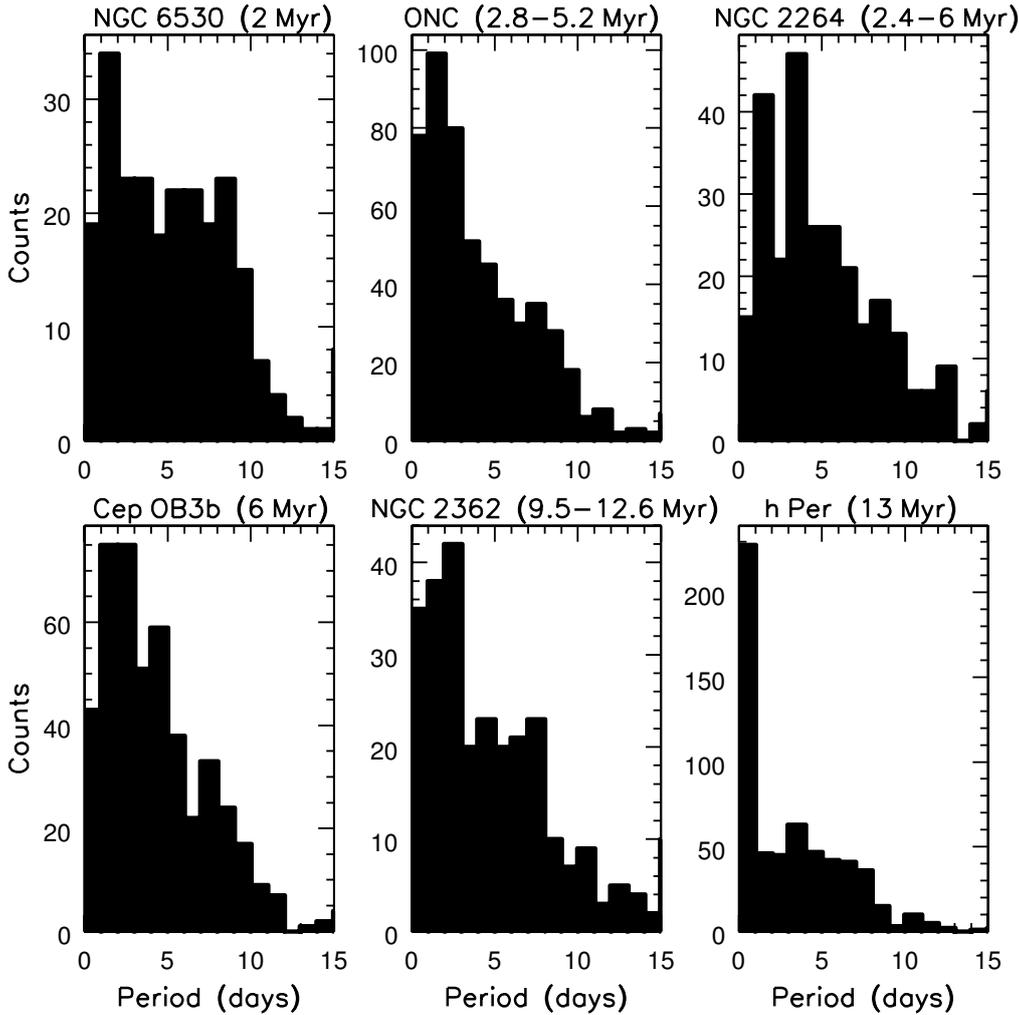}
\caption{Comparison of the period distributions obtained for various young clusters of different ages, in the mass range 0.1--2\,M$_\odot$. The clusters are, from left to right, from top to bottom: NGC~6530 (age of $\sim$2~Myr, \citealp{bell2013}; periods from \citealp{henderson2012}; 241 objects with measured rotation periods in the mass range of interest here); the Orion Nebula Cluster (ONC; age in the range 2.8--5.2~Myr, \citealp{naylor2009}; rotation periods from \citealp{rodriguez2009} for objects with M$_\star$$<$0.4~M$_\odot$, and from \citealp{irwin2009}, after \citealp{herbst2001, herbst2002} and \citealp{stassun1999}, for objects with M$_\star$$>$0.4~M$_\odot$; 528 objects with measured periods in the selected mass range); NGC~2264 (age in the range 2.4--6~Myr, \citealp{naylor2009}; period measurements reported in this study; 272 objects); Cepheus~OB3b (age of $\sim$6~Myr, \citealp{bell2013}; periods from \citealp{littlefair2010}; 460 objects with period measurements in the selected mass range); NGC~2362 (age in the range 9.5--12.6~Myr, \citealp{bell2013}; periods from \citealp{irwin2008}; 272 objects with periods in the mass range of interest); h~Persei (age of $\sim$13~Myr, \citealp{mayne2008}; periods from \citealp{moraux2013}; 586 objects with measured periods in the selected mass range).}
\label{fig:per_evol}
\end{figure*}

Recently, \citet{bouvier2014} have presented a complete review of all observational studies of rotation in young stars conducted so far on different clusters. To examine the rotation properties that we measure for NGC~2264 in the context of early stellar evolution, we select here all clusters from \citeauthor{bouvier2014}'s compilation, which have average ages between 1 and 15~Myr; this range defines the age scale of interest for the lifetimes of disks around young stars (e.g., \citealp{bell2013}). Since we are interested here in a statistical comparison of rotation properties as a function of age, we have retained only clusters, in the age range mentioned earlier, in which rotation periods are available for a large statistical sample of members (as detailed in Table~1 of \citealp{bouvier2014}). The selected clusters for our comparison are then, in order of age, NGC~6530, the Orion Nebula Cluster (ONC), Cepheus~OB3b, NGC~2362, and h~Persei. For these regions, we have then selected all members with known period and mass comprised between 0.1~M$_\odot$ and 2~M$_\odot$, i.e., approximately the mass range probed in our analysis of rotation in NGC~2264. The resulting period distributions for the various clusters, plus that obtained for NGC~2264 in this study and shown in Fig.\,\ref{fig:hist}, are shown in Fig.\,\ref{fig:per_evol} in an age-ordered sequence.

A striking difference can be observed between the first panel (depicting NGC~6530, the youngest cluster in our sample) and the last panel (depicting h~Per, the oldest cluster) on Fig.\,\ref{fig:per_evol}. As noted in \citet{henderson2012}, the period distribution obtained for NGC~6530, at an age of a few Myr and a disk fraction\footnote{This corresponds to the percentage of cluster members with signatures of disks in the near-IR (i.e., exhibiting an IR excess at $J,H,K$ wavelengths).} of about 50\% \citep{prisinzano2007}, appears to be uniform in the range P\,=\,0--10~days. Conversely, about 50\% of the cluster members with detected periods in h~Per \citep{moraux2013}, at an age where the disk fraction has dropped to a very small percentage \citep[$\sim$2-3\%;][]{cloutier2014}, exhibit rotation periods $\lesssim$1~day. For clusters of intermediate ages, the period distributions appear to exhibit transitional features between the first and the last panel in Fig.\,\ref{fig:per_evol}: a single or several peaks are observed in the P\,=\,0--5~day range, followed by a slow decline in number of objects toward larger periods. 

\subsubsection{A mass dependence on the observational picture of rotational evolution?}

\begin{figure}
\resizebox{\hsize}{!}{\includegraphics{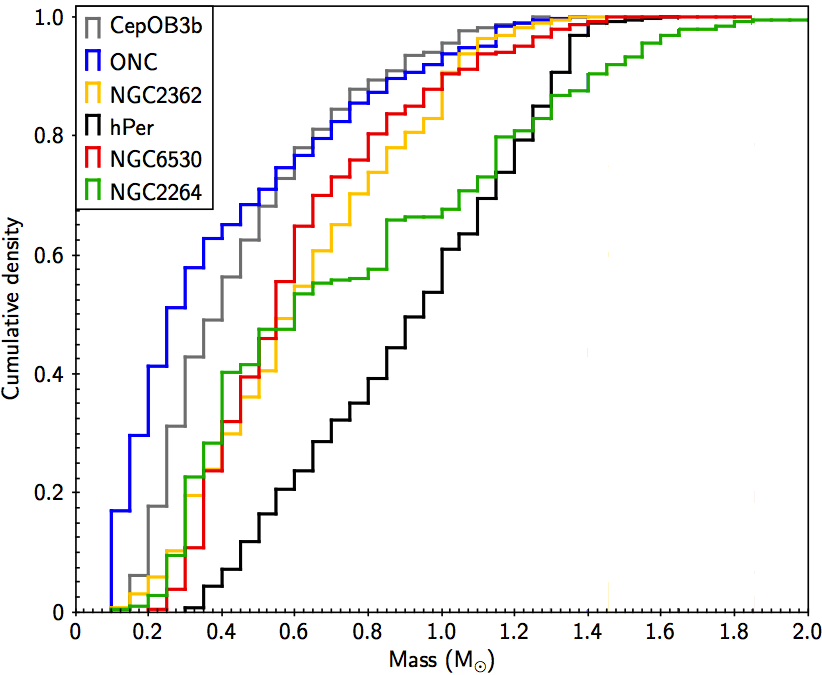}}
\caption{Cumulative distribution functions, in mass, of the young stellar populations whose rotation properties are used to build the histograms of periods shown in Fig.\,\ref{fig:per_evol} for the various clusters.}
\label{fig:mass_per_evol}
\end{figure}

As discussed in Sect.\,\ref{sec:mass_dependence}, the results we obtain for NGC~2264 suggest that the rotation properties of cluster members are somewhat dependent on stellar mass. Fast rotators seem to be more predominant among lower-mass stars (M$_\star$$<$0.4~M$_\odot$) than among higher-mass stars (M$_\star$$>$1~M$_\odot$; see right panel of Fig.\,\ref{fig:hist_tts_mass}), although our data does not allow us to draw conclusive evidence in this respect from a statistical point of view. A similar analysis is presented by \citet{littlefair2010} for the case of Cep~OB3b. On the other hand, several studies conducted on other clusters have found a strong dependence of stellar rotation on stellar mass. \citet{henderson2012}, for instance, reported a statistically significant difference in rotational periods between lower-mass (M$_\star$~$\leq$~0.5~M$_\odot$) and higher-mass (M$_\star$~$>$~0.5~M$_\odot$) stars in NGC~6530, with the latter rotating faster than the former. Conversely, a mass dependence in the opposite direction (lower-mass stars spinning on average faster than higher-mass stars) was reported for the ONC \citep{herbst2002} and for h~Per \citep{moraux2013}. As detailed in Table~1 of \citet{bouvier2014}, the various studies of rotation in young clusters, whose results we are comparing in Fig.\,\ref{fig:per_evol}, refer to mass regimes which may vary somewhat from case to case. This is illustrated in Fig.\,\ref{fig:mass_per_evol}, showing the cumulative distribution functions in mass of the rotation surveys used to build Fig.\,\ref{fig:per_evol}.
If the rotation properties of young stars are truly dependent on stellar mass, this diversity in mass properties between the different samples can have an impact on the evolutionary picture we may deduce from Fig.\,\ref{fig:per_evol}.

One way to circumvent this issue is to group cases with similar distributions in mass from Fig.\,\ref{fig:mass_per_evol}, and compare the rotation properties as a function of age within each individual group. One such group consists of the ONC and Cep~OB3b (masses from $\sim$0.1~M$_\odot$ to $\sim$1.3~M$_\odot$), whose rotation properties are illustrated respectively in the upper middle panel and in the lower left panel of Fig.\,\ref{fig:per_evol}. No definite qualitative difference is noted between the two period distributions. In both cases, the distribution has a single\footnote{As mentioned in Sect.\,\ref{sec:mass_dependence}, \citet{herbst2002} found the period distribution of the ONC to be unimodal when only objects less massive than 0.25~M$_\odot$ are considered, and double-peaked when only objects more massive than 0.25~M$_\odot$ are considered. The single-peaked nature of the period distribution shown here for the ONC reflects the fact that this is dominated by lower-mass stars: indeed, as illustrated in Fig.\,\ref{fig:mass_per_evol}, about 50\% of objects in the ONC sample considered here have mass below 0.3~M$_\odot$.} peak close to P$\sim$2~d, perhaps more sharp in the case of the ONC, and then declines steadily toward longer periods; in both regions, almost no objects exhibit rotational periods longer than $\sim$12~d. It is important to mention that, while the Cep~OB3b cluster \citep{kun2008} is a well defined subgroup of one of the three OB associations known in the Cepheus constellation, the ONC \citep{muench2008} likely comprises several different populations of stars, with a non-negligible age spread among cluster members. This ought to be taken into account when examining their respective period distributions as two distinct blocks in the picture of PMS rotational evolution. 

A second group of cases with similar distributions in mass from Fig.\,\ref{fig:mass_per_evol} is that including NGC~6530 and NGC~2362. The corresponding period distributions are shown on the upper left panel and on the lower middle panel of Fig.\,\ref{fig:per_evol}, respectively. Contrary to the case of the ONC and Cep~OB3b, a marked age difference exists between these two clusters: at an age of about 2~Myr, NGC~6530 is the youngest cluster among those shown in Fig.\,\ref{fig:per_evol} and infrared studies indicate that about half of its members are surrounded by dusty disks; conversely, at an average age of about 12~Myr, only 10--20\% of objects in the NGC~2362 cluster show evidence of dust in the circumstellar environment \citep{dahm2007}. The overall shape of the period distribution appears to evolve between the two: only a hint of a peak around P=1-2~d is present in the case of NGC~6530, and the distribution is fairly uniform in the $\sim$1-10~d period range; conversely, in the case of NGC~2362 a more definite peak around P$\sim$2~d is present and stands out against the flat segment of distribution in the 3--8~d period range, after which the distribution displays a rapidly declining tail toward longer periods.

\begin{figure*}
\centering
\includegraphics[scale=0.7]{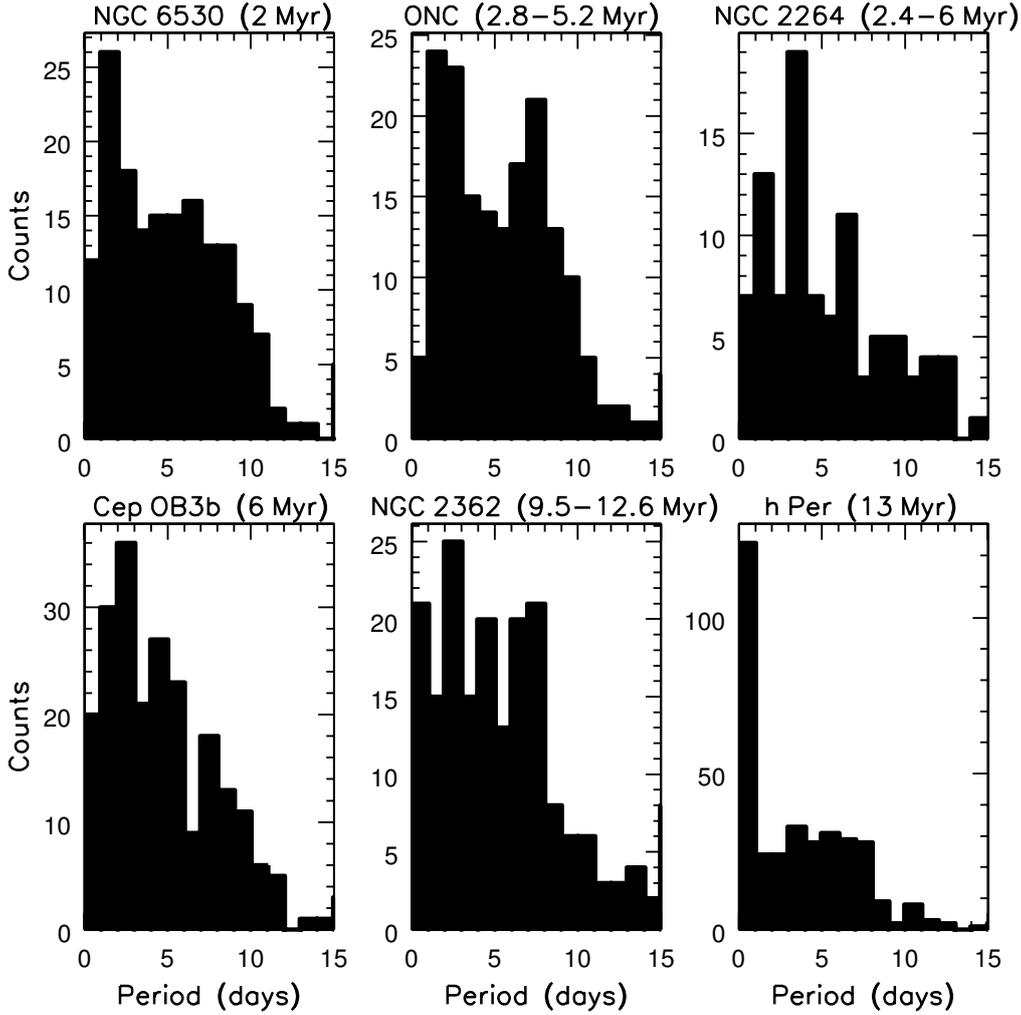}
\caption{Same as in Figure~\ref{fig:per_evol}, but including only objects in the mass range 0.4--1.1~M$_\odot$ for each cluster.}
\label{fig:per_evol_mass_restrict}
\end{figure*}

To investigate the impact of different mass regimes on the global picture of Fig.\,\ref{fig:per_evol}, we used Fig.\,\ref{fig:mass_per_evol} to identify a mass range common to all samples. Then, we selected, for each cluster, only objects with masses in this range, and used these mass-selected subsamples of objects to re-draw the period distributions in Fig.\,\ref{fig:per_evol}. The selected mass range goes from 0.4~M$_\odot$ (low-mass end of the h~Per sample) to 1.1~M$_\odot$ (chosen a bit smaller than the highest mass regime common to all samples, $\sim$1.3~M$_\odot$, to avoid the mass range where the so-called ``Kraft break'' in stellar rotation properties occurs; see \citealp{kraft1967}). The results of this exercise are illustrated in Fig.\,\ref{fig:per_evol_mass_restrict}.  

When comparing Fig.\,\ref{fig:per_evol} and Fig.\,\ref{fig:per_evol_mass_restrict} qualitatively, the following differences appear:
\begin{itemize}
\item in the case of NGC~6530 (2~Myr), the net effect of this mass selection on the global shape of the period distribution is a removal of objects from the intermediate (7-9~d) period range, with a steadier decline in number, instead of a uniform behavior, for P$\gtrsim$7~d;
\item in the case of the ONC (2.8-5.2~Myr), a second peak at P$\sim$7-8~d appears in the period distribution when M$_\star$$<$0.4~M$_\odot$ objects are excluded, in agreement with what reported by \citet{herbst2002} regarding a bimodal nature of the period distribution for the more massive population component of the star-forming region;
\item in the case of NGC~2264 (2.4-6~Myr), more substructures emerge, notably a hint of a peak at P$\sim$6-7~d, and in addition the first peak of the period distribution (P=1-2~d) becomes less pronounced with respect to its second peak (P=3-4~d) compared to the histogram shown in Fig.\,\ref{fig:hist};
\item in the case of Cep~OB3b (6~Myr), the net effect of this mass selection is sharpening the peak of the distribution, which becomes more populated in the higher-P (2-3~d) half than in the lower-P (1-2~d) half;
\item in the case of NGC~2362 (9.5-12.6~Myr), when removing the M$_\star$$<$0.4~M$_\odot$ objects, the peak at P$\sim$2~d disappears and the shape of the distribution evolves toward a uniform distribution in the P=0-8~d range;
\item in the case of h~Per (13~Myr), no appreciable changes in the shape of the period distribution appear when restricting the sample to more massive objects.
\end{itemize}

The above list suggests that, at least in the age range between 3 and 10~Myr, the rotation properties of young star clusters are somewhat dependent on the mass regimes probed in their populations: when restricting the sample to more massive TTS (0.4-1.1~M$_\odot$), the features at longer periods in the cluster distributions become more marked than when lower-mass TTS (M$_\star$$<$0.4~M$_\odot$) are considered. Conversely, no significant variations in the evolutionary picture emerge when comparing the age-ordered sequence of period histograms in Fig.\,\ref{fig:per_evol} to that in Fig.\,\ref{fig:per_evol_mass_restrict}. From the youngest (NGC~6530) to the most evolved (h~Per) case, the period distribution evolves from a uniform distribution in the $\sim$0-8~d period range (with perhaps a hint of a peak at a few days) to a distribution with a sharp peak at P$<$1~d superimposed on a flat continuum, about five times less strong than the peak, that extends down to P$\sim$8~d. At intermediate stages, the period distributions of young clusters exhibit a variety of features and substructures in the period range from 0 to 10~d, which may reflect a diversity in the specific environmental conditions or in the nature of the stellar populations probed in the different cases. Overall, the bulk of objects is found at periods shorter than $\sim$7~days, and the distribution declines more or less steadily toward longer periods.

\subsubsection{The rotational evolution of young stars: observations vs. simulations}

The wealth of observational data available to date for rotation periods of young stars in cluster of different ages have enabled a number of studies that follow a semi-empirical approach to model the rotational evolution of stars in the pre-main sequence. Namely, these models take a starting population of stars whose rotation properties are assigned based on observations of the youngest clusters, and follow their rotational evolution assuming that this is governed by specific physical mechanisms (e.g., disk-locking, angular momentum conservation) depending on the presence or absence of accretion disks. The comparison of the final period distribution simulated with observational data for clusters of similar age therefore enables investigating what processes regulate the spin evolution of young stars, especially in the earliest stages.

\citet{vasconcelos2015} presented Monte~Carlo simulations of the rotational evolution of a population of 280,000 young stars with mass between 0.3 and 1~M$_\odot$, from an age of 1~Myr to an age of 12~Myr. Their models assume that stars evolve at a constant angular velocity when they are coupled to an active accretion disk, and at a constant angular momentum when the disk is dissipated. Their Figure~6 illustrates how the shape of the global period distribution evolves from the beginning to the end of the simulated time span. It is assumed that, at an age of 1~Myr, a fraction of young stars have already lost their disks, and that the rotation properties of the latter are statistically distinct from those of disk-bearing sources; this is consistent with what observed in several young clusters \citep[e.g.,][]{xiao2012, henderson2012}. At an age of 2.1~Myr, the initial distribution, with a broad ``bump'' around a period of 3~d and a slow decline toward longer periods, has evolved into a distribution with a more pronounced peak around P~$\sim$2~d followed by a flatter region, about half as high as the peak, that extends from 5 to 10~d and then declines steadily toward longer periods. This behavior can reproduce qualitatively the shape of the period distribution observed for the ONC and illustrated in the upper-middle panel of Fig.\,\ref{fig:per_evol}. By an age of 12.1~Myr, the peak in the distribution has become sharp and shifted to P$\lesssim$1~d, while the number count in the successive histogram bin drops to about 0.4 times that in the peak and decreases steadily with increasing period. This trend is similar, at least on a qualitative basis, to what is observed for the NGC~2362 and the more evolved h~Per clusters (lower-middle and lower-right panel of Fig.\,\ref{fig:per_evol}, respectively). 

At an age of $\sim$3-5~Myr (on the order of disk lifetime; \citealp{haisch2001}), NGC~2264 is a benchmark cluster for PMS rotational evolution. By that time, about 50\% of young stars will have lost their disks and started to spin up toward the main sequence. An implication of this is that, at the evolutionary stage of NGC~2264, a fraction of objects will have just been released from their disks, and therefore will not have had enough time yet to spin up significantly. As discussed in \citet{vasconcelos2015}, during the first few Myr of evolution of a given ensemble of stars, the progressive release of YSOs from their disks results in a widening of the period distribution associated with the disk-free component of the population: newly released objects will rotate more slowly than stars that have lost their disks earlier and have thus already started to spin up freely. This may explain the tail of slow rotators observed among WTTS in NGC~2264 (Fig.\,\ref{fig:hist_tts_mass}). As the evolution continues, more and more sources are released from their disks and stars that had been released earlier keep spinning up; therefore, the bulk of the non-accreting population shifts toward shorter periods and the long-P region of the initial distribution is depleted. This is consistent with the picture shown in Fig.\,\ref{fig:per_evol}, where clusters at intermediate evolutionary stages exhibit wider distributions at the longer period end than the h~Per cluster.

As discussed above, the predicted trend of spin rate evolution of young stellar populations, simulated in \citet{vasconcelos2015}, shows an overall agreement with the evolutionary picture we may get from comparing the observed period distributions of clusters at different ages (Fig.\,\ref{fig:per_evol} of this study). This supports the view that young stars may be locked to their disks during the earliest stages of their evolution, and then spin up as they contract toward the ZAMS once the magnetic coupling with their accretion disks has ceased. Aligned with this interpretation are the conclusions of \citet{landin2016}, who tested the idea, put forward by \citet{lamm2005}, that NGC~2264 represents a later stage in the scenario of PMS rotational evolution than the ONC. The authors simulated the backward evolution of the NGC~2264 period distribution down to the age of the ONC, under the assumption that the spin rate of cluster members is governed by disk locking as long as the stars possess a disk, and by angular momentum conservation afterwards. The period distribution predicted by \citet{landin2016} for the younger NGC~2264, following this approach, would indeed show the overall features of the period distribution observed for the ONC.   
	
\section{Conclusions} \label{sec:conclusions}

In this study we have presented the most accurate and unbiased analysis of rotation properties available to date for the NGC~2264 cluster in the mass range $\sim$0.2--1.7~M$_\odot$. We examined a population of about 500 cluster members, whose optical light variations were monitored continuously for 38~days with the {\it CoRoT} space telescope in the framework of the CSI~2264 campaign (Dec.\,2011\,-\,Jan.\,2012). Light curves were searched for periodicity using three different methods: the Lomb-Scargle periodogram, the autocorrelation function, and the string-length method. A significant period was detected for about 62\% of sources in the sample; the period detection rate is lower among objects with active accretion disks (CTTS) than among objects that have already been released from their disks (WTTS).

The main results of this work can be summarized as follows.
\begin{enumerate}
\item The period distribution derived for the cluster consists of a smooth distribution centered on P$\sim$5.2~d with two peaks. The peaks are located at P=1-2~d and P=3-4~d. Although our dataset allows us to reliably measure rotation periods as long as 19~d, over 95\% of periodic sources in our sample have period shorter than 13~d.

\item A clear statistical distinction in rotation properties exists between WTTS and CTTS: although the respective period distributions overlap, the former spin on average faster than the latter. A typical period of 4.9~d is measured across the WTTS population of the cluster, while the mean period measured for the CTTS sample is of 6.1~d. The first peak in the NGC~2264 period distribution (P=1-2~d) is clearly associated with its WTTS population, whereas very few CTTS are found with P$<$2.5~d; conversely, the second peak, at P=3-4~d, takes contributions from both CTTS and WTTS. 

\item Our results suggest some mass dependence in the rotation properties of NGC~2264 members, in agreement with earlier findings. Lower-mass objects appear to exhibit rotation periods that are shorter on average than higher-mass objects, although our analysis does not allow us to reject the null hypothesis that lower-mass and higher-mass objects have similar period distributions to the 5\% significance level.

\item A clear connection is found between rotation and accretion; objects that exhibit large UV excesses (indicative of high mass accretion rates onto the stars) are typically associated with long rotation periods; conversely, a dearth of fast rotators with strong UV excesses is evident among disk-bearing objects. This supports the idea that magnetic star-disk coupling has an impact on the rotation properties of young stars.

\item No clear relationship emerges between the rotation period of the stars and their inner structure (notably the presence/absence of a radiative core); no evolution in rotation properties is observed along a given mass track on the H-R diagram of the cluster among CTTS or WTTS. This may indicate that the NGC~2264 population is still too young for the transition from fully convective to partly radiative stellar interiors to have a significant impact on the observed properties. 

\item The connection between rotation properties and accretion indicators (UV excess) that we find here for the NGC~2264 population is reminiscent of the connection between rotation and disk indicators (IR excess) in young stars reported in \citet{rebull2006}. Furthermore, it shows the same behavior of the distribution of accretion rates as a function of rotation period simulated in \citet{vasconcelos2015} in the hypothesis that the spin rate of young stars evolves at constant angular velocity in the presence of a disk and at constant angular momentum when the disk has disappeared. This would support a scenario in which young stars are locked to their disks during the accretion phase and then start to spin up to conserve angular momentum once disk accretion and star-disk coupling have ceased.    
\end{enumerate}

Thanks to the extensive obervational effort devoted to characterizing the period distributions of star clusters at different ages, and to the simultaneous modeling effort aimed at reproducing and interpreting those observations, we have now achieved a global understanding of how the stellar spin rate may evolve across the pre-main sequence. However, several open issues remain. For instance, it is not clear why the period distributions of some clusters exhibit two separate peaks, as is the case for NGC~2264. In this study, we show that objects in the two peaks observed for NGC~2264 are different in nature: the shorter-period peak consists of disk-free cluster members (WTTS), while the longer-period peak consists of both disk-accreting (CTTS) and disk-free cluster members. However, the origin of these peaks is not as clear. Notably, they appear to be additional features on top of an underlying, smooth distribution of periods; this behavior is observed both when considering the NGC~2264 cluster as a whole and when examining CTTS and WTTS separately. Are the initial rotation periods in a given population randomly distributed (in which case we would expect the period distribution to evolve with time into another random distribution where the center has shifted toward shorter values)? And, therefore, is the presence of multiple peaks indicative of the fact that the ensemble of stars under exam is a composite population (e.g., the result of distinct star formation episodes)? Or do the initial rotation periods in a given sample of objects cluster around a certain value, depending on, e.g., environmental conditions at birth? Significant contribution to this discussion may be provided in the near future by Gaia: data issued from the mission will help identify kinematical substructures and populations in NGC~2264, which may shed new light on the nature of the specific features seen in the rotational distribution of the cluster. 

\begin{acknowledgements}
We thank the referee for a prompt and constructive report. This work is based on data from the {\it CoRoT} space mission, which has been developed and operated by CNES, with the contribution of Austria, Belgium, Brazil, ESA (RSSD and Science Programme), Germany, and Spain. This publication also makes use of data from MegaPrime/MegaCam, a joint project of CFHT and CEA/DAPNIA, at the Canada-France-Hawaii Telescope (CFHT) which is operated by the National Research Council (NRC) of Canada, the Institut National des Sciences de l'Univers of the Centre National de la Recherche Scientifique (CNRS) of France, and the University of Hawaii. We thank Suzanne Aigrain for detrending the {\it CoRoT} light curves used in this work. L.V. acknowledges useful discussions on histogram statistics with Dipan Sengupta. The authors acknowledge support through the PRIN INAF 2014 funding scheme of the National Institute for Astrophysics (INAF) of the Italian Ministry of Education, University and Research (``The GAIA-ESO Survey'', P.I.: S. Randich). This study was also supported by the grant ANR 2011 Blanc SIMI5-6 020 01 ``Toupies: Towards understanding the spin evolution of stars''. S.H.P.A. acknowledges financial support from CNPq, CAPES and Fapemig.
\end{acknowledgements}

\bibliographystyle{aa}
\bibliography{references}

\onecolumn

\longtab{
\label{tab:periods}
\begin{longtable}{lrrccccclrc}
\caption{Periodicity of \emph{CoRoT} light curves for NGC~2264 members.}\\
\hline \hline
\multicolumn{1}{l}{{\small CSIMon-\#}} &
\multicolumn{1}{c}{RA\tablefootmark{1}} &
\multicolumn{1}{c}{Dec\tablefootmark{1}} &
\multicolumn{1}{c}{Class\tablefootmark{2}} &
\multicolumn{1}{c}{{\it CoRoT}\_2011} &
\multicolumn{1}{c}{LC\tablefootmark{3}} &
\multicolumn{1}{c}{M$_\star$(M$_\odot$)} &
\multicolumn{1}{c}{P$_{2011}^{CoRoT}$(d)} &
\multicolumn{1}{c}{FAP\tablefootmark{4}} &
\multicolumn{1}{c}{Q\tablefootmark{5}} &
\multicolumn{1}{c}{P$_{2008}^{CoRoT}$(d)\tablefootmark{6}} \\
\hline
\endfirsthead

\caption{\emph{Continued from previous page}}\\
\hline \hline
\multicolumn{1}{l}{{\small CSIMon-\#}} &
\multicolumn{1}{c}{RA\tablefootmark{1}} &
\multicolumn{1}{c}{Dec\tablefootmark{1}} &
\multicolumn{1}{c}{Class\tablefootmark{2}} &
\multicolumn{1}{c}{{\it CoRoT}\_2011} &
\multicolumn{1}{c}{LC\tablefootmark{3}} &
\multicolumn{1}{c}{M$_\star$(M$_\odot$)} &
\multicolumn{1}{c}{P$_{2011}^{CoRoT}$(d)} &
\multicolumn{1}{c}{FAP\tablefootmark{4}} &
\multicolumn{1}{c}{Q\tablefootmark{5}} &
\multicolumn{1}{c}{P$_{2008}^{CoRoT}$(d)\tablefootmark{6}} \\
\hline
\endhead

\hline
\multicolumn{11}{r}{{\emph{Continued on next page}}}
\endfoot

\hline
\endlastfoot

  000005 & 100.32145 & 9.89435 & w & 616849481 & QPS & 0.31 & 3.998 & 5.0E-3& 1.13 & -- \\
  000006 & 100.52982 & 9.89571 & & 223998980 & N & & &  & &  \\
  000007\tablefootmark{*} & 100.47095 & 9.96739 & c & 223994721 & B & 0.69 & 3.192 & $<$1E-4& 0.75 &  \\
  000008 & 100.45248 & 9.90322 & w & 616826337 & & 0.80 &  &  & & -- \\
  000009 & 100.53812 & 9.80132 & w & 223999591 & U & 1.30 &  &  & &  \\
  000011 & 100.32187 & 9.90900 & c & 223985009 & S & 0.70 &  &  & &  \\
  000012 & 100.28892 & 9.93559 & & 602099712 & QPS & 1.13 & 2.961 & $<$1E-4& 0.20 & -- \\
  000014 & 100.52775 & 9.69215 & w & 602087973 & U & 0.66 &  &  & & -- \\
  000015 & 100.53798 & 9.98410 & w & 223999581 & N & 0.65 & &  & & \\
  000017 & 100.38329 & 10.00677 & c & 223988827 & P & 1.13 & 4.771 & $<$1E-4& -0.03 & 4.767 \\
  000018\tablefootmark{\#,§} & 100.30510 & 9.91908 & w & 223983925 & Be? & 1.47 & 4.26 & $<$1E-4 & & 3.704 \\
  000020 & 100.53847 & 9.73428 & w & 602091922 & QPS & 0.71 & 5.179 & $<$1E-4& 0.27 & -- \\
  000021 & 100.24771 & 9.99595 & c & 223980412 & QPD & 1.20 & 3.147 & $<$1E-4& 0.15 & 6.39 \\
  000022 & 100.26904 & 10.01185 & w & 603809295 & U & 0.24 &  &  & & -- \\
  000024 & 100.48687 & 9.79589 & w & 603414392 & N & 0.44 &  &  & & -- \\
  000029 & 100.26367 & 9.96528 & w & 223981349 & QPS & 0.70 & 8.012 & $<$1E-4& 0.13 & 8.014 \\
  000033 & 100.28027 & 9.97533 & w & 223982407 & P & 1.20 & 2.586 & $<$1E-4& 0.09 & 2.582 \\
  000035 & 100.44896 & 9.86731 & w & 223993180 & P & 1.69 & 2.413 & $<$1E-4& 0.56 & 2.411 \\
  000038 & 100.29532 & 10.01137 & w & 223983310 & QPS & 1.40 & 3.615 & $<$1E-4& -0.55 & 3.589 \\
  000045 & 100.47106 & 10.00069 & w & 616803281 & QP & 0.28 & 0.85 & 5.0E-4 & -- & -- \\
  000047 & 100.38365 & 10.01796 & w & 223988855 & P & & 1.342 & $<$1E-4& 0.16 & -- \\
  000048 & 100.46442 & 9.89518 & & 223994268 & QPS & 1.40 & 3.631 & $<$1E-4& 0.47 & 3.762 \\
  000050 & 100.25639 & 10.01014 & w & 223980944 & QPS & 1.22 & 3.554 & $<$1E-4& 0.10 & -- \\
  000051 & 100.30753 & 9.92890 & w& 223984075 & Be & 1.58 & 3.673 & $<$1E-4 & & 3.793 \\
  000051 & 100.30753 & 9.92890 & w& 223984075 & Be & 1.58 & 3.221 & 2.5E-3 & & "" \\
  000055 & 100.39549 & 10.02980 & w& 616803382 &  & 0.36 &  & & & -- \\
  000056 & 100.47150 & 9.84649 & c& 223994760 & QPD & 1.16 & 5.833 & $<$1E-4& 0.14 & 5.634 \\
  000057 & 100.26642 & 9.96930 & w& 223981535 & P & 0.30 & 4.544 & $<$1E-4& 0.01 & 4.557 \\
  000058 & 100.53625 & 9.68922 & c& 616895632 & QPS & 1.29 & 2.142 & $<$1E-4& 0.22 & -- \\
  000060 & 100.31388 & 9.91415 & w& 616826525 & N & 0.29 &  &  & & -- \\
  000062 & 100.59126 & 9.80918 & c& 224003566 & N & 0.90 & &  & &  \\
  000063 & 100.29972 & 9.99479 & c & 616826518 & N & 0.32 &  &  & & -- \\
  000066 & 100.26490 & 10.00982 & w & 603431452 & QPS & 0.45 & 11.24 & $<$1E-4& 0.07 & 11.25 \\
  000067 & 100.48470 & 9.83495 & & 603420197 & N & &  &  & &  \\
  000069 & 100.53066 & 9.82972 & & 223999063 & N & &  &  & &  \\
  000071 & 100.25105 & 9.98046 & w& 223980621 & QPS & 1.69 & 5.41 & $<$1E-4& 0.32 & 3.049 \\
  000075 & 100.29829 & 10.03990 & w& 223983509 & P & 1.48 & 2.385 & $<$1E-4& 0.09 & 2.39 \\
  000080 & 100.32480 & 10.06725 & c& 616803514 & U & 0.28 &  &  & & -- \\
  000087 & 100.27743 & 9.59585 & w& 602083898 & L & 1.07 &  &  & & -- \\
  000088 & 100.39181 & 9.35371 & w& 616969757 & U & 0.32 &  & &  & -- \\
  000090\tablefootmark{§} & 100.28733 & 9.56278 & c& 616919796 & Be & 0.30 & 4.115 & $<$1E-4 & & 4.042 \\
  000095 & 100.18384 & 9.39872 & w& 223976494 & P & 1.00 & 2.256 & $<$1E-4& 0.10& 2.267 \\
  000096 & 100.24432 & 9.76515 & w& 616872594 &  & 0.63 & &  & & -- \\
  000098 & 100.22645 & 9.33462 & & 223979114 &  & & 0.767 & 1.0E-2\tablefootmark{a}& -0.20 & -- \\
  000102 & 100.18016 & 9.52084 & w& 400007786 & QPS & 0.30 & 8.9 & $<$1E-4& -0.07 & -- \\
  000103 & 100.24807 & 9.58637 & c& 223980447 & QPS & 0.93 & 3.348 & $<$1E-4& 0.21 & 1.675 \\
  000105 & 100.21000 & 9.81390 & w& 616849431 & N & 1.28 &  &  & & -- \\
  000108 & 100.31183 & 9.54330 & w& 616919655 & QPS & 0.30 & 4.058 & 5.0E-1\tablefootmark{b}& -0.17 & -- \\
  000109 & 100.40601 & 9.62440 & c& 223990338 & N & &  &  & & -- \\
  000111 & 100.24379 & 9.55883 & & 223980201 & U & &  &  & & -- \\
  000117 & 100.22555 & 9.81206 & c& 602095753 & B & 0.32 & &  & & -- \\
  000119 & 100.33749 & 9.56006 & c& 223985987 & QP & 0.91 & 3.31 & $<$1E-4& 0.63 & 3.308 \\
  000122 & 100.44630 & 9.63463 & w& 223993002 & QPS & 0.95 & 5.319 & $<$1E-4& 0.05 & -- \\
  000123 & 100.28419 & 9.56926 & w& 616919795 & N & 0.31 &  &  & & -- \\
  000126 & 100.24099 & 9.68894 & c& 616895876 & D & 0.62 &  & & & -- \\
  000131 & 100.20535 & 9.39732 & c& 616969822 & U & 0.78 & 12.867 & $<$1E-4& 0.44 & -- \\
  000134 & 100.31009 & 9.44953 & c& 603808964 & QP & 0.28 & 3.017 & 3.0E-2& -- &  \\
  000139 & 100.22364 & 9.96668 & w& 616826651 & QPS & 0.45 & 8.042 & $<$1E-4& 0.26 & -- \\
  000141 & 100.26129 & 9.38862 & & 603808908 & QPS & & 4.652 & $<$1E-4& 0.24 & -- \\
  000145 & 100.34776 & 9.76631 & w& 616872632 & U & 0.71 & & & & -- \\
  000146 & 100.19060 & 9.97463 & w& 223976871 & N & 0.85 &  &  & & -- \\
  000153 & 100.24963 & 9.78457 & c& 400007889 & QPS & 0.29 & 1.896 & $<$1E-4& 0.35 & -- \\
  000158 & 100.18684 & 9.77732 & w& 605538580 &  & 0.95 & 10.108 & $<$1E-4& -0.45 & -- \\
  000159 & 100.21445 & 9.62068 & w& 602087947 & QPS & 0.45 & 8.798 & $<$1E-4& 0.14 & -- \\
  000160 & 100.24929 & 9.86359 & w& 605538519 & QPS & 0.36 & 1.792 & $<$1E-4& 0.18 & 1.805 \\
  000164 & 100.26880 & 9.50376 & w& 616919778 & MP & 0.29 & 0.669 & $<$1E-4&  & -- \\
  000164 & 100.26880 & 9.50376 & w& 616919778 & MP & 0.29 & 0.904 & 7.0E-4 & & -- \\
  000168 & 100.42866 & 9.41899 & c& 223991832 & QP & 0.90 & 10.019 & $<$1E-4& 0.6 & 8.608 \\
  000169 & 100.45027 & 9.71203 & w& 223993277 & QPS & 1.55 & 1.173 & $<$1E-4& 0.21 & 1.184 \\
  000172 & 100.29297 & 9.36376 & w& 616969725 & U & 0.33 &  &  & & 8.08 \\
  000176 & 100.21752 & 9.87531 & w& 602266743 & QPS & 0.36 & 7.694 & $<$1E-4& 0.03 & -- \\
  000177 & 100.27584 & 9.60638 & c& 223982136 & QPS & 1.48 & 3.029 & $<$1E-4& 0.22 & 3.018 \\
  000183 & 100.31879 & 9.43564 & & 602079851 & U & & 6.273 & $<$1E-4& 0.52 & -- \\
  000184 & 100.33018 & 9.51354 & & 616919664 & N & 0.16 &  &  & & -- \\
  000185 & 100.41154 & 9.53663 & c& 616919566 & S & 1.22 &  &  & &  \\
  000188 & 100.25719 & 9.93097 & w& 602099710 & QPS & 0.28 & 1.74 & 1.0E-2& -- & -- \\
  000192 & 100.20837 & 9.74840 & c& 616872583 & N & 0.30 &  &  & & -- \\
  000198 & 100.33183 & 9.52900 & w& 223985611 & QPS & 1.19 & 4.996 & 2.0E-1\tablefootmark{b}& 0.80 & 4.94 \\
  000200 & 100.28339 & 9.51120 & w& 616919794 & QPS & 1.12 & 1.929 & $<$1E-4& & -- \\
  000204 & 100.19670 & 9.88588 & & 602095749 & N & 0.37 &  &  & & -- \\
  000206 & 100.24747 & 9.95985 & w& 616826502 & N & 0.70 &  &  & & -- \\
  000207 & 100.24598 & 9.81841 & w& 602095756 & QPS & 0.34 & 1.996 & $<$1E-4& 0.28 & -- \\
  000216 & 100.26538 & 9.47233 & w& 616944007 &  & 0.45 &  &  & & -- \\
  000217\tablefootmark{§} & 100.27903 & 9.68180 & w& 400007956 & MP? & 0.27 & 1.262 & 9.0E-2 & & 1.26 \\
  000219 & 100.32868 & 9.59839 & c& 616919663 & N & 0.23 &  &  & & -- \\
  000220\tablefootmark{§} & 100.35228 & 9.62653 & c& 616895930 & MP? & 0.30 & 0.75 & $<$1E-4 & 0.52 &  \\
  000223 & 100.23094 & 9.62326 & w& 602087949 & P & 0.91 & 1.9 & $<$1E-4& 0.10 & -- \\
  000226 & 100.27236 & 9.55374 & w& 603402475 & P & 1.20 & 1.206 & $<$1E-4& 0.00 & -- \\
  000227 & 100.22477 & 9.84948 & w& 605538529 & N & 0.26 &  &  & &  \\
  000235 & 100.24226 & 9.87655 & & 605538496 & U & &  &  & & -- \\
  000236 & 100.26056 & 9.58217 & w& 223981174 & P & 1.37 & 1.979 & $<$1E-4& 0.11 & 1.974 \\
  000237 & 100.29033 & 9.41520 & w& 616944029 & P & 0.45 & 3.381 & $<$1E-4& 0.13 & -- \\
  000241 & 100.34598 & 9.45741 & w& 223986498 & QPS & 1.85 & 3.25 & $<$1E-4& 0.16 & 3.206 \\
  000242 & 100.29940 & 9.44206 & c& 602079796 & D & 0.45 &  &  & &  \\
  000247 & 100.28035 & 9.83240 & w& 616849465 &  & 0.23 &  &  & & -- \\
  000250 & 100.25206 & 9.75086 & c& 223980688 & QPD & 1.35 & 8.929 & $<$1E-4& 0.37 &  \\
  000253 & 100.30370 & 9.76689 & w& 616872613 & N & 0.25 &  &  & & -- \\
  000255 & 100.42801 & 9.71574 & w& 223991789 & P & 0.62 & 3.927 & $<$1E-4& 0.14 & 3.956 \\
  000256 & 100.43427 & 9.41733 & w& 223992193 & EB & 0.36 & 3.874 & 6.0E-3\tablefootmark{b}& 0.45 & -- \\
  000263 & 100.26081 & 9.58698 & w& 602083896 & QPS & 1.20 & 4.287 & $<$1E-4& -0.07 & -- \\
  000273 & 100.32653 & 9.66143 & c& 616895921 & N & 0.45 &  &  & & -- \\
  000274 & 100.27864 & 9.38924 & w& 602075361 & QPS & 2.20 & 12.123 & $<$1E-4& -- & 11.92 \\
  000279 & 100.33823 & 9.53743 & c& 603402480 & QPS & 0.32 & 7.935 & $<$1E-4& 0.45 & -- \\
  000280 & 100.17088 & 9.46509 & c& 616944098 & N & 0.99 &  &  & &  \\
  000289 & 100.19650 & 9.48049 & & 616943962 & U & &  &  & &  \\
  000290 & 100.24440 & 9.60366 & c& 223980233 & QPS & 0.25 & 5.940 & $<$1E-4& &  \\
  000292 & 100.44757 & 9.70010 & w& 223993084 & QPS & 0.45 & 6.573 & $<$1E-4& 0.11 & 6.456 \\
  000294 & 100.26819 & 9.45852 & w& 616944010 &  & 0.28 & &  & & -- \\
  000296\tablefootmark{7} & 100.21079 & 9.91592 & c& 602099706 & QPD & 1.42 & 7.83 & $<$1E-4& 0.57 &  \\
  000297 & 100.18817 & 9.47901 & c& 223976747 & D & 1.42 &  &  & & 3.173 \\
  000298 & 100.27368 & 9.90520 & w& 605538656 & Be & 0.45 & 1.308 & $<$1E-4 & 0.33 & 1.289 \\
  000298 & 100.27368 & 9.90520 & w& 605538656 & Be & 0.45 & 1.246 & $<$1E-4 & & "" \\
  000305 & 100.23951 & 9.91596 & & 605538647 & N & &  &  & & -- \\
  000306 & 100.30207 & 9.77236 & w& 616872612 & Be & 0.23 & 0.452 & $<$1E-4& 0.14 & -- \\
  000306 & 100.30207 & 9.77236 & w& 616872612 & Be & 0.23 & 0.425 & 1.0E-2& & -- \\
  000311 & 100.48245 & 9.66614 & w& 616895733 & QPS & 0.32 & 6.497 & $<$1E-4& -0.15 & -- \\
  000314 & 100.18579 & 9.54061 & c& 616919732 & QPD & 0.29 & 3.279 & 2.0E-4 & 0.80 & 6.3 \\
  000319 & 100.21445 & 9.52969 & w& 616919745 & N & 0.33 &  &  & & -- \\
  000325 & 100.24726 & 9.92227 & c& 605538641 & D & 1.99 &  &  & & -- \\
  000326 & 100.24511 & 9.65520 & c& 223980258 & QP & 0.66 & 6.642 & $<$1E-4& 0.6 & 6.99 \\
  000328 & 100.36252 & 9.50364 & c& 400007735 & N & 0.45 &  &  & & 9.996 \\
  000330 & 100.38172 & 9.80911 & w& 223988742 & QPS & 1.55 & 5.054 & $<$1E-4& 0.02 & 5.025 \\
  000335 & 100.40540 & 9.75182 & c& 223990299 & P & 1.40 & 4.577 & $<$1E-4& 0.00 & 4.469 \\
  000338 & 100.47703 & 9.48775 & & 223995167 & N & &  &  & &  \\
  000339 & 100.39308 & 9.43150 & w& 603396408 & U & 0.30 &  &  & & -- \\
  000340 & 100.48303 & 9.67968 & c& 616895736 & QP & 0.33 & 1.225 & 3.0E-3& 0.53 & -- \\
  000341 & 100.22608 & 9.82232 & c& 616849439 & B & 0.53 &  &  & &  \\
  000342 & 100.23227 & 9.77934 & c& 616872592 & D & 0.25 &  &  & & -- \\
  000344 & 100.28963 & 9.59041 & w& 602083902 & P & 0.36 & 0.856 & $<$1E-4& 0.10 & -- \\
  000345 & 100.18776 & 9.85769 & w& 616849563 &  & 0.35 &  &  & & -- \\
  000346 & 100.28789 & 9.50255 & c& 603402478 & S & 0.70 &  &  & & -- \\
  000348 & 100.20871 & 9.87432 & w& 616849429 & QPS & 0.25 & 5.189 & 1.0E-4 & 0.62 & -- \\
  000351 & 100.34502 & 9.49416 & w& 616943891 & MP & 0.93 & 10.488 & $<$1E-4& & -- \\
  000351 & 100.34502 & 9.49416 & w& 616943891 & MP & 0.93 & 1.064 & 4.0E-2& 0.93 & -- \\
  000354 & 100.28680 & 9.39528 & w& 223982779 & QPS & 0.55 & 1.727 & $<$1E-4& 0.46 & 1.882 \\
  000356 & 100.27111 & 9.82300 & c& 616849458 & N & 0.20 &  &  & & 4.364 \\
  000357 & 100.27396 & 9.51698 & c& 616919781 & N & 1.20 &  &  & & -- \\
  000358 & 100.27803 & 9.79099 & c& 400007959 & QPD & 0.29 & 5.821 & $<$1E-4& 0.15 & 5.738 \\
  000368 & 100.24334 & 9.45697 & w& 616943994 & U & 0.24 & 1.031 & 2.0E-1& & 1.029 \\
  000370 & 100.23663 & 9.63025 & c& 223979728 & QPS & 1.20 & 11.838 & $<$1E-4& 0.32 &  \\
  000372 & 100.33561 & 9.75990 & w& 223985845 & P & 1.26 & 2.567 & $<$1E-4& 0.12 & 2.604 \\
  000375 & 100.42884 & 9.39107 & w& 616969611 & QPS & 0.31 & 8.725 & $<$1E-4& 0.43 & -- \\
  000377 & 100.45361 & 9.72037 & w& 223993499 & U & 1.98 &  & & &  \\
  000378 & 100.22048 & 9.74841 & c& 616872605 & QPS & 1.06 & 11.029 & $<$1E-4& 0.21 & -- \\
  000379\tablefootmark{8} & 100.27069 & 9.84613 & c& 223981811 & D & 1.60 & 3.66 & $<$1E-4& 0.78 &  \\
  000383 & 100.23242 & 9.67172 & w& 400007851 & QP & 0.32 & 1.027 & 2.0E-1& -- & -- \\
  000389 & 100.20319 & 9.54513 & w& 616919741 & U & 1.25 &  &  & & -- \\
  000394 & 100.28234 & 9.68749 & w& 616895903 & QPS & 0.62 & 1.963 & $<$1E-4& 0.36 & -- \\
  000395 & 100.29228 & 9.52463 & w& 616919644 & QPS & 0.28 & 1.654 & 2.0E-2& -- & -- \\
  000397 & 100.29497 & 9.77812 & w& 400007809 & U & 0.27 &  &  & &  \\
  000406 & 100.24864 & 9.47881 & c& 616943998 & B & 1.13 & 6.631 & $<$1E-4& 0.50 & -- \\
  000407 & 100.34179 & 9.85350 & w& 616849492 & QPS & 0.72 & 4.504 & 1.0E-4 & 0.41 & -- \\
  000410 & 100.27674 & 9.90298 & & 605538659 & P & 0.32 & 0.833 & $<$1E-4& -1.29 & -- \\
  000412 & 100.19630 & 9.54449 & c& 616919737 & B & 0.45 & 6.679 & $<$1E-4& 0.51 & -- \\
  000413 & 100.44350 & 9.71856 & w& 603414387 & P & 0.95 & 4.281 & $<$1E-4& -0.01 & -- \\
  000415 & 100.18274 & 9.80848 & w& 616849557 & P & 0.45 & 0.961 & $<$1E-4& 0.42 & -- \\
  000420 & 100.22579 & 9.93111 & w& 602099707 & QPS & 0.66 & 8.1 & $<$1E-4& -0.42 & -- \\
  000423 & 100.31191 & 9.43199 & c& 602079850 & D & 0.26 &  &  & & -- \\
  000424 & 100.31334 & 9.63267 & c& 616895917 & N & 0.45 &  &  & & -- \\
  000425 & 100.31951 & 9.49786 & c& 616943882 & S & 1.22 & 7.51 & 8.0E-3& 0.5 & -- \\
  000426 & 100.33108 & 9.50799 & c& 616919665 & U & 1.20 &  &  & & -- \\
  000427 & 100.17683 & 9.53906 & w& 223976099 & QPS & 1.12 & 14.813 & $<$1E-4& 0.22 & 14.17 \\
  000430 & 100.23655 & 9.50419 & w& 223979719 & Be & 0.52 & 0.533 & $<$1E-4& 0.27 & -- \\
  000430 & 100.23655 & 9.50419 & w& 223979719 & Be & 0.52 & 0.552 & $<$1E-4& & -- \\
  000433 & 100.25462 & 9.58117 & c& 616919770 & QPD & 0.44 & 9.798 & $<$1E-4& 0.50 & -- \\
  000434 & 100.26691 & 9.58924 & c& 616919776 & MP & 0.33 & 7.485 & $<$1E-4& & -- \\
  000434 & 100.26691 & 9.58924 & c& 616919776 & MP & 0.33 & 0.725 & 1.0E-4 & 0.83 & -- \\
  000438 & 100.26427 & 9.50139 & w& 616919773 & P & 0.34 & 1.308 & $<$1E-4& -0.33 & -- \\
  000440 & 100.41271 & 9.49394 & w& 223990764 & U & 0.74 & & & &  \\
  000441 & 100.24206 & 9.61485 & c& 223980048 & QPD & 0.36 & &  &  & 12.5 \\
  000443 & 100.36481 & 9.53213 & w& 223987667 & U & 1.30 &  &  & & -- \\
  000444 & 100.43574 & 9.70346 & w& 223992277 & QPS & 0.36 & 10.246 & $<$1E-4& 0.00 & -- \\
  000445 & 100.51066 & 9.61458 & w& 223997570 & MP? & 0.95 & 3.651 & $<$1E-4& & 3.66 \\
  000448\tablefootmark{9} & 100.26502 & 9.50808 & c& 602083897 & U & 0.25 & 4.731 & 5.0E-4 & 0.78 & 9.73 \\
  000450 & 100.49184 & 9.71841 & w& 602091914 & P & 0.94 & 2.102 & $<$1E-4& 0.07 & -- \\
  000451 & 100.46447 & 9.73602 & w& 616872431 & QPS & 0.45 & 4.515 & $<$1E-4& 0.21 & -- \\
  000454 & 100.26513 & 9.60130 & & 616895890 & N & &  &  & & -- \\
  000456 & 100.21475 & 9.72339 & c& 616872585 & QPD & 1.41 & 5.054 & $<$1E-4& 0.62 & -- \\
  000457 & 100.27805 & 9.57935 & c& 616919789 & S & 1.82 &  &  & &  \\
  000461 & 100.47901 & 9.50181 & w& 616919638 & QPS & 0.35 & 3.429 & $<$1E-4& 0.26 & -- \\
  000462\tablefootmark{10} & 100.17576 & 9.56040 & c& 223976028 & D & 1.60 & 1.913 & 5.0E-2& &  \\
  000468\tablefootmark{\#} & 100.28694 & 9.76696 & w& 223982794 & Be & 1.30 & 4.146 & $<$1E-4& & -- \\
  000468\tablefootmark{\#} & 100.28694 & 9.76696 & w& 223982794 & Be & 1.30 & 4.933 & $<$1E-4& & -- \\
  000469 & 100.17142 & 9.56607 & c& 602083890 & S & 0.71 &  &  & & -- \\
  000470 & 100.18431 & 9.89426 & & 602095747 & N & &  & &  & -- \\
  000474 & 100.27837 & 9.45892 & c& 603396438 & B & &  &  & & -- \\
  000477 & 100.31560 & 9.43806 & w& 223984608 & P & 1.22 & 6.227 & $<$1E-4& 0.02 & 6.098 \\
  000484 & 100.27258 & 9.56930 & c& 616919779 &  & 0.13 & 19.5 & $<$1E-4& & -- \\
  000486 & 100.26684 & 9.81911 & w& 602095758 & QPS & 0.69 & 12.385 & $<$1E-4& 0.02 & -- \\
  000488 & 100.24094 & 9.94167 & w& 223979980 &  & 1.60 & 0.583 & $<$1E-4& -- & 0.577 \\
  000491 & 100.23401 & 9.60857 & c& 616895873 & L & 1.70 &  &  & & -- \\
  000493 & 100.35452 & 9.60004 & w& 400008004 &  & 0.27 &  & & & 3.19 \\
  000497 & 100.25919 & 9.86446 & w& 616849449 & P & 2.09 & 9.988 & $<$1E-4& 0.06 & 10.0 \\
  000498 & 100.19793 & 9.82470 & c& 616849574 & QPD & 1.90 & 4.3 & $<$1E-4& 0.15 & 8.53 \\
  000499 & 100.26896 & 9.42175 & w& 616944012 & N & 0.25 &  &  & & -- \\
  000500 & 100.23279 & 9.85847 & & 616849441 & U & 0.38 &  &  & & -- \\
  000510 & 100.26787 & 9.41449 & c& 602079845 & B & 0.62 &  &  & & 14.99 \\
  000515 & 100.40097 & 9.65568 & w& 223989989 & P & 0.72 & 6.542 & $<$1E-4& -0.08 & 6.547 \\
  000517 & 100.26964 & 9.60742 & w& 223981753 & MP & 2.40 & 2.772 & 4.0E-4 & 0.36 & 2.971 \\
  000517 & 100.26964 & 9.60742 & w& 223981753 & MP & 2.40 & 3.004 & 1.0E-3& 0.36 & "" \\
  000518 & 100.25705 & 9.80614 & w& 223980989 & QPS & 1.21 & 6.546 & $<$1E-4& 0.51 & -- \\
  000519 & 100.29133 & 9.50560 & w& 616919643 & QPS & 0.34 & 6.008 & 6.0E-4 & -- & -- \\
  000524 & 100.26806 & 9.80614 & w& 616849453 & P & 1.31 & 5.152 & $<$1E-4& 0.06 & -- \\
  000525\tablefootmark{§} & 100.21323 & 9.74612 & c& 223978308 & {\small QPD,MP?} & 1.80 & 1.992 & $<$1E-4& & 5.374 \\
  000529 & 100.21334 & 9.46610 & w& 616943973 & QP & 0.31 & 7.163 & $<$1E-4& -- & -- \\
  000530 & 100.33079 & 9.36309 & c& 602075358 & N & 0.82 &  &  & & -- \\
  000536 & 100.21666 & 9.75132 & w& 400007394 & QPS & 0.45 & 3.402 & $<$1E-4& -- & 3.443 \\
  000548 & 100.28568 & 9.71432 & w& 603414376 & QPS & 0.45 & 10.7 & $<$1E-4& -1.46 & -- \\
  000555 & 100.35191 & 9.54589 & w& 616919676 & QP & 0.19 & 1.048 & 8.0E-4 & 0.62 & -- \\
  000558 & 100.41561 & 9.67442 & c& 223990964 & QPS & 1.60 & 11.708 & $<$1E-4& 0.43 & 10.17 \\
  000559 & 100.35105 & 9.53172 & w& 223986811 &  & 0.95 & 7.956 & $<$1E-4& 0.47 & 7.92 \\
  000560 & 100.31547 & 9.63841 & w& 223984600 & QPS & 1.34 & 5.313 & $<$1E-4& -0.13 & 5.343 \\
  000563 & 100.42436 & 9.55060 & w& 616919579 & N & 0.23 &  &  & & -- \\
  000565 & 100.25325 & 9.85621 & w& 616849446 & P & 1.35 & 4.375 & $<$1E-4& 0.02 & 4.344 \\
  000566 & 100.21982 & 9.71678 & c& 400007955 & U & 0.26 &  &  & &  \\
  000567 & 100.23500 & 9.59813 & c& 616919752 & B & 1.45 &  & & & -- \\
  000568 & 100.22405 & 9.51084 & w& 223978947 & U & 1.80 &  & & & 8.5 \\
  000574 & 100.21192 & 9.93140 & w& 223978227 & P & 0.70 & 3.794 & $<$1E-4& 0.15 & 3.779 \\
  000577 & 100.43258 & 9.68055 & c& 616895846 & S & 1.62 & &  & & -- \\
  000578 & 100.45846 & 9.49228 & c& 223993840 & P & 0.94 & 3.269 & $<$1E-4& 0.13 & 3.25 \\
  000581 & 100.20040 & 9.89427 & w& 602095750 & QPS & 1.00 & 5.521 & 5.0E-4 & 0.36 & -- \\
  000583 & 100.22171 & 9.49839 & w& 400008031 &  & 0.28 &  & & & -- \\
  000586 & 100.24166 & 9.69209 & & 605538448 & U & &  &  & & -- \\
  000590 & 100.24549 & 9.48131 & c& 603396403 & S & 0.45 &  &  & & -- \\
  000596 & 100.44768 & 9.63129 & w& 602087963 & P & 0.45 & 9.142 & $<$1E-4& 0.08 & -- \\
  000598 & 100.24721 & 9.77128 & c& 616872597 & B & 0.45 &  &  & & -- \\
  000604 & 100.27124 & 9.86236 & w& 616849455 & P & 1.20 & 9.958 & $<$1E-4& 0.03 & -- \\
  000606 & 100.31028 & 9.55595 & w& 223984253 & P & 1.20 & 10.798 & $<$1E-4& 0.05 & 10.42 \\
  000607 & 100.31516 & 9.44262 & w& 223984572 & Be & 0.28 & 1.581 & $<$1E-4& 0.31 &  \\
  000607 & 100.31516 & 9.44262 & w& 223984572 & Be & 0.28 & 1.277 & 2.0E-2& &  \\
  000609 & 100.41503 & 9.55553 & w& 603402484 & N & 0.29 &  &  & & -- \\
  000610 & 100.18907 & 9.63095 & w& 616896016 & N & 0.67 &  &  & & -- \\
  000612 & 100.24250 & 9.92905 & w& 602266739 & P & 0.30 & 4.304 & $<$1E-4& 0.05 & -- \\
  000613 & 100.27406 & 9.80486 & c& 616849463 & S & 0.78 &  &  & & -- \\
  000614 & 100.17435 & 9.69406 & w& 616896008 & N & 0.27 &  &  & & 1.805 \\
  000617 & 100.28953 & 9.86389 & w& 602095761 & QPS & 0.30 & 5.956 & $<$1E-4& 0.15 & -- \\
  000619 & 100.31142 & 9.57033 & c& 603402479 & D & 0.69 & 6.404 & 5.0E-2& 0.57 & -- \\
  000620 & 100.33643 & 9.50333 & w& 602083907 & QPS & 0.36 & 8.6 & $<$1E-4& 0.16 & -- \\
  000622 & 100.34081 & 9.75860 & w& 602091881 & QPS & 1.16 & 12.629 & $<$1E-4& -0.05 & -- \\
  000624 & 100.26835 & 9.86390 & w& 616849454 & QPS & 1.85 & 7.47 & 8.0E-1\tablefootmark{b}& 0.50 & 7.5 \\
  000629 & 100.27497 & 9.59762 & w& 616919784 & U & 1.20 & & & & -- \\
  000631 & 100.27462 & 9.38222 & & 223982068 & N & &  &  & & -- \\
  000636 & 100.20351 & 9.72379 & c& 616872578 & S & 0.62 &  &  & & -- \\
  000637 & 100.20504 & 9.96071 & c& 616826638 & L & 0.45 & 12.306 & $<$1E-4& & 12.5 \\
  000638 & 100.26900 & 9.91213 & & 602266765 & N & &  &  & & -- \\
  000646 & 100.19197 & 9.82152 & w& 616849567 & P & 2.72 & 0.746 & $<$1E-4& 0.06 & 0.747 \\
  000647 & 100.26233 & 9.79842 & w& 223981285 & MP & 1.41 & 1.181 & $<$1E-4& & 1.152 \\
  000647 & 100.26233 & 9.79842 & w& 223981285 & MP & 1.41 & 1.073 & 3.0E-3& & "" \\
  000647 & 100.26233 & 9.79842 & w& 223981285 & MP & 1.41 & 1.119 & 3.0E-3& & "" \\
  000647 & 100.26233 & 9.79842 & w& 223981285 & MP & 1.41 & 1.231 & 5.0E-3& & "" \\
  000650 & 100.25409 & 9.54568 & c& 223980807 & D & 2.50 &  &  & &  \\
  000656 & 100.33554 & 9.79165 & w& 616872623 & P & 0.36 & 4.442 & $<$1E-4& 0.14 & -- \\
  000657 & 100.32378 & 9.49060 & w& 400007328 & QPS & 0.30 & 4.829 & 8.0E-1\tablefootmark{b}& 0.37 & 2.434 \\
  000660 & 100.25215 & 9.48775 & c& 223980693 & QPD & 1.40 & 5.125 & $<$1E-4& 0.34 & 5.282 \\
  000667 & 100.36989 & 9.64413 & c& 223987997 & D & 1.53 & 5.897 & 1.0E-3& 0.62 & 6.456 \\
  000676 & 100.44917 & 9.56935 & c& 223993199 & D & 1.20 & 9.404 & $<$1E-4& 0.69 &  \\
  000677 & 100.24718 & 9.37275 & w& 616969698 & N & 0.23 &  &  & & -- \\
  000679 & 100.26681 & 9.39229 & & 223981550 & U & &  &  & & 14.58 \\
  000680 & 100.30555 & 9.46872 & w& 603396406 & P & 1.09 & 5.768 & $<$1E-4& 0.06 & -- \\
  000681 & 100.37965 & 9.44949 & c& 603808965 & D & 1.69 &  &  & &  \\
  000688 & 100.27123 & 9.81331 & w& 616849460 & QPS & 0.62 & 3.756 & $<$1E-4& 0.02 & 3.748 \\
  000689 & 100.21422 & 9.89006 & & 605538681 & N & &  &  & & -- \\
  000695 & 100.38543 & 9.63537 & w& 223988965 & P & 0.90 & 3.235 & $<$1E-4& -0.19 & 9.5 \\
  000697 & 100.39398 & 9.60902 & c& 223989567 & U & 0.45 &  &  & &  \\
  000705\tablefootmark{§} & 100.29283 & 9.55697 & w& 616919645 & MP? & 0.30 & 0.679 & 2.0E-2& &  \\
  000714 & 100.34228 & 9.35863 & w& 616969724 &  & 0.30 &  &  & & 3.405 \\
  000717 & 100.31298 & 9.44565 & c& 616943877 & QPD & 0.53 & 8.558 & $<$1E-4& 0.49 & -- \\
  000718 & 100.26733 & 9.34564 & c& 616969715 & N & 0.34 &  &  & & -- \\
  000719 & 100.26848 & 9.85725 & w& 602095759 & QPS & 1.20 & 3.975 & $<$1E-4& 0.28 & 4.024 \\
  000723 & 100.21888 & 9.84961 & w& 616849436 & U & 0.32 &  &  & & -- \\
  000724 & 100.21947 & 9.73917 & w& 400007529 & QPS & 0.36 & 5.127 & $<$1E-4& 0.18 & 4.842 \\
  000728 & 100.28241 & 9.73417 & w& 223982535 & QPS & 1.58 & 5.158 & $<$1E-4& 0.17 & 5.052 \\
  000733 & 100.20749 & 9.79200 & c& 616872582 & N & 0.25 &  &  & & -- \\
  000743 & 100.28086 & 9.51970 & w& 603402476 & N & 0.32 &  &  & & -- \\
  000744 & 100.18384 & 9.83265 & & 616849558 & N & &  &  & & -- \\
  000745 & 100.20112 & 9.61073 & w& 603408576 & QPS & 0.45 & 1.669 & $<$1E-4& 0.37 & -- \\
  000747 & 100.34176 & 9.72021 & w& 616872626 & P & 0.95 & 6.61 & $<$1E-4& 0.00 & -- \\
  000749 & 100.30890 & 9.44460 & w& 616943875 & P & 0.45 & 1.436 & $<$1E-4& 0.06 & -- \\
  000753 & 100.27486 & 9.65392 & c& 616895898 & N & 0.21 &  &  & & 7.67 \\
  000754 & 100.22094 & 9.88318 & c& 605538488 & P & 0.64 & 0.969 & $<$1E-4& 0.13 & -- \\
  000755 & 100.19562 & 9.81333 & w& 605538556 & Be & 0.95 & 3.5 & $<$1E-4&  & -- \\
  000755 & 100.19562 & 9.81333 & w& 605538556 & Be & 0.95 & 4.054 & $<$1E-4& & -- \\
  000756 & 100.18974 & 9.85641 & & 616849565 & U & & & & & -- \\
  000765 & 100.22349 & 9.55688 & c& 223978921 & QPS & 1.45 &  2.371 & $<$1E-4& 0.22 &  \\
  000766\tablefootmark{8} & 100.20158 & 9.81069 & c& 602095741 & S & 0.53 & 2.798 & $<$1E-4& 0.63 & -- \\
  000770 & 100.31032 & 9.62065 & w& 616895918 & QPS & 0.66 & 5.442 & $<$1E-4& -0.05 & 5.405 \\
  000771 & 100.32610 & 9.56489 & c& 223985261 & U & 1.38 & & & & 18.08 \\
  000774\tablefootmark{*} & 100.24519 & 9.51592 & c& 223980264 & S & 1.83 & 3.49 & 2.0E-4 & 0.82 & 3.482 \\
  000777 & 100.26328 & 9.43417 & w& 400007811 & U & 0.67 &  &  & & -- \\
  000779\tablefootmark{*} & 100.20080 & 9.45026 & w& 602079840 & U & 0.21 & 3.131 & 1.0E-3& 1.27 & -- \\
  000784 & 100.25591 & 9.56895 & w& 616919771 & P & 1.20 & 10.098 & $<$1E-4& -0.01 & -- \\
  000794 & 100.30442 & 9.38455 & w& 616969735 &  & 0.47 & 4.031 & 2.0E-1\tablefootmark{a}& -- & -- \\
  000797 & 100.20071 & 9.40329 & & 603808962 & N & &  &  & & -- \\
  000798 & 100.23731 & 9.81134 & w& 223979759 & QPS & 1.57 & 3.808 & $<$1E-4& -0.33 & 3.84 \\
  000804 & 100.23216 & 9.85385 & c& 616849440 & B & 1.06 & 3.271 & $<$1E-4& 0.41 & 3.217 \\
  000805 & 100.43724 & 9.74455 & w& 223992383 &  & 0.71 & 3.425 & 9.0E-3& 0.51 & 3.38 \\
  000808 & 100.21494 & 9.47905 & c& 603396401 & B & 1.24 &  &  & & -- \\
  000809 & 100.26456 & 9.52181 & w& 223981406 & P & 1.58 & 2.167 & $<$1E-4& 0.09 & 2.157 \\
  000810 & 100.29096 & 9.45339 & c& 605538241 & P & 1.20 & 2.925 & $<$1E-4& 0.06 & 2.914 \\
  000811 & 100.18004 & 9.78534 & c& 605538574 & QPD & 0.91 & 7.844 & $<$1E-4& 0.41 & 8.49 \\
  000819 & 100.37159 & 9.65997 & w& 223988099 & QPS & 1.65 & 3.333 & $<$1E-4& 0.23 & 3.273 \\
  000823 & 100.20889 & 9.95111 & c& 603809233 & U & 0.30 &  &  & & -- \\
  000826 & 100.21700 & 9.87052 & w& 616849435 & U & 0.67 & 14.531 & $<$1E-4& 0.55 & -- \\
  000842\tablefootmark{\#} & 100.24214 & 9.85375 & w& 605538526 & MP & 0.33 & 11.094 & $<$1E-4&  & -- \\
  000842\tablefootmark{\#} & 100.24214 & 9.85375 & w& 605538526 & MP & 0.33 & 1.917 & 2.0E-4 & & -- \\
  000843 & 100.21404 & 9.50371 & w& 400007916 & N & 0.21 &  &  & & -- \\
  000848 & 100.35306 & 9.43983 & w& 223986923 & QPS & 1.35 & 8.506 & $<$1E-4& -0.14 & 8.3 \\
  000869 & 100.23942 & 9.48981 & w& 603808963 & P & 0.32 & 8.898 & $<$1E-4& 0.42 & 8.854 \\
  000870 & 100.20729 & 9.88296 & & 616849427 & U & 0.30 &  &  & & -- \\
  000872 & 100.19208 & 9.79727 & w& 602266744 & QPS & 0.36 & 5.927 & $<$1E-4 & 1.36 & -- \\
  000876 & 100.30052 & 9.49774 & w& 616944031 & U & 0.21 &  &  & & -- \\
  000877 & 100.31993 & 9.45839 & c& 616943883 & B & 1.33 & 5.177 & $<$1E-4& 0.50 & -- \\
  000878 & 100.34572 & 9.49187 & w& 616943892 & N & 0.28 &  &  & & -- \\
  000879 & 100.26409 & 9.67912 & c& 603408580 & S & 0.45 & 11.398 & $<$1E-4& 0.25 & -- \\
  000881 & 100.28715 & 9.68745 & w& 616895909 & P & 1.16 & 3.919 & $<$1E-4& 0.18 & -- \\
  000886 & 100.18767 & 9.76161 & w& 616872573 & P & 0.71 & 4.612 & $<$1E-4& 0.05 & 4.625 \\
  000890 & 100.22993 & 9.84716 & w& 616849420 & P & 1.05 & 1.165 & $<$1E-4& 0.05 & 1.165 \\
  000892 & 100.25646 & 9.47031 & & 223980955 & QPS & 2.26 & 2.415 & 1.0E-2& 0.54 & -- \\
  000894 & 100.31422 & 9.77766 & w& 223984520 & P & 1.60 & 1.463 & $<$1E-4& 0.14 & 1.469 \\
  000901 & 100.18062 & 9.84986 & w& 616849547 & QPS & 0.69 & 9.031 & $<$1E-4& 0.21 & 9.114 \\
  000906 & 100.27711 & 9.89594 & & 605538675 & N & 0.37 &  &  & & -- \\
  000907 & 100.39952 & 9.67830 & w& 616895801 & N & 0.30 &  &  & & -- \\
  000910 & 100.19965 & 9.55087 & w& 603809014 & QPS & 0.31 & 2.581 & $<$1E-4& 0.57 & 2.568 \\
  000913 & 100.24680 & 9.88552 & & 605538690 & U & &  &  & & -- \\
  000914 & 100.21629 & 9.63219 & c& 602087948 & U & &  &  & & -- \\
  000919 & 100.30541 & 9.53064 & c& 616919654 & B & 0.25 &  &  & & -- \\
  000923 & 100.42167 & 9.54519 & & 223991355 & U & & & & &  \\
  000925 & 100.19408 & 9.36140 & & 223977092 & U & &  &  & &  \\
  000926 & 100.27678 & 9.47746 & c& 400007687 & QPS & 0.40 & 12.323 & $<$1E-4& -- &  \\
  000927 & 100.41256 & 9.55449 & w& 616919568 & QPS & 0.25 & 5.912 & $<$1E-4& 0.05 & -- \\
  000928 & 100.35674 & 9.57862 & c& 223987178 & N & 0.63 &  &  & & 9.84 \\
  000931 & 100.32534 & 9.64042 & c& 616895920 & N & 0.29 &  &  & &  \\
  000932 & 100.18689 & 9.96229 & w& 223976672 & P & 0.90 & 15.373 & $<$1E-4& 0.09 & 15.0 \\
  000936 & 100.27988 & 9.45816 & c& 605538236 & B & 1.20 & & & & -- \\
  000937 & 100.21896 & 9.86833 & c& 603809175 & N & 0.69 &  &  & &  \\
  000938 & 100.17636 & 9.57362 & w& 616919726 & QP & 0.29 & 7.037 & 1.0E-3& 0.44 & -- \\
  000941 & 100.36975 & 9.45299 & w& 400007743 & P & 0.30 & 1.313 & $<$1E-4& 0.61 & -- \\
  000945 & 100.20787 & 9.61375 & c& 223977953 & B & 1.40 &  &  & & 4.919 \\
  000948 & 100.36316 & 9.58504 & w& 223987553 & QPS & 1.49 & 1.546 & $<$1E-4& 0.08 & 1.544 \\
  000949 & 100.29927 & 9.39239 & & 616969729 & N & 0.31 & &  & & -- \\
  000951 & 100.32468 & 9.48365 & c& 602079852 & QP & 0.33 & 2.913 & 1.0E-4 & 0.61 & 10.44 \\
  000954 & 100.28063 & 9.43196 & w& 603396405 & P & 1.40 & 7.352 & $<$1E-4& 0.04 & -- \\
  000958 & 100.23104 & 9.15800 & w& 603808801 &  & 0.42 & 0.712 & 1.7E-1\tablefootmark{d}& -- & -- \\
  000964 & 100.27966 & 9.21065 & c& 223982375 & QPS & 0.95 & 3.289 & $<$1E-4& 0.54 & 3.32 \\
  000965 & 100.19170 & 9.29947 & c& 616996507 & QPS & 0.36 & 9.688 & $<$1E-4& 0.01 & 9.786 \\
  000967 & 100.30676 & 9.23151 & w& 223984026 & Be & 1.20 & 3.352 & $<$1E-4&  & -- \\
  000967 & 100.30676 & 9.23151 & w& 223984026 & Be & 1.20 & 2.781 & $<$1E-4& & -- \\
  000977 & 100.24625 & 9.28319 & w& 616996570 & P & 0.25 & 0.719 & $<$1E-4& 0.66 & -- \\
  000981 & 100.29208 & 9.24239 & & 223983112 & U & &  &  & & -- \\
  000985 & 100.06316 & 10.03274 & c& 223969098 & D & 0.90 &  &  & &  \\
  000989 & 100.03549 & 9.73707 & w& 223967301 & P & 0.45 & 0.958 & $<$1E-4& 0.14 & 0.957 \\
  000990 & 100.15702 & 9.66106 & w& 616896002 & N & 0.75 &  &  & & -- \\
  000991 & 100.08045 & 9.80829 & w& 616849658 & P & 0.62 & 1.033 & $<$1E-4& 0.06 & -- \\
  000995 & 100.10707 & 9.97660 & w& 223971866 & QPS & 1.75 & 9.284 & 1.0E-4 & 0.55 & 7.015 \\
  000996 & 100.17216 & 9.85068 & c& 616849542 & S & 0.69 &  &  & & 7.812 \\
  001000 & 100.04867 & 9.76557 & w& 602091887 & P & 0.36 & 1.433 & $<$1E-4& 0.31 & -- \\
  001001 & 99.98408 & 9.51282 & & 223963815 & MP & & 2.971 & $<$1E-4& &  \\
  001001 & 99.98408 & 9.51282 & & 223963815 & MP & & 2.706 & $<$1E-4& &  \\
  001001 & 99.98408 & 9.51282 & & 223963815 & MP & & 3.269 & 7.0E-3& &  \\
  001002 & 100.10938 & 9.63374 & w& 223971984 & P & 0.61 & 6.298 & $<$1E-4& 0.06 & 6.281 \\
  001003 & 100.10686 & 9.99994 & c& 616803611 & U & 0.24 & 3.454 & 2.0E-4 & 0.55 & 3.469 \\
  001005 & 100.02174 & 9.84904 & w& 616849610 & QPS & 0.33 & 1.769 & $<$1E-4& 0.45 & -- \\
  001009 & 99.94280 & 9.80551 & w& 223960995 & N & 0.65 &  &  & &  \\
  001012 & 100.13880 & 9.98137 & w& 223973818 & U & 0.65 & 4.465 & $<$1E-4& 2.21 & -- \\
  001015 & 99.91381 & 9.93322 & w& 223958963 & MP & 0.62 & 4.246 & 7.0E-1\tablefootmark{b}& & 0.859 \\
  001015 & 99.91381 & 9.93322 & w& 223958963 & MP & 0.62 & 0.858 & $<$1E-4& & "" \\
  001016 & 100.08762 & 9.60888 & w& 223970694 & P & 1.67 & 1.483 & $<$1E-4& 0.27 & 1.467 \\
  001017 & 100.09888 & 9.92329 & c& 223971383 & B & 0.30 &  &  & & 4.648 \\
  001022 & 100.16297 & 9.84962 & c& 616849543 & B & 1.50 &  &  & & 13.88 \\
  001023\tablefootmark{\#} & 100.04293 & 9.64862 & w& 223967803 & MP & 0.69 & 3.827 & $<$1E-4& & 3.841 \\
  001023\tablefootmark{\#} & 100.04293 & 9.64862 & w& 223967803 & MP & 0.69 & 0.629 & $<$1E-4& & "" \\
  001027 & 100.15500 & 9.51941 & w& 616919872 & P & 0.33 & 7.217 & $<$1E-4& 0.04 & -- \\
  001031 & 100.16687 & 9.58415 & c& 616919878 & QPS & 1.46 & 4.448 & $<$1E-4& 0.42 & -- \\
  001037 & 100.12858 & 9.57792 & c& 223973200 & QPS & 1.45 & 8.877 & $<$1E-4& 0.55 &  \\
  001038 & 100.09427 & 9.82952 & c& 602095739 & D & 0.66 & 6.383 & 5.0E-3& 0.62 & -- \\
  001047 & 99.98171 & 9.79216 & w& 223963678 & MP & 0.30 & 0.677 & $<$1E-4& & 0.676 \\
  001047 & 99.98171 & 9.79216 & w& 223963678 & MP & 0.30 & 0.914 & 2.0E-3& & "" \\
  001048 & 100.01110 & 9.59007 & c& 616919938 & B & 0.30 &  &  & & -- \\
  001053 & 100.17140 & 9.88236 & c& 602095745 & S & 0.91 & 11.838 & $<$1E-4& 0.52 & -- \\
  001054 & 100.15217 & 9.84600 & c& 400007538 & S & 0.36 & 8.142 & 2.0E-2\tablefootmark{c}& 0.49 &  \\
  001055 & 100.16845 & 9.84734 & c& 616849545 & QPS & 0.68 & 3.731 & $<$1E-4& 0.23 & 3.748 \\
  001056 & 100.02624 & 9.59904 & w& 616919952 & P & 0.36 & 1.519 & 5.0E-1\tablefootmark{d}& 0.44 & -- \\
  001059 & 100.08598 & 9.68055 & c& 616895951 & QPS & 0.22 & 5.819 & $<$1E-4& 0.20 & -- \\
  001061 & 100.10065 & 9.57013 & c& 616919835 & S & 0.64 &  &  & & -- \\
  001062 & 100.10348 & 9.88652 & & 616849505 & U & & & & & -- \\
  001064 & 100.14658 & 9.86579 & c& 616849538 & P & 0.45 & 2.698 & $<$1E-4& -0.06 & -- \\
  001067\tablefootmark{§} & 100.04552 & 9.90725 & w& 602099690 & MP? & 0.29 & 1.148 & 6.0E-4 & & -- \\
  001072 & 100.13061 & 9.69147 & w& 223973318 & N & 1.01 &  &  & &  \\
  001074 & 100.03512 & 9.99052 & w& 616826786 &  & 0.28 &  & & & -- \\
  001076 & 100.15916 & 9.49792 & c& 605424384 & U & 0.45 &  &  & & 5.884 \\
  001077 & 100.17760 & 9.89659 & & 616849552 & QPS & 1.00 & 8.746 & $<$1E-4& -0.05 & -- \\
  001081 & 100.08732 & 9.39841 & w& 223970675 & N & 0.76 &  &  & & -- \\
  001082 & 100.00467 & 9.59265 & w& 616919920 & QPS & 0.33 & 9.185 & $<$1E-4& -- & 9.114 \\
  001085 & 100.13667 & 9.85815 & w& 223973692 & P & 0.91 & 3.452 & $<$1E-4& 0.09 & 3.456 \\
  001087 & 100.00504 & 9.71013 & w& 616872743 &  & 0.30 & 3.931 & 1.0E-2& & -- \\
  001089 & 100.12452 & 9.83622 & w& 223972960 & U & 1.55 &  &  & &  \\
  001092 & 100.16125 & 9.61581 & w& 602087946 & QPS & 0.30 & 1.317 & $<$1E-4& -- & -- \\
  001094 & 100.13183 & 9.80649 & c& 603420177 & QPS & 1.14 & 4.246 & $<$1E-4& 0.39 & -- \\
  001099 & 100.17235 & 9.90385 & c& 223975844 & QPS & 1.52 & 3.419 & $<$1E-4& 0.36 & 3.332 \\
  001100 & 100.16393 & 9.57930 & c& 616919877 & U & 0.90 &  &  & & -- \\
  001101 & 100.04533 & 9.64467 & w& 603408567 & U & 0.53 &  & & & -- \\
  001102 & 100.05041 & 9.82508 & w& 602095736 & U & 1.20 & & & & -- \\
  001105 & 100.12026 & 9.55154 & w& 616919856 & Be & 0.30 & 0.758 & $<$1E-4& & -- \\
  001105 & 100.12026 & 9.55154 & w& 616919856 & Be & 0.30 & 0.958 & $<$1E-4& & -- \\
  001114 & 99.88913 & 9.86714 & c& 223957142 & QPS & 0.40 & 2.579 & $<$1E-4& 0.41 & 2.568 \\
  001115 & 100.00896 & 9.75395 & w& 223965459 & P & 0.83 & 1.352 & $<$1E-4& 0.03 & 1.351 \\
  001126 & 100.04007 & 9.69540 & w& 223967602 &  & 1.62 & 1.233 & $<$1E-4& 0.19 & 1.236 \\
  001131 & 99.89339 & 9.91422 & c& 223957455 & QPD & 0.36 & 5.144 & $<$1E-4& 0.44 & 10.16 \\
  001132 & 100.10781 & 9.84933 & c& 602095740 & QPS & 0.33 & 2.958 & $<$1E-4& 0.51 & -- \\
  001133 & 99.99820 & 9.94008 & w& 223964746 & P & 0.95 & 1.25 & $<$1E-4& 0.10 & -- \\
  001140 & 99.92278 & 9.77213 & c& 223959618 & QPD & 1.31 & 3.917 & $<$1E-4& 0.48 & 3.922 \\
  001142 & 100.14361 & 9.58839 & w& 616919870 & N & 0.36 & & & & -- \\
  001144 & 100.09620 & 9.46176 & c& 223971231 & QP & 1.12 & 4.106 & 1.0E-4 & 0.75 &  \\
  001147 & 99.95761 & 9.93939 & w& 223962024 & N & 0.88 &  &  & &  \\
  001149 & 100.12741 & 9.83736 & c& 603420176 & N & 0.30 &  &  & & -- \\
  001152 & 100.16453 & 9.81102 & w& 605539518 & U & 1.30 &  & &  & -- \\
  001156 & 100.13984 & 9.56013 & c& 616919866 & U & 0.30 & 16.335 & $<$1E-4& 0.24 & -- \\
  001157 & 100.08369 & 9.47460 & c& 223970440 & QPS & 0.95 & 3.813 & $<$1E-4& 0.43 &  \\
  001158 & 99.85939 & 9.68634 & w& 223955032 & QPS & 0.62 & 5.546 & $<$1E-4& -0.06 & 5.436 \\
  001163 & 100.12398 & 9.99260 & w& 223972918 & U & 0.65 &  &  & &  \\
  001167 & 100.15781 & 9.58166 & c& 400007528 & QPS & 0.30 & 8.804 & $<$1E-4& 0.22 & 9.42 \\
  001171 & 99.89160 & 9.82245 & c& 223957322 & D & 0.70 &  & & & 18.05 \\
  001172 & 100.02658 & 9.65932 & w& 616896077 & QPS & 0.65 & 6.492 & $<$1E-4& 0.06 & -- \\
  001174 & 100.05711 & 9.94183 & c& 616826810 & B & 0.35 & 6.84 & 4.0E-3& 0.80 &  \\
  001181 & 100.07507 & 9.83947 & w& 616849652 & QPS & 0.35 & 6.1 & $<$1E-4& 0.64 & -- \\
  001187 & 100.05904 & 9.57453 & c& 602083884 & B & 0.40 & 3.102 & 3.0E-4 & 0.86 & -- \\
  001189 & 100.16258 & 9.60000 & w& 223975253 & P & 0.63 & 4.581 & $<$1E-4& 0.55 &  \\
  001193 & 99.89478 & 9.78170 & w& 602091877 & U & 0.66 &  &  & & -- \\
  001194 & 100.15287 & 9.36811 & w& 223974689 & N & 0.65 &  &  & &  \\
  001195 & 100.14539 & 9.90199 & w& 400007919 & U & 0.25 &  &  & &  \\
  001196 & 99.98682 & 9.74033 & & 223963994 & U & & & & &  \\
  001197 & 100.10900 & 9.61923 & c& 603408572 & QPS & 0.30 & 4.027 & $<$1E-4& 0.50 & -- \\
  001199 & 100.17434 & 9.86241 & c& 602095746 & QPS & 1.20 & 3.617 & $<$1E-4& 0.57 & 3.674 \\
  001200 & 100.05182 & 9.73973 & w& 223968398 & P & 0.45 & 2.694 & $<$1E-4& 0.38 & 2.702 \\
  001201 & 100.15263 & 9.80638 & w& 605539512 & P & 1.16 & 16.435 & $<$1E-4& $<$0 & 15.25 \\
  001204 & 99.95670 & 9.55611 & w& 223961941 & QPS & 1.12 & 6.552 & $<$1E-4& 0.04 & 6.52 \\
  001205 & 100.12003 & 9.80664 & w& 400007344 & P & 1.13 & 6.861 & $<$1E-4& 0.02 & -- \\
  001209 & 100.00597 & 9.51826 & w& 223965280 & N & 0.64 &  &  & &  \\
  001217 & 100.15274 & 9.86756 & c& 616849540 & B & 1.30 & 7.865 & $<$1E-4& 0.59 & -- \\
  001218 & 100.11910 & 9.59657 & w& 616919855 & U & 0.62 & 4.592 & $<$1E-4& -- & -- \\
  001219 & 100.11435 & 9.87510 & c& 616849516 & N & 0.23 &  &  & & -- \\
  001221 & 100.09760 & 9.91543 & c& 616826682 &  & 0.22 & 8.221 & $<$1E-4 & 0.56 & -- \\
  001222 & 100.17577 & 9.81426 & & 605538554 & U & & &  & & -- \\
  001223 & 100.02502 & 9.82851 & c& 616849613 & QPS & 0.36 & 8.113 & $<$1E-4& 0.27 & -- \\
  001226 & 100.06400 & 9.71178 & w& 602091889 & N & 0.70 &  &  & & -- \\
  001232 & 100.15773 & 9.83083 & w& 602095744 & U & 0.31 &  &  & & -- \\
  001234 & 100.04634 & 9.63499 & c& 223968039 & QP & 0.94 & 9.606 & $<$1E-4& 0.62 &  \\
  001236\tablefootmark{\#} & 100.09258 & 9.90801 & w& 223971008 & MP & 0.61 & 7.204 & $<$1E-4& & 7.38 \\
  001236\tablefootmark{\#} & 100.09258 & 9.90801 & w& 223971008 & MP & 0.61 & 2.396 & $<$1E-4& & "" \\
  001238 & 99.96290 & 9.60913 & w& 616896186 & P & 0.43 & 1.363 & $<$1E-4& 0.08 & -- \\
  001239 & 99.91183 & 9.86440 & w& 223958794 & N & 1.47 &  &  & &  \\
  001240 & 100.10996 & 9.94970 & c& 616826701 &  & & 1.792 & 2.0E-2& -- & -- \\
  001247 & 100.12060 & 9.70475 & w& 223972691 & QPS & 0.93 & 7.344 & $<$1E-4& 0.28 & 7.206 \\
  001248 & 99.94468 & 9.68164 & w& 223961132 & P & 1.39 & 3.842 & $<$1E-4& 0.02 & 3.839 \\
  001249 & 100.08446 & 9.93510 & c& 616826670 & QP & 0.30 & 1.954 & $<$1E-4& 0.42 & -- \\
  001250 & 100.08611 & 9.57623 & w& 603402460 & QPS & 0.75 & 2.385 & $<$1E-4& 0.60 & -- \\
  001251 & 100.13262 & 9.60018 & w& 616895982 & U & 0.33 &  &  & & -- \\
  001254 & 100.07084 & 9.77590 & w& 602091890 & P & 0.63 & 0.779 & $<$1E-4& 0.11 & -- \\
  001256 & 99.92321 & 9.57794 & w& 223959652 & P & 1.36 & 3.65 & $<$1E-4& 0.10 & 3.732 \\
  001259 & 100.14511 & 9.76257 & w& 616872718 & N & 0.32 &  &  & & -- \\
  001261\tablefootmark{11} & 100.07197 & 9.42894 & w& 602079822 &  & 0.25 &  & & & -- \\
  001264 & 100.12756 & 9.76961 & w& 602091893 & P & 1.32 & 7.171 & $<$1E-4& -0.12 & 7.151 \\
  001265 & 100.04535 & 9.66871 & w& 616896095 & U & 0.28 & 12.261 & $<$1E-4& 1.43 & -- \\
  001271 & 100.01113 & 9.69690 & w& 616896061 & N & 0.32 &  &  & & 10.0 \\
  001274 & 100.17260 & 9.80267 & w& 605539508 & P & 1.20 & 4.712 & $<$1E-4& 0.08 & 4.743 \\
  001275 & 100.11978 & 9.51669 & c& 223972652 & S & 1.60 &  &  & &  \\
  001277 & 100.11050 & 9.58946 & w& 602083887 &  & 0.44 & 12.206 & $<$1E-4& 4.99 & -- \\
  001278 & 100.17415 & 9.83120 & w& 616849549 & P & 0.63 & 1.05 & $<$1E-4& 0.41 & 1.049 \\
  001279 & 100.13016 & 9.51862 & w& 223973292 & P & 0.91 & 1.975 & $<$1E-4& 0.04 & 1.974 \\
  001281 & 99.84457 & 9.28438 & w& 223953966 & MP & 1.60 & 4.132 & $<$1E-4& & 3.987 \\
  001281 & 99.84457 & 9.28438 & w& 223953966 & MP & 1.60 & 0.254 & 5.0E-1\tablefootmark{d}& & "" \\
  001281 & 99.84457 & 9.28438 & w& 223953966 & MP & 1.60 & 0.084 & & & "" \\
  001284 & 100.05605 & 9.32427 & w& 223968646 & N & 1.10 &  &  & &  \\
  001286 & 100.04946 & 9.35282 & w& 223968235 & P & 0.45 & 5.423 & $<$1E-4& 0.11 & -- \\
  001290 & 99.87659 & 9.56040 & w& 223956264 & P & 0.70 & 2.248 & $<$1E-4& 0.10 & 2.229 \\
  001291 & 99.96803 & 9.31930 & w& 223962712 & N & 0.45 &  &  & &  \\
  001292 & 99.94881 & 9.43520 & w& 223961409 & MP & 1.48 & 1.025 & 9.0E-3& 1.07 & 1.104 \\
  001292 & 99.94881 & 9.43520 & w& 223961409 & MP & 1.48 & 1.15 & 1.0E-2& & "" \\
  001294 & 100.02303 & 9.37390 & c& 616970063 & QP & 0.92 & 6.723 & 1.0E-4 & 0.48 & -- \\
  001295 & 100.06584 & 9.35924 & w& 602075331 & P & 0.45 & 0.643 & $<$1E-4& 0.45 & -- \\
  001296 & 99.76562 & 9.67317 & c& 223948127 & QPD & 0.69 & 9.725 & $<$1E-4& 0.30 &  \\
  001298 & 100.05850 & 9.34126 & w& 223968804 & QPS & 1.45 & 1.292 & $<$1E-4& 0.23 & 1.295 \\
  001300 & 99.90011 & 9.40728 & w& 223957908 & N & 0.68 &  &  & &  \\
  001302 & 99.81543 & 9.49149 & w& 223951822 & N & 0.68 &  &  & &  \\
  001303\tablefootmark{§} & 100.01678 & 9.45196 & w& 223965989 & MP? & 0.64 & 0.821 & $<$1E-4& & 0.819 \\
  001304 & 100.05672 & 9.41372 & c& 223968688 & Be & 2.17 & 1.081 & $<$1E-4& & 1.117 \\
  001304 & 100.05672 & 9.41372 & c& 223968688 & Be & 2.17 & 1.127 & $<$1E-4& & "" \\
  001306 & 99.86619 & 9.47752 & & 223955517 & U & &  &  & &  \\
  001307 & 99.84556 & 9.60470 & w& 223954040 & QPS & 1.17 & 9.585 & $<$1E-4& 0.06 & 9.684 \\
  001308 & 99.99685 & 9.45679 & c& 223964667 & QPD & 0.63 & 6.717 & $<$1E-4& 0.40 & 6.456 \\
  001309 & 99.92771 & 9.53093 & w& 223959949 & N & 0.45 &  &  & &  \\
  001310 & 99.87232 & 9.34970 & w& 223955994 & U & 1.29 &  &  & &  \\
  001312 & 99.86491 & 9.38590 & & 223955438 & N & &  &  & &  \\
  001313 & 100.15139 & 9.31597 & w& 223974593 & Be & 0.45 & 1.156 & $<$1E-4& 0.28 & 1.156 \\
  001313 & 100.15139 & 9.31597 & w& 223974593 & Be & 0.45 & 0.906 & $<$1E-4& & "" \\
  001359 & 100.27631 & 9.49189 & w& 223982169 & QPS & 0.29 & 3.181 & $<$1E-4& -0.17 & 3.162 \\
  001386 & 99.82098 & 9.97093 & w& 223952236 & N & 1.20 & &  & &  \\
  001388 & 99.88739 & 9.94156 & w& 223957004 & N & 1.10 & &  & &  \\
  001389 & 99.95082 & 9.98490 & & 223961560 & U & & &  & &  \\
  001573 & 100.05239 & 10.09457 & c& 223968439 & S & 0.62 &  &  & & 8.688 \\
  001579 & 100.14630 & 10.07272 & w& 223974272 & QPS & 0.69 & 1.34 & $<$1E-4& 0.23 & -- \\
  001581 & 100.39892 & 10.07105 & w& 602103885 &  & 0.27 & & & & -- \\
  001588 & 100.37020 & 10.15404 & & 223988020 & U & &  &  & &  \\
  001590 & 100.41213 & 10.15986 & & 605537061 & P & & 3.175 & $<$1E-4& 0.36 & -- \\
  001594 & 100.07184 & 10.22646 & & 223969672 & Be & & 0.800 & $<$1E-4& &  \\
  001594 & 100.07184 & 10.22646 & & 223969672 & Be & & 0.902 & $<$1E-4& &  \\
  001596 & 100.28734 & 10.23947 & & 223982807 & N & &  &  & &  \\
  001597 & 100.25637 & 10.24891 & & 223980941 & U & & 3.819 & $<$1E-4& 0.50 & 3.794 \\
  001598 & 100.35009 & 10.24228 & c& 602113781 & U & &  &  & & -- \\
  001599 & 100.06796 & 10.31211 & c& 616735324 & U & & 0.946 & 6.0E-3& -- & -- \\
  001610 & 99.97960 & 9.36463 & w& 602075320 & N & 0.44 &  &  & & -- \\
  001612 & 100.24800 & 9.49767 & c& 616943997 &  & 0.30 & &  & & 9.34 \\
  001618 & 100.44000 & 9.65865 & w& 603408592 &  & 0.30 &  &  & & -- \\
  001627 & 100.28042 & 10.22539 & & 223982423 &  & &  &  & & 9.026 \\
  001628 & 100.17600 & 9.81436 & & 605538554 & N & & &  & & -- \\
  005009 & 100.54130 & 9.79835 & c& 602091907 & U & 0.32 & &  & & -- \\
  005589 & 100.16910 & 9.46370 & w& 603396398 &  & 0.30 & 12.519 & $<$1E-4& $\sim$0 & -- \\
  005664 & 100.22704 & 9.15886 & c& 223979150 & QP & 0.45 & 1.192 & $<$1E-4& $<$0 & -- \\
  005745 & 100.51808 & 9.16136 & c& 617022483 &  & 0.32 & 15.946 & $<$1E-4& & -- \\
  005836 & 100.37126 & 9.30428 & c& 602075360 &  & 0.34 &  &  & & -- \\
  006037 & 99.87232 & 9.72772 & w& 223955976 &  & 0.45 & 3.261 & $<$1E-4& 0.16 & -- \\
  006079 & 99.87131 & 9.71071 & c& 223955882 & EB & 0.45 & 0.511 & $<$1E-4& 0.05 & -- \\
  006324 & 99.95312 & 9.29311 & c& 616996779 &  & 0.36 &  &  & & -- \\
  006325 & 100.06033 & 9.22703 & c& 616996720 &  & 0.54 & 0.956 & 7.0E-3& -- & -- \\
  006465 & 99.85485 & 9.54393 & c& 223954720 & EB & 0.88 & 2.829 & 9.0E-1\tablefootmark{b}& 0.29 & -- \\
  006491\tablefootmark{*} & 99.85626 & 9.52761 & c& 616920065 & Be? & 1.08 & 2.452 & 3.0E-4 & & -- \\
  006491\tablefootmark{*} & 99.85626 & 9.52761 & c& 616920065 & Be? & 1.08 & 2.75 & 5.0E-4 & & -- \\
  006491\tablefootmark{*} & 99.85626 & 9.52761 & c& 616920065 & Be? & 1.08 & 2.271 & 2.5E-3& & -- \\
  006930 & 99.76704 & 9.27055 & c& 223948224 &  & 0.65 &  &  & & -- \\
  006991 & 99.84170 & 10.10648 & c& 223953770 &  & 0.53 &  &  & & -- \\
  007004 & 99.85813 & 10.08542 & c& 223954942 &  & 0.80 &  &  & & -- \\
  014132\tablefootmark{12} & 99.76481 & 9.27110 & c& 602070634 & QPD & 0.28 & 9.102 & $<$1E-4& 0.24 & -- \\
\end{longtable}
\tablefoot{
\tablefoottext{1}{Coordinates from the 2MASS survey.}
\tablefoottext{2}{``c'' = CTTS; ``w'' = WTTS.}
\tablefoottext{3}{{\it CoRoT} light curve morphology class (cf. Appendix~\ref{app:light_curve_types}; \citealp{cody2014}): ``B'' = burster; ``U'' = unclassifiable variable type;``S'' = stochastic; ``N'' = non-variable; ``D'' = dipper; ``QPS'' = quasi-periodic symmetric; ``QPD'' = quasi-periodic dipper; ``P'' = periodic; ``MP'' = multi-periodic; ``EB'' = eclipsing binary; ``L'' = long-timescale variable; ``Be'' = beats.}
\tablefoottext{4}{False Alarm Probability = fraction of times a periodogram power higher than that corresponding to the extracted period occurs, at the same frequency, among 10\,000 ``noise-like'' light curves built by dividing the original light curve in 12\,h-long segments and reassembling them in random order.}
\tablefoottext{5}{Ratio of the amounts of effective light curve rms (i.e., $rms^2 - \sigma^2$) measured after and before subtracting the periodic trend from the light curve (see definition in \citealp{cody2014}). A ``--'' indicates that the computed value of $Q$, associated with the period listed, is not reported because affected by systematics or by an erroneous estimate of the photometric uncertainty on the light curve. $Q$ is not reported for multi-periodic (MP) objects.}
\tablefoottext{6}{From \citet{affer2013}. Blank space = aperiodic; ``--'' = object not present in \citeauthor{affer2013}'s sample. N.B.: multiple periods are not investigated in the study of \citet{affer2013}.}
\tablefoottext{7}{A period of 3.91~d is reported for this object in the analysis of \citet{mcginnis2015} and adopted in \citet{sousa2016}. The LSP obtained for this object presents a first peak at P=3.91~d and a second, slightly lower peak at P=7.83~d. Conversely, the ACF analysis presents a slightly higher feature close to P=8~d, and a strong indication of periodicity at P$\sim$7.8~d is conveyed by the SL method, whereas no significant indication of periodicity at P$\sim$4~d appears in the latter. The light curve of this object appears as a sequence of dips which vary considerably in shape and especially depth along the monitored interval of time; however, the phased light curve at P=7.83~d suggests that dips might come in pairs, alternating deeper and shallower minima, which may correspond, e.g., to a primary and a secondary opposite warps in the inner disk. Based notably on the SL result, we report here the longer periodicity of 7.83~d, although no decisive evidence in either direction can be achieved from our data.}
\tablefoottext{8}{A periodic signal is present, but the waveform is strongly variable.}
\tablefoottext{9}{The periodicity is only detected in part of the light curve.}
\tablefoottext{10}{The results of the statistical tests do not strongly support the period detection here, but a visual inspection of the light curve indeed suggests the presence of a periodic pattern.}
\tablefoottext{11}{The light curve seems to suggest a long periodicity of about 24.5~d, but this is beyond what can be accurately probed here with our time coverage.}
\tablefoottext{12}{Case similar, and more evident, to that of CSIMon-000296. The light curve exhibits a clear alternation of deeper and shallower minima, with the former being more jagged and the latter being sharper in shape; each pair of dips may represent two opposite warps in the inner disk. The periodogram provides a period indication at P=4.48~d, which is the value reported in \citet{mcginnis2015} for this object; conversely, a clear indication of periodicity at P=9.102~d is provided by both the ACF and the SL analysis. We therefore consider this latter period estimate to provide a better match to the observed light curve for CSIMon-014132.}
\tablefoottext{*}{More uncertain period estimate.}
\tablefoottext{\#}{Spectroscopic binary.}
\tablefoottext{§}{Cases with possible (uncertain) additional periodicities: CSIMon-000018 (3.612~d); CSIMon-000090 (3.571~d, 4.783~d); CSIMon-000217 (7.463~d); CSIMon-000220 (7.675~d); CSIMon-000525 (hint of an additional long periodicity at 24~d, but this is beyond what can be accurately probed here with our time coverage); CSIMon-000705 (9.389~d, only seen in the first part of the light curve); CSIMon-001067 (3.581~d); CSIMon-001303 (7.771~d).}
\tablefoottext{a}{Jumps in the light curves affect the periodogram analysis.}
\tablefoottext{b}{The periodogram peaks at a value corresponding to the half-period, whereas the correct periodicity is identified using the ACF and the SL methods.}
\tablefoottext{c}{The periodogram peaks at a value corresponding to half that reported as period here; more ambiguous case.}
\tablefoottext{d}{The light curve shows a spurious long-term trend that severely affects the LSP diagnostics for the detection of the short periodicity.}
}
}

\twocolumn

\appendix

\section{Classes of variables among NGC~2264 members} \label{app:light_curve_types}

\citet{cody2014} provide a nice illustration of the diversity of light curve morphologies observed across the disk-bearing sample of NGC~2264 members. Here we extended their classification to the full set of members. A brief description of different morphology classes and of possible physical interpretations is presented in the following.

\begin{itemize}
\item {\it Burster (B)}: light curve exhibiting sudden, rapid (0.1--1 day) rises in flux, followed by decreases on comparable timescales. Accretion instabilities appear to be driving the behavior displayed by this group of objects \citep{stauffer2014}.
\item {\it Dipper (D)}: light curve characterized by transient optical fading events, possibly linked with extinction by circumstellar material. In some cases (e.g., AA~Tau; \citealp{bouvier2007a}), these events may recur periodically (likely resulting from warps located in the inner disk), although displaying changes in depth and/or shape (quasi-periodic dipper, QPD) from one cycle to the next.
\item {\it Spotted (P)}: flux variations dominated by surface spot modulation. Light curves may exhibit a definite, stable pattern over thousands of rotational cycles \citep{rotor_wtts, venuti2015}. 
\item {\it Multi-periodic (MP)}: beating-like light curves or superposition of separate timescales of modulation. Several physical processes, such as differential rotation, spot evolution/migration, stellar pulsations may contribute to objects in this class.
\item {\it Eclipsing binary (EB)}: the light variation trend due to spot modulation and/or disk occultation is interspersed with periodic eclipses when one of the two stars in the binary system passes in front of the other  during its orbital motion \citep[e.g.,][]{gillen2014}.
\item {\it Stochastic (S)}: light curves exhibiting prominent flux changes on a variety of timescales, with no preference for fading or brightening events and no obvious periodicity. Time-dependent accretion onto the star, resulting in transient hot spots, may drive the observed variability for this class of objects \citep{stauffer2016}.
\item {\it Long-timescale variable (L)}: variability for these objects grows or declines systematically up to the longest timescale of observation. These long timescales of variability may reflect disk dynamics beyond the inner edge.
\end{itemize}

\section{Impact of bin size and phase on the shape and features of the period histogram} \label{app:hist_bin}

\begin{figure*}[t]
\centering
\includegraphics[width=\textwidth]{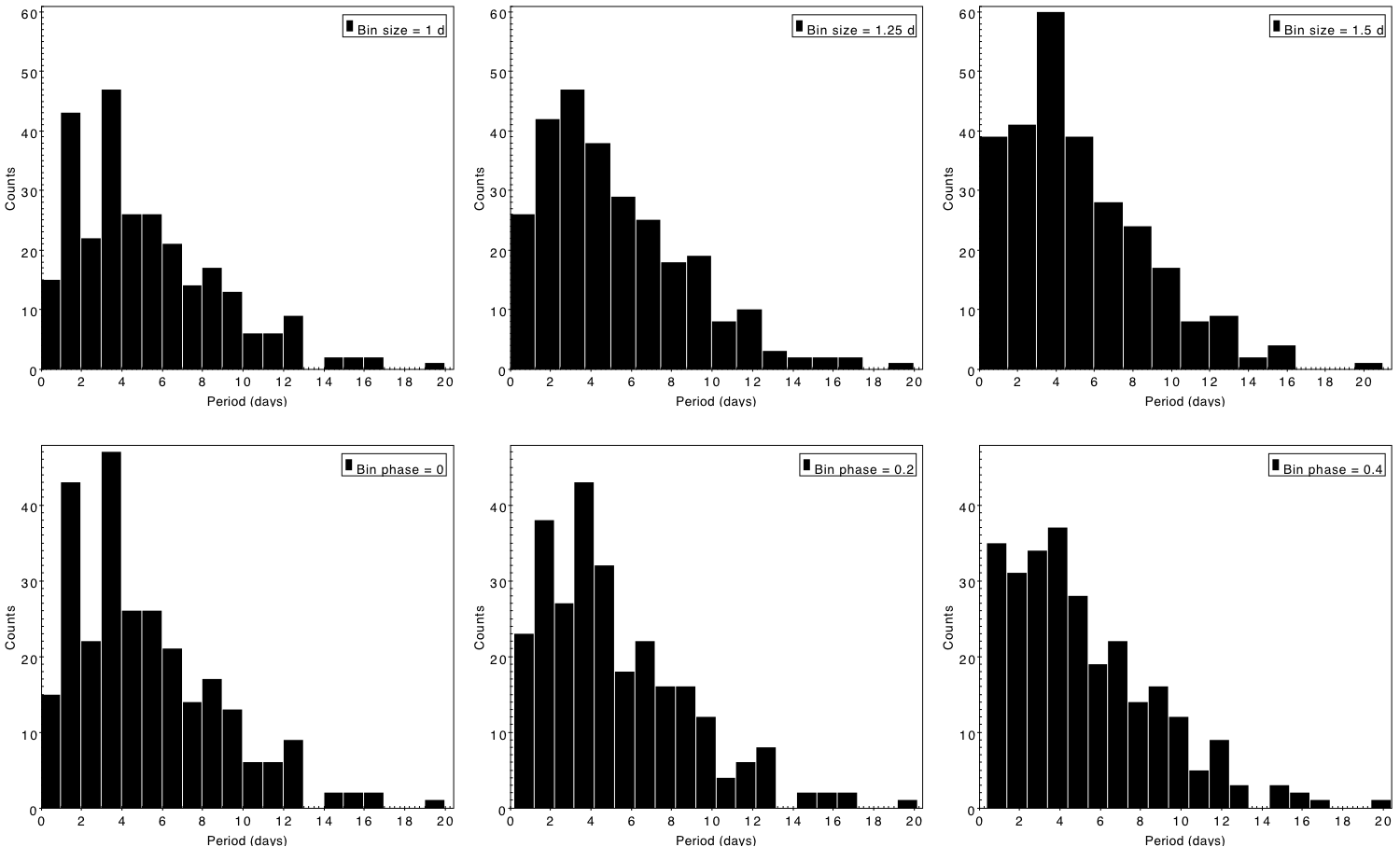}
\caption{Illustration of how the shape and features of the period histogram in Fig.\,\ref{fig:hist} would evolve when varying the bin size (upper panels) or the bin phase (lower panels). The starting bin size and phase (of 1~d and 0, respectively; these are shown on the upper left panel and on the lower left panel of this Figure) correspond to those adopted in Fig.\,\ref{fig:hist} and considered for the subsequent analysis. In the upper panel, the bin size is progressively increased by 0.25~d at each step from the left diagram to the right diagram; in the lower panel, a fixed bin size of 1~d is adopted and the bin center is progressively shifted by 0.2~d at each step from the left diagram to the right diagram.}
\label{fig:hist_bin_size_phase}
\end{figure*}

While in Figure~\ref{fig:hist} and throughout the main paper text we use the commonly adopted bin size of 1~d and bin phase of 0 (i.e., bin scheme starting from 0.0~d) for the period histogram, in this Appendix we explore how the histogram shape would be affected by different choices in bin size and phase. This is illustrated in Fig.\,\ref{fig:hist_bin_size_phase}.

The upper panel of Fig.\,\ref{fig:hist_bin_size_phase} shows how the observed histogram shape and features evolve when varying the bin size. A smaller bin size of 0.5~d (not shown here) would produce an histogram with the same broad features and better resolved peaks than that shown in Fig.\,\ref{fig:hist} and reported in the upper left diagram of Fig.\,\ref{fig:hist_bin_size_phase}. Conversely, increasing the bin size would determine the peak at shorter periods to progressively merge with the peak at longer periods; this trend can be seen when reading the upper panel of Fig.\,\ref{fig:hist_bin_size_phase} from the left to the right.

Similarly, the lower panel of Fig.\,\ref{fig:hist_bin_size_phase} shows the change in shape of the period histogram when shifting the bin centers along the period axis. Small shifts (from the left to the middle panel in the lower part of Fig.\,\ref{fig:hist_bin_size_phase} do not have a strong impact on the global properties of the period histogram: hints of two peaks can still be observed when we adopt a bin phase of 0.2. However, as discussed in Sect.\,\ref{sec:per_dist}, larger shifts, comparable to the width of a single peak, would determine these to redistribute into two neighboring histogram channels, hence transforming the distribution into a flatter distribution with no significant peaks between 0 and 5~d (lower right panel of Fig.\,\ref{fig:hist_bin_size_phase}.

\section{Discrepant period estimates between this work and \citet{affer2013}} \label{app:V16_A13}

\begin{figure*}[t]
\centering
\includegraphics[width=\textwidth]{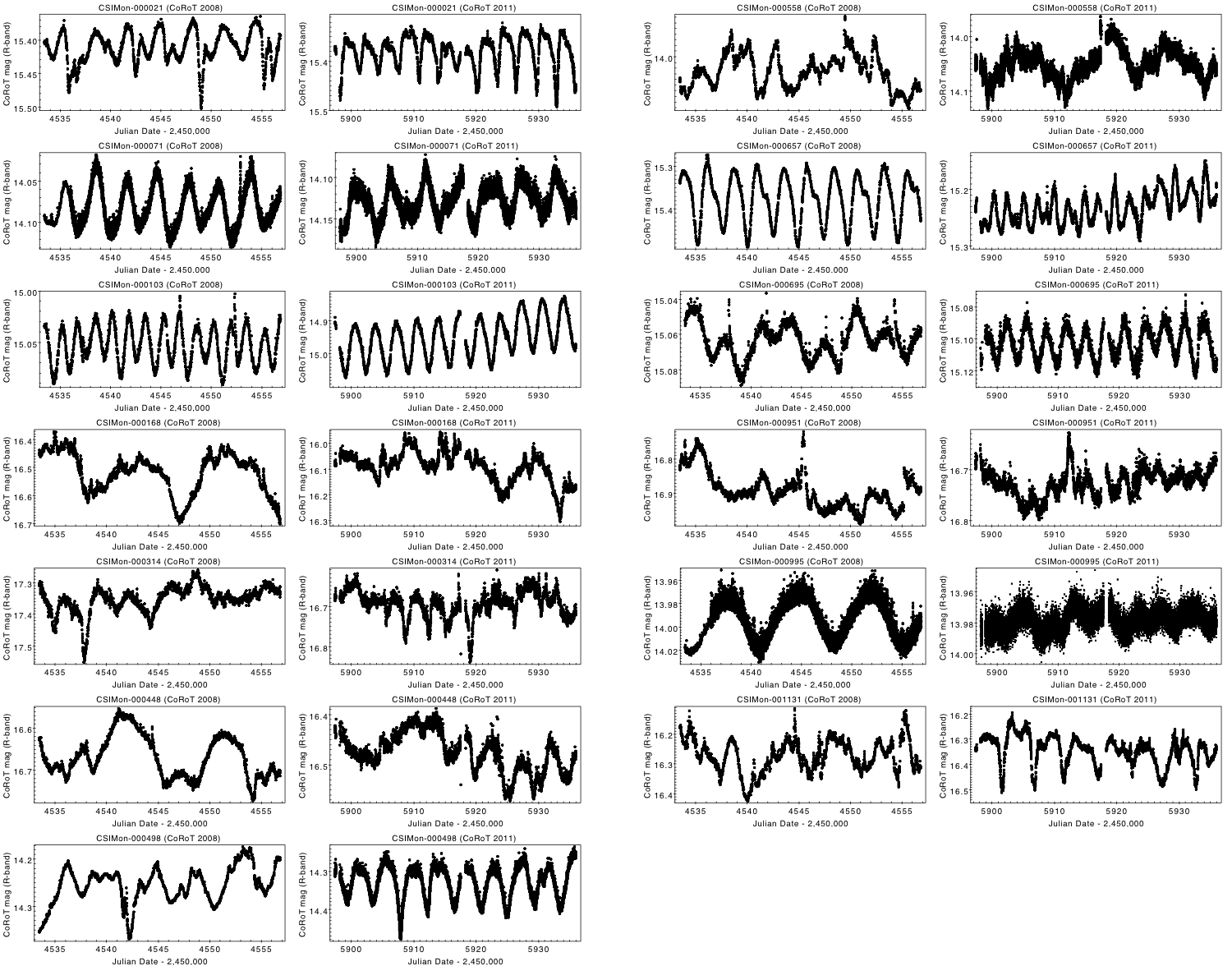}
\caption{Comparison of the light curves obtained with {\it CoRoT} in 2008 (left) and 2011 (right) for objects with discrepant period estimates between \citet{affer2013} and this study.}
\label{fig:A13_V16_outliers}
\end{figure*}

\begin{figure*}[t]
\centering
\includegraphics[width=\textwidth]{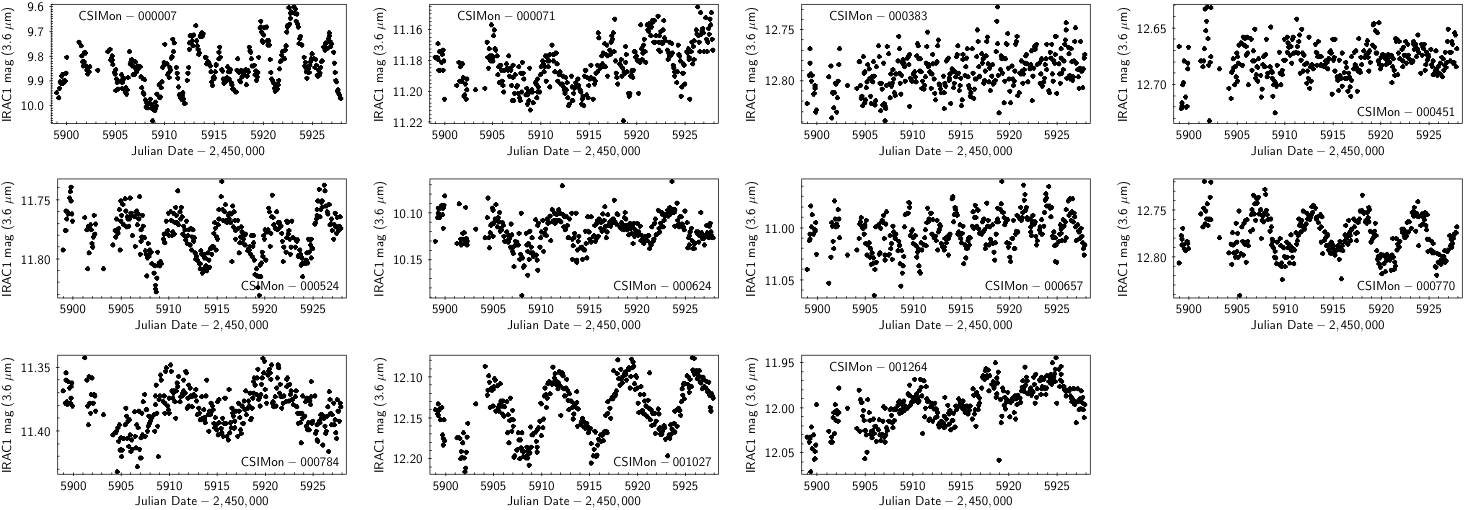}
\caption{Mid-IR light curves, obtained with the {\it Spitzer}/IRAC 3.6~$\mu$m channel, for some of the sources discussed in Appendices~\ref{app:V16_A13} and \ref{app:V16_C07}. A detailed description of {\it Spitzer}/IRAC data obtained within the CSI~2264 campaign is presented in \citet{cody2014}.}
\label{fig:IRAC1_lc_app}
\end{figure*}

As shown in Fig.\,\ref{fig:P_comp}, in a small number of cases the period estimates derived in the present study are in contrast with previous period estimates reported in the literature. In this Appendix, we illustrate and discuss the objects individually that are common to our sample and \citeauthor{affer2013}'s (\citeyear{affer2013}) sample, that are outliers with respect to the equality line on the left panel of Fig.\,\ref{fig:P_comp}. A detailed comparison of the light curves on which the period analysis was performed in the two studies is shown in Fig.\,\ref{fig:A13_V16_outliers}. For comparison purposes, in some of the cases examined in this and in the next Appendix we will also refer to the preliminary results of a similar analysis of rotation that is being conducted on the mid-IR time series photometry obtained with {\it Spitzer}/IRAC during the CSI~2264 campaign (Rebull et al., in prep.). The relevant light curves at 3.6~$\mu$m are shown in Fig.\,\ref{fig:IRAC1_lc_app}. 

\paragraph{CSIMon-000021} 
The period reported in \citet{affer2013} is about twice that reported in the present study. This object is classified as a narrow dipper in \citet{stauffer2015}, and fading events recur in a regular pattern at a period of 3.15~d. The {\it CoRoT} light curve obtained in 2008 instead appears as a superposition of a spot modulation component, which has a visual period close to the value measured here, and of a dip component, which seems to occur at a longer periodicity of about twice the spot modulation periodicity. 

\paragraph{CSIMon-000071}
The period reported in \citet{affer2013} is about half that found in the present study. The light curve ``unit'' from the {\it CoRoT} run of 2008 appears to be M-shaped, with the secondary minimum only slightly less deep than the first minimum. Conversely, the 2011 light curve has a single minimum and a period of 5.41~d, consistent with the preliminary results obtained from the period analysis of CSI~2264 {\it Spitzer/IRAC} light curves (Rebull et al., in preparation).

\paragraph{CSIMon-000103}
The period reported here is about twice the period in \citet{affer2013}, but is consistent with the period estimate reported in \citet{lamm04}. Similar to the case of CSIMon-000071, the {\it CoRoT} light curve observed in 2008 for this object had two minima in a period unit; as illustrated in Fig.\,\ref{fig:half_per}, a periodogram-based analysis of such cases may incur in an erroneous period detection at the half period.

\paragraph{CSIMon-000168}
This is a long-period object, with a period estimate of 10.02~d from this study and of 8.61~d from \citet{affer2013}. The two period estimates are only marginally inconsistent when we consider the associated uncertainty from Eq.\,\ref{eqn:P_err}; note that no error estimate is reported in \citet{affer2013}. As shown in Fig.\,\ref{fig:A13_V16_outliers}, at both epochs the light curve is not entirely smooth.

\paragraph{CSIMon-000314}
The period reported in \citet{affer2013} is about twice the period reported here. As illustrated in Fig.\,\ref{fig:A13_V16_outliers}, the light curve of this object is not very well behaved at any epochs. No other period estimates are available for this object from other datasets in the CSI~2264 campaign or from previous studies in the literature.

\paragraph{CSIMon-000448}
The period reported in \citet{affer2013} is about twice the period reported here. As in the case on CSIMon-000314, the light curve is not entirely smooth.

\paragraph{CSIMon-000498}
The period reported in \citet{affer2013} is about twice the period reported here. As shown in Fig.\,\ref{fig:A13_V16_outliers}, the flux modulation appears to be more regular at the 2011 epoch than at the 2008 epoch. \citet{lamm04} also found a period value close to the one reported here.

\paragraph{CSIMon-000558}
Case similar to CSIMon-000168: long-period object, with measured period of 11.71~d in this study and of 10.17~d in \citet{affer2013}. The two estimates are only marginally inconsistent within the error estimated on our derivation of period.

\paragraph{CSIMon-000657}
The period reported in \citet{affer2013} is about half the period reported here. \citet{lamm04} report a period estimate consistent with that of \citet{affer2013}, and preliminary results from the period analysis of {\it Spitzer/IRAC} light curves appear to agree with these. Fig.\,\ref{fig:A13_V16_outliers} shows that the 2008 {\it CoRoT} light curve may actually consist of two separate, alternating features of slightly different shape; this would imply that the periodogram peak indicated in \citet{affer2013} corresponds to half the actual rotation rate.

\paragraph{CSIMon-000695}
The period reported in \citet{affer2013} is about thrice that found here. Fig.\,\ref{fig:A13_V16_outliers} shows that a modulation is well seen in the {\it CoRoT} 2011 light curve, whereas the pattern is more fragmented on the 2008 light curve. 

\paragraph{CSIMon-000951}
The period reported in \citet{affer2013} is significantly larger than that found in this study. As can be observed on Fig.\,\ref{fig:A13_V16_outliers}, the light curve is partly irregular at any epochs. We note that the period estimate reported in \citet{lamm04} is consistent with the one that we report here.

\paragraph{CSIMon-000995}
Another object with long periodicity; \citet{affer2013} report P = 7.02~d, while we derive here P = 9.28~d. The light curve comparison shown in Fig.\,\ref{fig:A13_V16_outliers} illustrates that the modulated pattern was better traced in 2008.

\paragraph{CSIMon-001131}
The period reported in \citet{affer2013} is about twice the period we report here. The light curve comparison shown in Fig.\,\ref{fig:A13_V16_outliers} illustrates that the flux variations were better behaved in 2011.

\section{Discrepant period estimates between this work and \citet[][after \citealp{lamm04} and \citealp{makidon04}]{cieza2007}} \label{app:V16_C07}

\begin{figure*}[t]
\centering
\includegraphics[width=0.77\textwidth]{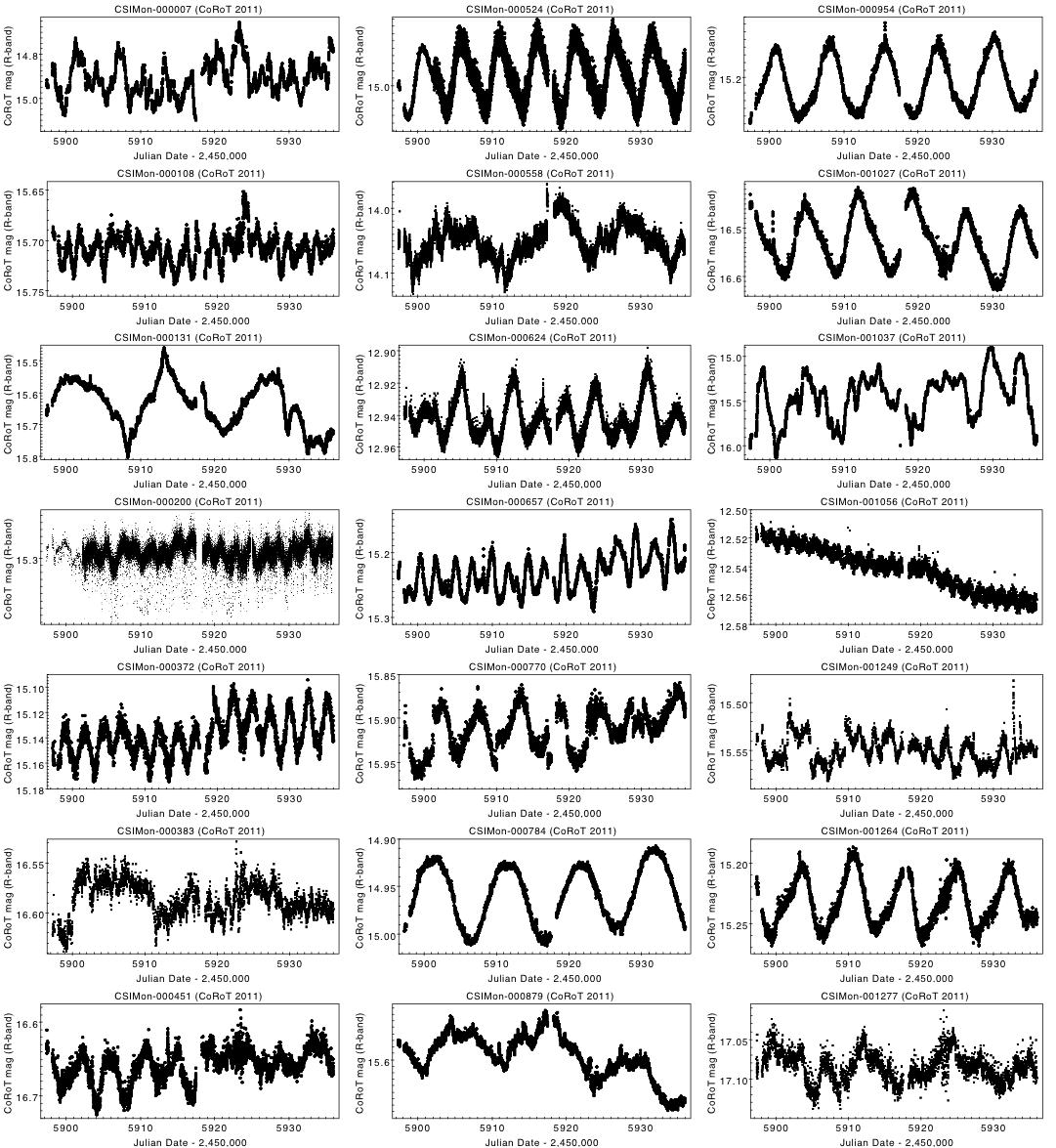}
\caption{{\it CoRoT} 2011 light curves for objects whose period estimate derived in this study are in disagreement with the periods reported in \citet{cieza2007}.}
\label{fig:CB07_V16_outliers}
\end{figure*}

Fig.\,\ref{fig:CB07_V16_outliers} illustrates the {\it CoRoT} light curves analyzed in this study for objects whose derived period is in disagreement with the estimate reported in \citet{cieza2007}. In the following, these discrepant cases are discussed individually. For some of them, we also discuss indications of periodicity deduced from CSI~2264 {\it Spitzer}/IRAC photometry; the relevant light curves at 3.6~$\mu$m are shown in Fig.\,\ref{fig:IRAC1_lc_app}. 

\paragraph{CSIMon-000007}
The period reported in \citet{cieza2007} for this object is close to 1~d (0.93~d), whereas it has a period of 3.19~d in our study (see Table \ref{tab:periods}). The light curve profile is partly irregular (this is one of the objects classified as dominated by stochastic accretion bursts by \citealp{stauffer2014}); preliminary results from the analysis of the {\it Spitzer/IRAC} dataset (Rebull et al., in preparation) would also suggest a periodicity of 3.17~d.

\paragraph{CSIMon-000108}
The period reported in \citet{cieza2007} for this object is of 1.37~d, whereas it has a period of 4.06~d in this study. The light curve unit appears to have two maxima; indication for the period reported in this study derives primarily from the ACF and SL methods.

\paragraph{CSIMon-000131}
The period reported in \citet{cieza2007} is about half that found in this study (6.4~d and 12.87~d, respectively). A long-period modulation can be seen in the {\it CoRoT} 2011 light curve, although it has an irregular variability component superimposed. 

\paragraph{CSIMon-000200}
The period reported in \citet{cieza2007} is about twice that found in this study (3.88~d and 1.92~d, respectively). The light curve exhibits a fairly small amplitude of variability, but a modulation effect is clearly seen in the {\it CoRoT} light curve. The light curve unit might be M-shaped (with two maxima), but that is not entirely evident from the light variation pattern and the derived period diagrams; for this reason, we opted here for the shorter period.

\paragraph{CSIMon-000372}
The period reported in \citet{cieza2007} is about half that found in this study (1.3~d and 2.57~d, respectively). A modulation at a period of about 2.5~d is clearly seen on the {\it CoRoT 2011} light curve shown in Fig.\,\ref{fig:CB07_V16_outliers}, and this result is supported by the analysis of the former {\it CoRoT} dataset by \citet{affer2013}.

\paragraph{CSIMon-000383}
The period reported in \citet{cieza2007} is about half that found in this study (0.51~d and 1.03~d, respectively). The {\it CoRoT} 2011 light curve for this object is affected by instrument systematics, which may impact the results of the period analysis. Preliminary results of the analysis of {\it Spitzer/IRAC} light curves (Rebull et al., in preparation) seem to support the period value reported in \citet{cieza2007}, after \citet{makidon04}.

\paragraph{CSIMon-000451}
The period reported in \citet{cieza2007} for this object is 0.68~d, whereas it has a period of 4.52~d in our study (see Table \ref{tab:periods}). A modulation effect of several days is well observed during the first half of the {\it CoRoT} 2011 light curve, although it becomes more irregular during the second fraction of the monitored time span. A close periodicity to the one we report here is suggested by preliminary results of the {\it Spitzer/IRAC} light curve analysis (Rebull et al., in prep.).

\paragraph{CSIMon-000524}
The period reported in \citet{cieza2007} for this object is 1.23~d, whereas a period of 5.15~d is found here. The periodic pattern is clearly outlined in the {\it CoRoT} light curve, and the period found here is supported by the {\it Spitzer/IRAC} light curves (Rebull et al., in prep.)

\paragraph{CSIMon-000558}
The period reported in \citet{cieza2007} for this object is close to 1~d (0.88~d), whereas it has a period of 11.71~d in our study (see Table \ref{tab:periods}). A long-term modulation can be clearly observed on the {\it CoRoT} 2011 light curve, and is supported by \citeauthor{affer2013}'s (\citeyear{affer2013}) results. No significant evidence of shorter periodicities results from our analysis.

\paragraph{CSIMon-000624}
The period reported in \citet{cieza2007} is about half that found in this study (3.73~d and 7.47~d, respectively). The {\it CoRoT} 2011 light curve appears to be the alternation of taller and shorter maxima; the indication for the period reported here derives from the ACF and SL methods, whereas the periodogram peaks at half that value (as discussed for the case in Fig.\,\ref{fig:half_per}). A similar period to the one reported here is suggested by the preliminary results of the {\it Spitzer/IRAC} light curve analysis (Rebull et al., in prep.).

\paragraph{CSIMon-000657}
The period reported in \citet{cieza2007} is about half that found in this study (2.43~d and 4.83~d, respectively). \citet{affer2013} also report a period consistent with that listed in \citet{cieza2007} (see discussion about this object in Appendix \ref{app:V16_A13}). The light curve unit in the {\it CoRoT} 2011 dataset appears to have a complex and time-varying shape, with several maxima. The value of period reported here is based on the ACF and SL diagnostic tools; the periodogram peaks at half its value.

\paragraph{CSIMon-000770}
The period reported in \citet{cieza2007} for this object is close to 1~d (0.84~d), whereas it has a period of 5.44~d in our study (see Table \ref{tab:periods}). The same value of period that we find here is suggested by preliminary {\it Spitzer/IRAC} results (Rebull et al., in prep.), and this was also reported in \citet{affer2013} from the analysis of the previous {\it CoRoT} run on NGC~2264. 

\paragraph{CSIMon-000784}
The period reported in \citet{cieza2007} for this object is close to 1~d (0.91~d), whereas it has a period of 10.10~d in our study (see Table \ref{tab:periods}). The modulation is clearly seen in the {\it CoRoT} 2011 light curve (see Fig.\,\ref{fig:CB07_V16_outliers}; the same value of period is suggested by {\it Spitzer/IRAC} data.

\paragraph{CSIMon-000879}
The period reported in \citet{cieza2007} for this object is close to 1~d (0.91~d), whereas it has a period of 11.40~d in our study (see Table \ref{tab:periods}). The light curve for this object is not very regular, but a long-period modulation can be detected fairly clearly by eye (see Fig.\,\ref{fig:CB07_V16_outliers}).

\paragraph{CSIMon-000954}
The period reported in \citet{cieza2007} for this object is close to 1~d (0.88~d), whereas it has a period of 7.35~d in our study (see Table \ref{tab:periods}). The modulation pattern with a periodicity of several days is very clearly seen in the {\it CoRoT} 2011 light curve.

\paragraph{CSIMon-001027}
The period reported in \citet{cieza2007} for this object is close to 1~d (1.15~d), whereas it has a period of 7.22~d in our study (see Table \ref{tab:periods}). As in the previous case, the modulation pattern with a periodicity of several days is very clearly seen in the {\it CoRoT} 2011 light curve. A very similar result is derived from the analysis of {\it Spitzer/IRAC} light curves (Rebull et al., in prep.).

\paragraph{CSIMon-001037}
The period reported in \citet{cieza2007} for this object is 12.09~d, whereas a period of 8.88~d is found here. The actual light curve pattern is fairly irregular, which may affect the accuracy of the derived period for this object, although a global effect of modulation with long periodicity can be detected. No periodicity was detected for this object in the study of \citet{affer2013}.

\paragraph{CSIMon-001056}
The period reported in \citet{cieza2007} is about half that found in this study (0.78~d and 1.52~d, respectively). The {\it CoRoT} 2011 light curve has a small amplitude, superimposed over a spurious long-term trend, but the modulated pattern can be clearly seen.

\paragraph{CSIMon-001249}
The period reported in \citet{cieza2007} is about twice that found in this study (3.87~d and 1.95~d, respectively). The {\it CoRoT} 2011 light curve is partly affected by instrument systematics.

\paragraph{CSIMon-001264}
The period reported in \citet{cieza2007} for this object is close to 1~d (0.88~d), whereas it has a period of 7.17~d in our study (see Table \ref{tab:periods}). A modulated pattern with a periodicity of several days is clearly seen for this object on the {\it CoRoT} 2011 light curve. A close periodicity to the one we report here was found by \citet{affer2013}, and it also appears from the preliminary results from {\it Spitzer/IRAC} light curves (Rebull et al., in prep.).

\paragraph{CSIMon-001277}
The period reported in \citet{cieza2007} for this object is 5.29~d, whereas a period of 12.21~d is found here. The light curve in the {\it CoRoT} 2011 dataset is not very well behaved, which may impact our period determination for this object.

\section{Objects which appear to have evolved from periodic to aperiodic or vice versa between the two {\it CoRoT} runs} \label{app:per_aper}

About 5\% of objects in our sample (Table~4), found to be periodic here, were reported as non-periodic in the study of \citet{affer2013}. Similarly, about 5\% of objects in our sample, classified as non-periodic in Table~4, had a periodicity assigned in \citet{affer2013}. 

\begin{figure*}[t]
\centering
\includegraphics[width=\textwidth]{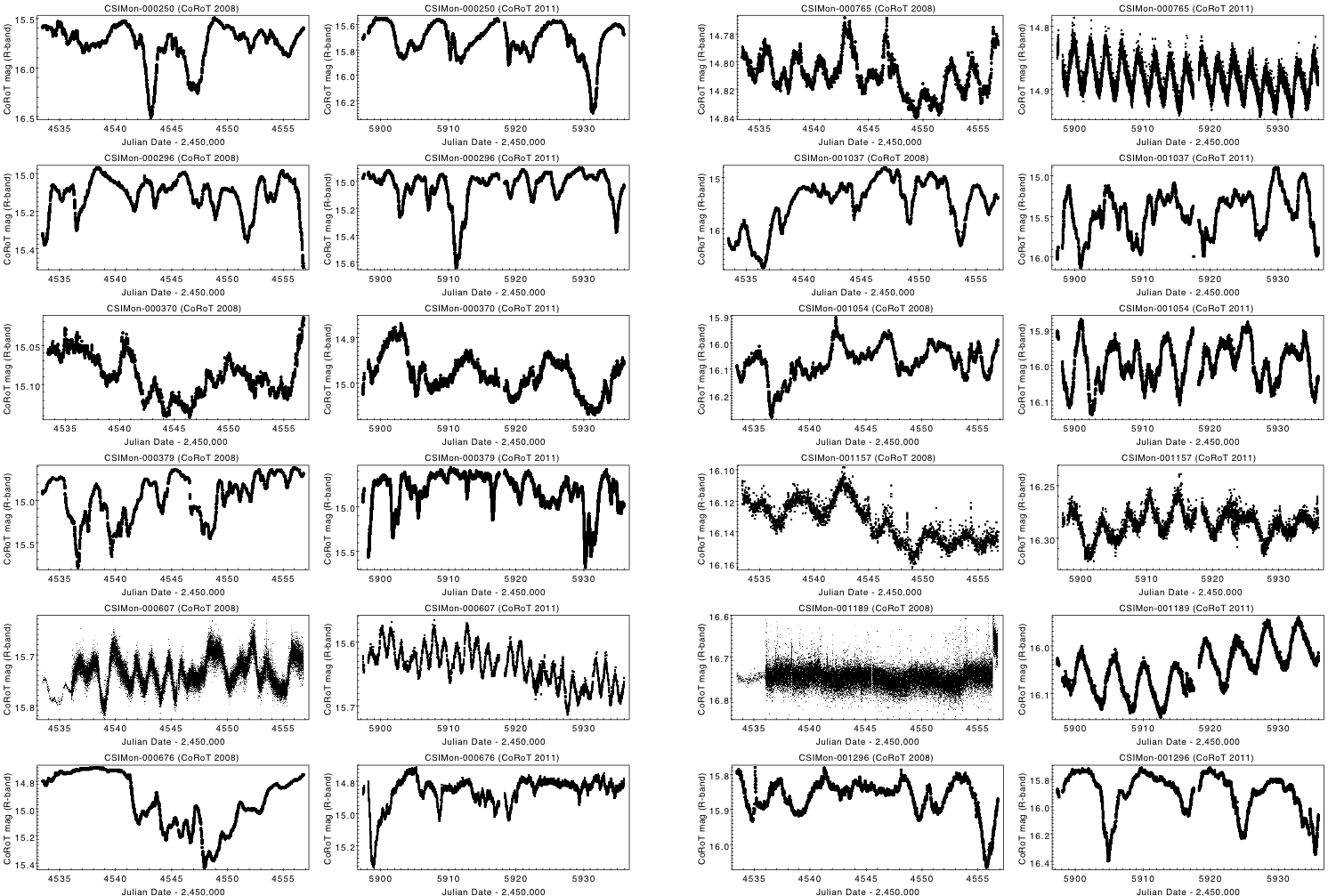}
\caption{Twelve cases of objects (six on the left side, six on the right side of the picture) that had aperiodic {\it CoRoT} light curves in 2008 (periodicity data from \citealp{affer2013}) but are assigned a periodicity here based on their 2011 {\it CoRoT} light curves. Each object corresponds to a pair of panels: the first illustrates the {\it CoRoT} 2008 light curve; the second illustrates the {\it CoRoT} 2011 light curve. Cases shown are, from top to bottom: CSIMon-250, 296, 370, 379, 607, 676 (left side); CSIMon-765, 1037, 1054, 1157, 1189, 1296 (right side).}
\label{fig:aper_per_fin}
\end{figure*}

\begin{figure*}[t]
\centering
\includegraphics[width=\textwidth]{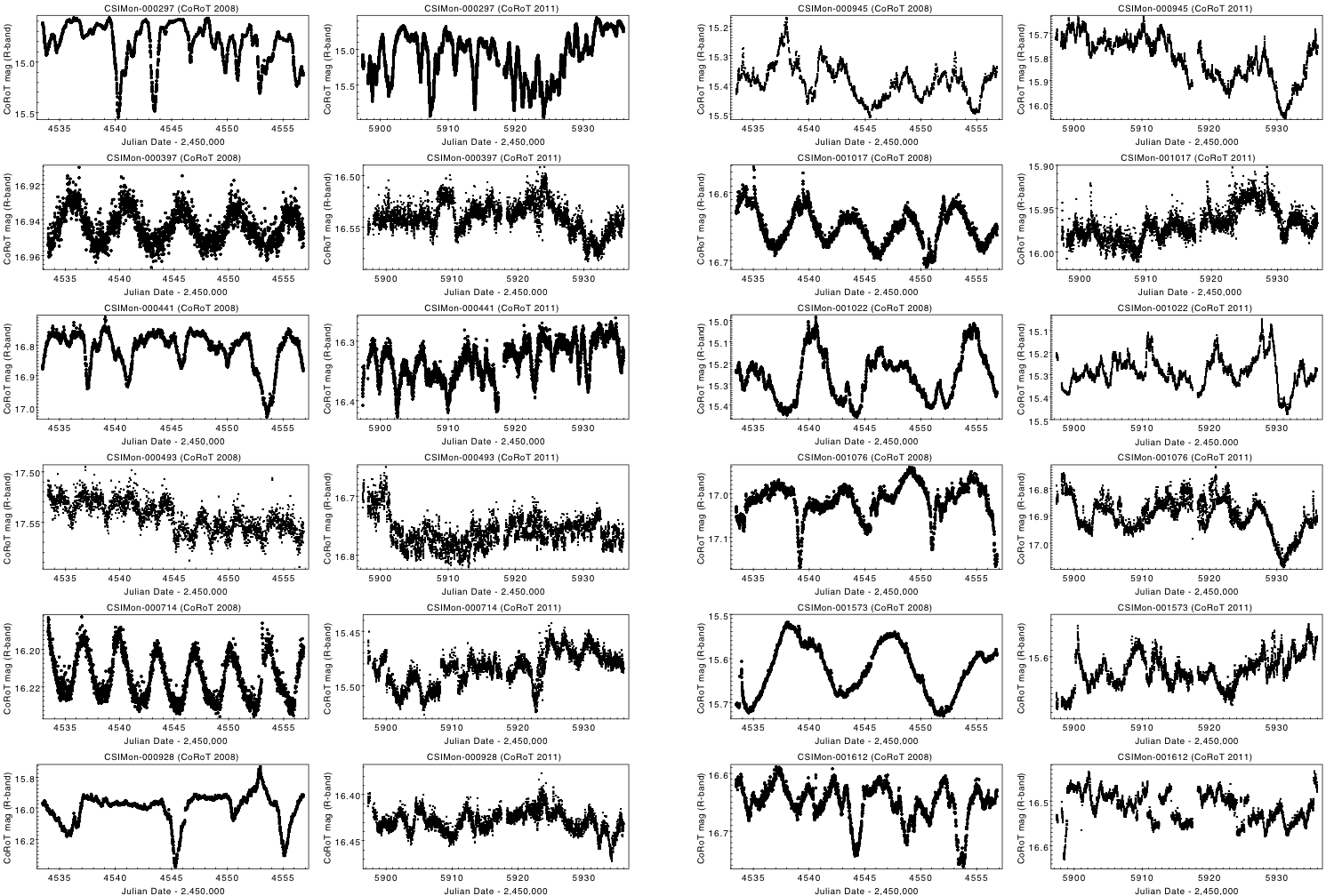}
\caption{Twelve cases of objects (six on the left side, six on the right side of the picture) that had periodic {\it CoRoT} light curves in 2008 (periodicity data from \citealp{affer2013}) but are classified as aperiodic here based on their 2011 {\it CoRoT} light curves. Each object corresponds to a pair of panels: the first illustrates the {\it CoRoT} 2008 light curve; the second illustrates the {\it CoRoT} 2011 light curve. Cases shown are, from top to bottom: CSIMon-297, 397, 441, 493, 714, 928 (left side); CSIMon-945, 1017, 1022, 1076, 1573, 1612 (right side).}
\label{fig:per_aper_fin}
\end{figure*}

Some of these cases can be attributed to the fact that their light curves exhibit a certain degree of irregularity, or that the observed flux variations develop on fairly long timescales ($\sim$weeks); these situations translate to more unclear period assessment. In other cases, the light curves at one or the other epoch are affected by systematics, that hampers the period analysis. For a fraction of cases, however, the discrepant result on the light curve periodicity/non-periodicity at the two epochs reflects a real evolution in photometric behavior between the two epochs.

Figure~\ref{fig:aper_per_fin} illustrates twelve cases of objects that exhibited aperiodic light curves in 2008 but have a period measured from the 2011 epoch and reported in Table~4. We can identify two main groups among the examples shown. The first (e.g., the CTTS CSIMon-000370 and CSIMon-000765, respectively third panel on the left and first panel on the right side of the picture, from the top) consists of objects with nicely modulated light curves at one epoch and more irregular light curve shapes at the other epoch. These light curve changes may be driven by a variation in the accretion activity of the objects: more intense in the first epoch, with light curves dominated by a changing mix of cold magnetic spots and hot accretion spots at the stellar surface \citep[e.g.,][]{herbst1994}; more moderate at the second epoch, when well-behaved cold spot modulation prevails in the light curve. The second group of cases is exemplified by objects CSIMon-000296 and CSIMon-001296 (CTTS; second panel on the left and last panel on the right side of the picture, from the top); these exhibit a dipper-like light curve (flat luminosity maximum interspersed by flux dips associated with extinction events from circumstellar material; \citealp{mcginnis2015}). The extinction events, possibly linked to inner disk warps at the base of the accretion funnels, occur aperiodically at one epoch and periodically at the other; as discussed in \citet{mcginnis2015}, this may be due to a transition between unstable and stable accretion regimes. Finally, special mention goes to the object CSIMon-001189 (fifth panel from the top on the right side of Fig.\,\ref{fig:aper_per_fin}, a WTTS whose nearly flat-band light curve observed in 2008 (uniform spot distribution + instrumental noise?) evolved into a smooth modulated pattern recorded in 2011.

Similarly, Figure~\ref{fig:per_aper_fin} illustrates twelve cases of objects with detected periodicity in \citet{affer2013} from 2008 light curves, that appear to be aperiodic in the 2011 epoch. Again, we can identify the two main types of photometric behaviors and evolution discussed in the previous paragraph. In some cases, e.g., the CTTS CSIMon-001573 (fifth panel from the top on the right side of the picture), the predominantly modulated light curve pattern observed in 2008 evolved into a more irregular, possibly hot-spot dominated flux variation trend in 2011. In other cases, e.g., the CTTS CSIMon-000928 (last panel from the top on the right side of the figure), the periodic, AA~Tau-like dipper profile observed at the 2008 epoch evolved into an aperiodic light curve trend in 2011.

\end{document}